%Paper: hep-th/9209066
%From: sumathi%theory%tifrvax.BITNET@cearn.cern.ch
%Date: Fri, 18 Sep 92 22:31:54 IST
%Date (revised): Mon, 21 Sep 92 10:52:23 IST

%This file needs to be processed using the macro phyzzx.tex

\input phyzzx
\input epsf

\voffset = 24pt
\overfullrule=0pt
\scrollmode
\hsize = 6.4in
\vsize = 8.0in

%{}~\hfill\vbox{\hbox{TIFR/TH/92-18}\hbox{hepth@xxx/9209066}
%\hbox{April,1992}}
%\hfill {April,1992}

\title{AN ANYON PRIMER}
%\foot{based on lectures delivered at the VII
%SERC school at Physical Research Laboratory, Ahmedabad, 30 December 1991
%- 18 January 1992.}

\author{Sumathi Rao\foot{E-mail address : sumathi@mri.ernet.in}
\foot{On leave from Institute of Physics, Sachivalaya Marg, Bhubaneswar,
751005, India}}
\address{Mehta Research Institute, 10 Kasturba Gandhi Marg,\break 
Old Kutchery Road, Allahabad 211002, India}

\let\refmark=\NPrefmark \let\refitem=\NPrefitem
\def\define#1#2\par{\def#1{\Ref#1{#2}\edef#1{\noexpand\refmark{#1}}}}
\def\con#1#2\noc{\let\?=\Ref\let\<=\refmark\let\Ref=\REFS
         \let\refmark=\undefined#1\let\Ref=\REFSCON#2
         \let\Ref=\?\let\refmark=\<\refsend}

\define\ANYSUP
Y. H. Chen, F. Wilczek, E. Witten and B. I. Halperin, Int. Jour. Mod. Phys.
{\bf B3}, 1001 (1989).

\define\LEINMYR
J. M. Leinaas and J.Myrrheim, Il Nuovo Cimento,
{\bf 37},1 (1977).

\define\FHL
A. L. Fetter, C. B. Hanna and R. B. Laughlin, Phys. Rev. {\bf B39}, 9679
(1989).

\define\WU
Y. S. Wu, Phys. Rev. Lett. {\bf 52}, 2103 (1984).

\define\WILCZEK
F. Wilczek, Phys. Rev. Lett. {\bf 48}, 1144 (1982); ibid,{\bf 49}, 957 (1982).

\define\SMHO
A. Comtet, Y. Georgelin and S. Ouvry, J. Phys. {\bf A22}, 3917 (1989).

\define\HIGHTC
J. G. Bednorz and K. A. Muller, Z. Phys. {\bf B64}, 189 (1986).

\define\AROVAS
D. P. Arovas, R. Schrieffer, F. Wilczek and A. Zee, Nucl. Phys. {\bf B251},
117 (1985).

\define\EXAMPLE
S. K. Paul and A. Khare, Phys. Lett. {\bf B174}, 420 (1986).

 \define\LGCS
 S. M. Girvin and A. H. Macdonald, Phys. Rev. Lett. {\bf 58}, 1252 (1987);
 S. C. Zhang, T. H. Hannson and S. A. Kivelson, Phys. Rev. Lett. {\bf 62},
82 (1989); N. Read, ibid {\bf 62}, 86 (1989).

\define\WUWU
Y. S. Wu, Phys. Rev. Lett. {\bf 53}, 111(1984); M. D. Johnson and 
G. S. Canright, Phys. Rev.{\bf B41}, 6870 (1990); A.P. Polychronokas,
Phys. Lett.{\bf B264}, 362 (1990); C. Chou, Phys. Rev.{\bf D44}, 2533 (1991),
Erratum,{\it ibid} {\bf D45}, 1433(1992); G. Dunne, A. Lerda, S. Sciuto
and C. A. Trugenberger, Nucl. Phys. {\bf B 370}, 601 (1992); J. Grundberg,
T. H. Hannson, A. Karlhede and E. Westerberg, Phys. Rev. {\bf B44}, 8373
(1991); K. H. Cho and C. H. Rim, Ann. Phys. ({\it N.Y.} {\bf 213}, 295
(1992); S. V. Mashkevich, Int. Jnl of Mod. Phys. {\bf A7}, 7931 (1992).  

\define\KHARE
J. Law, M. K. Srivastava, R. K. Bhaduri and A. Khare, Jnl. of Phys.
{\bf A25}, L183 (1992).

\define\KIVELSON
S. Kivelson, Phys. Rev. Lett. {\bf 65}, 3369 (1990).

\define\MURTHYETAL
M. Sporre, J. J. M. Verbaarschot and I. Zahed, Phys. Rev. Lett. {\bf 67},
1813 (1991);  M. V. N. Murthy, J. Law, M. Brack and R. K. Bhaduri,
Phys. Rev. Lett. {\bf 67}, 1817 (1991);  M. V. N. Murthy et al, Phys. Rev.
{\bf B45}, 4289 (1992);
M. Sporre, J. J. M. Verbarschot and I. Zahed, Phys. Rev.  {\bf B46}, 5738
(1992).

\define\DIPTIMAN
D. Sen, Phys. Rev. Lett.{\bf 68}, 2977 (1992);  Phys. Rev.{\bf D46}, 1846
(1992);  R. Chitra, C. Nagaraja Kumar and D. Sen, Mod. Phys. Lett. {\bf
A7}, 855 (1992);  Ranjan K. Ghosh and S. Rao, Institute of Physics
preprint, IP/BBSR/95-29. 

\define\HUANG
See any book on stastical mechanics, -$e.g.$, Statistical Mechanics by K.Huang,
(John Wiley \& Sons, Inc.,1963).

\define \THIRDVIRIAL
R. K. Bhaduri et al, Phys. Rev. Lett. {\bf 66}, 523 (1991);
A. D. de Veigy and S. Ouvry, Nucl. Phys. {\bf B388}, 715 (1991);
R. Emparan and M. A. Valle, Mod. Phys. Lett. {\bf A8},3291 (1993).

\define\EXACT
D. M. Gaitonde and S. Rao, Int. Jnl of Mod. Phys. {\bf B6}, 3543 (1992). 

\define\WENZEE
X. G. Wen and A. Zee, Phys. Rev. {\bf B41}, 240 (1990).

\define\BANKSLYKKEN
T. Banks and J. Lykken, Nucl. Phys. {B 336}, 500 (1990).

\define\FTAS
S. Randjbar-Daemi, A. Salam and J. Strathdee,  Nucl. Phys. {\bf B340}, 403
(1990);  J. E. Hetrick, Y. Hosotani and B. -H. Lee, Ann. Phys. (N.Y.) {\bf
209} 151 (1991).

\define\RATSTAT
D. M. Gaitonde and S. Rao,  Phys. Rev. {\bf B44}, 929 (1991).

\define\ASTESTS
X. G. Wen and A. Zee, Phys. Rev. Lett. {\bf 62}, 2873 (1989);
B. I. Halperin, J. March-Russell and F. Wilczek, Phys. Rev. {\bf B40}, 8726
(1990);
P. Lederer, D. Poilblanc and T. M. Rice, Phys. Rev. Lett. {\bf 63},
1519 (1989);
D. M. Gaitonde and S. Rao, Mod. Phys. Lett. {\bf B4}, 1143 (1990);
J. E. Hetrick and Y. Hosotani, Phys. Rev. {\bf B45}, 2981(1992).

\define\ASFQHE
F. Wilczek, Fractional Statistics and Anyon Superconductors, (World
Scientific, Singapore 1990).

\define\LAYER
A. G. Rojo and G. S. Canright, Phys. Rev. Lett. {\bf 66}, 949 (1991);
A. G. Rojo and A. J. Leggett, Phys. Rev. Lett. {\bf 67}, 3614 (1991).

\define\CSL
D. M. Gaitonde, D. P. Jatkar and S. Rao,
Phys. Rev. {\bf B46}, 12026 (1992).

\define\PTEXPT
R. Kiefl et al, Phys. Rev. Lett. {\bf 64}, 2082 (1990);
K. Lyons et al, Phys. Rev. Lett. {\bf 64}, 2949 (1990);
S. Spielman et al, Phys. Rev. Lett. {\bf 65}, 123 (1990);
H. J. Weber et al, Solid State Comm. {\bf 76}, 511 (1990).

\define\FROHLICH
J. Frohlich  and P. A. Marchetti, Commun. Math. Phys. {\bf 121}, 177
(1989).

\define\CSREFS
S. Deser, R. Jackiw and S. Templeton, Phys. Rev. Lett. {\bf 48}, 975
(1982); Ann. Phys. {\bf 140}, 372 (1984); J. Schonfeld, Nuc. Phys. {\bf
B156}, 135 (1982);
A. J. Niemi and G. W. Semenoff, Phys. Rev. Lett. {\bf 51}, 2077(1983);
A. N. Redlich, Phys. Rev. Lett. {\bf 52}, 18 (1984);
ibid, Phys. Rev. {\bf D29}, 2366 (1984); M. B. Paranjape, Phys. Rev. Lett.
{\bf 55}, 2390 (1985);
R. D. Pisarski and S. Rao, Phys. Rev. {\bf D32}, 2081
(1986); S. Rao and R. Yahalom, Phys. Lett. {\bf B172}, 227
(1986);
R. Blankenbecler and D. Boyanovsky, Phys. Rev. {\bf D34}, 612 (1986).

\define\WITTEN
 E. Witten, Commun. Math. Phys. {\bf 117}, 353 (1988);ibid {\bf 118}, 411
(1988); ibid {\bf 121}, 351 (1989).

\define\NIELSEN
H. Nielsen and P. Olesen, Nucl. Phys. {\bf B61}, 45 (1973).

\define\DILEEP
D. P. Jatkar and A. Khare, Phys. Lett. {\bf B236}, 283 (1990).

\define\MOOREREAD
G. Moore and N. Read, Nucl. Phys. {\bf B360}, 362 (1991); M. Greiter, X.
G. Wen and F. Wilczek, Nucl. Phys. {\bf B374}, 567 (1992); B. Blok and X.
G. Wen, Nucl. Phys. {\bf B374}, 615 (1992).

\define\VONKLITZING
K. von Klitzing, G. Dorda and M. Pepper, Phys. Rev. Lett. {\bf 45}, 494
(1980).

\define\FQHEBOOK
The Quantum Hall Effect (Edited by R. Prange and S. Girvin), Springer, New
York (1987).

\define\TSUI
D. C. Tsui,H. L. Stormer and A. C. Gossard, Phys. Rev. Lett. {\bf 48},
1559 (1982).

\define\ASW
D. P. Arovas, J. R. Schrieffer and F. Wilczek, Phys. Rev. Lett. {\bf 53},
722 (1984).

\define\LAUGHLIN
R. B. Laughlin, Phys. Rev. Lett. {\bf 50}, 1395 (1983); See also R. B.
Laughlin in \FQHEBOOK.

\define\JAIN
J. K. Jain, Phys. Rev. Lett, {\bf 63}, 199 (1989); J. K. Jain, J. Phys.
Chem. Solids, {\bf 51}, 889 (1990).

\define\HALDHALP
F. D. M. Haldane, Phys. Rev. Lett. {\bf 51}, 605 (1983); B. I. Halperin,
Phys. Rev. Lett. {\bf 52}, 1583 (1984); {\bf 52}, 2390 (E).

\define\READ
S. M. Girvin and A. H. Macdonald, Phys. Rev. Lett. {\bf 58}, 1252 (1987);
S. C. Zhang, T. H. Hansson and S. Kivelson, Phys. Rev. Lett. {\bf 62}, 82
(1989); N. Read, Phys. Rev. Lett. {\bf 62}, 86 (1989).

\define\LGCSTWO
B. Blok and X. G. Wen, Phys. Rev. {\bf B42}, 8133 (1990);
ibid, {\bf B42}, 8145 (1990);
D. H. Lee and S. C. Zhang, Phys. Rev. Lett. {\bf 66}, 1220 (1191); D. H.
Lee, S. Kivelson and S. C. Zhang, Phys. Rev. Lett. {\bf 67}, 3302 (1991).

\define\CHITRA
R. Chitra and D. Sen, Phys. Rev. {\bf B46}, 10923 (1992).

\define\MCCABE
J. McCabe and S. Ouvry, Phys. Lett. {\bf B260}, 113 (1991); A. Khare 
and J. McCabe, Phys. Lett. {\bf B269}, 330 (1991); G. Amelino-Camelia,
Phys. Lett. {\bf B} 286, 329 (1992).

\define\JAINRAO
J. K. Jain and S. Rao, Mod. Phys. Lett. {\bf B9}, 611 (1995).

\define\GODDARD
P. Goddard and D. I. Olive, Rep. Prog. Phys. {\bf 41}, 91 (1978).

\define\DYWITTEN
E. Witten, Phys. Lett. {\bf 86B}, 283 (1979).

\define\PRANGE
R. E. Prange, Phys. Rev. {\bf B23}, 4802 (1981).

\def\bpone{{\bf p}_1}
\def\bptwo{{\bf p}_2}
\def\brone{{\bf r}_1}
\def\brtwo{{\bf r}_2}
\def\baone{{\bf a}_1}
\def\batwo{{\bf a}_2}
\def\barel{{\bf a}_{{\rm rel}}}
\def\psirel{\psi_{\rm rel}}
\def\psirelprime{\psi'_{\rm rel}}
\def\ba{{\bf a}}
\def\baprime{{\bf a}'}
\def \bnabla{{\bf \nabla}}

\def\pThetasquared{{\partial^2\over\partial\Theta^2}}
\def\qphi{{q\phi\over 2\pi}}
\def\lqphi{{l +\qphi}}
\def\prsquared{{{\partial^2}\over{\partial r^2}}}
\def\proverr{{1\over r} {\partial\over\partial r}}
\def\pRsquared{{{\partial^2}\over{\partial R^2}}}
\def\pRoverR{{1\over R} {\partial\over\partial R}}

\def\psicm{\psi_{\rm CM}}
\def\Rrelr{{\cal R}_{\rm rel}(r)}

 \def\b{\beta}
 \def\ep{\epsilon_{\bf p}}
 \def\np{n_{\bf p}}
 \def\zbp{z e^{-\b\ep}}
 \def\sump{\sum_{\bf p}}
 \def\prodp{\prod_{\bf p}}
 \def\sumN{\sum_{N=0}^{\infty}}
 \def\dtwor{d^2 {\bf r}_1 .....d^2  {\bf r}_N}
 \def\dtwop{d^2 {\bf p}_1 .....d^2  {\bf p}_N}
 \def\norm{{1\over N!\lambda^{2N}}}
 \def\br{{\bf r}}
 \def\bp{{\bf p}}
 \def\bR{{\bf R}}
 \def\bP{{\bf P}}
 \def\e{{\epsilon}}
 \def\a{{\alpha}}
 \def\zbpm{z e^{-\b\bp^2/2m}}
 \def\d{\delta}

\def\bp{{\bf p}}
\def\ba{{\bf a}}
\def\br{{\bf r}}
\def\pqa{{(\bp_i - q\ba_i)^2 \over 2m}}
\def\rij{(\br_i - \br_j)}
\def\sij{\sum_{i\ne j}}
\def\s1N{\sum_{i=1}^N}
\def\mrij{|\br_i - \br_j|^2}
\def\ptp{{\phi\over 2\pi}}

\def\rhobar{{\bar \rho}}
\def\psixy{\psi_{p_x,p_y}(x,y)}
\def\Exy{E_{p_x,p_y}}
\def\qbptp{{q(b+B)\over 2\pi}}

 \def\emna{\epsilon_{\mu\nu\alpha}}
\def\jmam{j_{\mu}a^{\mu}}
\def\CS{(\mu/2) \epsilon_{\mu\nu\alpha} a^{\mu}\partial^{\nu} a^{\alpha}}
\def\e0ij{\epsilon_{0ij}}
\def\p{\partial}
\def\Sex{S_{\rm ex}}
\def\Srot{S_{\rm rot}}
\def\SAB{S_{\rm AB}}
\def\xss{{\dot {\bf r}}_{\alpha}^2}
\def\xs{{\dot {\bf r}}_{\alpha}}
\def\pa{{\bf p}_{\alpha}}

\def\ba{{\bf a}}
\def\bx{{\bf x}}
\def\suma{\sum_{\alpha}}
\def\brprime{{\bf r}'}
\def\rtheta{(r \rightarrow \infty, \theta)}
\def\bA{{\bf A}}
\def\br{{\bf r}}

\def\rxx{\rho_{xx}}
\def\ryy{\rho_{yy}}
\def\rxy{\rho_{xy}}
\def\ryx{\rho_{yx}}

\def\sxx{\sigma_{xx}}
\def\syy{\sigma_{yy}}
\def\sxy{\sigma_{xy}}
\def\syx{\sigma_{yx}}

\def\denom{(\omega^2 \tau^2 + 1)}

\def\exp{e^{-\Sigma_i |z_i|^2/4l^2}}
\def\expone{e^{-(|q|/e)\Sigma_i |z_{0i}|^2/4l^2}}
\def\br{{\bf r}}
\def\dis{\displaystyle}

\def\bai{{\bf a}_i}
\def\bpi{{\bf p}_i}
\def\bA{{\bf A}}
\def\ba{{\bf a}}
\def\bri{{\bf r}_i}
\def\brj{{\bf r}_j}
\def\brij{(\bri - \brj)}
\def\brmodij{|\bri -\brj|}
\def\phid{\phi^{\dagger}}

\def\sumi{\sum_{i=1}^N}
 \def\sumij{\sum_{i<j}}

 \def\brone{{\bf r}_1}
 \def\brtwo{{\bf r}_2}
 \def\brN{{\bf r}_N}

\def\psiN{\psi(\brone,\brtwo,...\brN)}
\def\phiN{\phi(\brone,\brtwo,...\brN)}
\def\phitN{{\tilde \phi}(\brone,\brtwo,...\brN)}
\def\p{\partial}

\abstract

In this set of lectures, we give a pedagogical introduction to the subject
of anyons. We discuss 1) basic concepts in anyon physics, 2) quantum
mechanics of two anyon systems, 3) statistical mechanics of many anyon
systems, 4) mean field approach to many anyon systems and anyon
superconductivity , 5) anyons in field theory and 6) anyons in the
Fractional Quantum Hall Effect (FQHE).

\endpage

This set of lectures is aimed at an audience who may be hearing the
term `anyon' for the first time. We shall start by explaining what
the term `anyon' means and why they are interesting. Just as fermions
are spin 1/2, 3/2,...., particles obeying Fermi-Dirac statistics
and bosons are spin 0,1,....,particles obeying Bose-Einstein statistics,
`anyons' are particles with `any' spin obeying `any' statistics. One may
wonder why such particles have not been seen until now. After all, had
they really existed in nature, they would have been just as familiar as
the usual fermions and bosons. The explanation is quite simple. As will be
seen in the course of the lectures, even theoretically, anyons can only
exist in two space dimensions, whereas the real world is three dimensional.
This naturally leads to the next question, `why bother to study them at
all ?'.
The answer is
that there do exist phenomena in our three dimensional world that are
planar - systems where motion in the third dimension is essentially
frozen. Anyons are relevant in the explanation of such phenomena. Besides,
the study of anyons has led to a considerable improvement in our theoretical
understanding of concepts like quantum statistics.

The theoretical possibility of anyons was put forward as early as
1977\LEINMYR. However, anyons shot into prominence and became a major
field of research only in the last few years. One reason for this upsurge
of interest  was the discovery that the experimentally observed FQHE\TSUI\
had a natural explanation\LAUGHLIN\HALDHALP\
in terms of anyons. An even more dramatic rise in its
popularity occured when it was discovered that a gas of anyons
superconducts\FHL\ANYSUP,
when it is coupled to electromagnetism. In fact, for a while,
`anyon superconductivity'  was one of the top "candidate" theories to
explain high $T_c$ superconductors\HIGHTC.
Now, due to lack of experimental
confirmation, the theory is no longer a `hot' candidate, but interest in
the field of anyons remains as high as ever.
The rest of these lectures will involve a more detailed elaboration on the
theme of anyons.

We shall first start in Sec.(1)
with basic notions of spin and statistics and understand why
anyons can only exist in two spatial dimensions\LEINMYR\WU.
Then we shall study a
simple physical model of an anyon that incorporates fractional spin and
statistics\WILCZEK. In Sec.(2), using this model, we shall solve some
simple two anyon quantum
mechanics problems, and see that the anyon energy eigenvalues
actually interpolate
between fermionic and bosonic eigenvalues. We shall also discover that
even non-interacting two anyon states are not simple products of single
anyon states\LEINMYR.
This is the crux of the problem in handling many anyon
systems. Here, we shall concentrate on two approximations
in which the many anyon
problem has been tackled. The quantum statistical mechanics of a many
anyon system has been studied\AROVAS\
via the virial expansion of the equation of
state, which is valid in the high temperature, low density limit.
In Sec.(3), we shall first briefly review the classical and quantum cluster
expansions and the derivation of the virial coefficients in terms of the
cluster integrals. Then,
using the results of the two anyon problems discussed
in Sec.(2), we shall derive the second virial coefficient of the anyon
gas\SMHO.
The other approximation in which the many anyon system has been
studied is the mean field approach which is valid in the high density, low
temperature limit. In this approximation, every anyon sees an `average' or
`mean' field due to the presence of all the other anyons. Thus, the many
body problem is reduced to the problem of a single particle moving in an
`average' potential. It is in this mean field approach that anyon
superconductivity has been established\FHL\ANYSUP.
In Sec.(4), we shall study the
mean field approach and derive anyon superconductivity in a heuristic way.

Finally, we shall briefly touch upon two slightly more advanced topics,
just to give a flavour as to why the study of anyons form such an
interesting and relevant field of research today. The more formal topic
deals with the formulation of a field theory of anyons.
In Sec.(5), we shall introduce anyons in a
field theory formalism using a Chern-Simons construction\AROVAS\
and study an explicit example\EXAMPLE\ of a Lagrangian field theory whose
topological excitations are anyons. The second topic deals with the more
physical question of applicability of anyon physics to condensed matter
systems. In Sec.(6), we shall briefly indicate how anyons arise in the
FQHE, which is a system of
two dimensional electrons at low temperatures and in strong magnetic
fields and show
how the idea of statistics transmutation is used in novel
explanations\JAIN\READ\ of the effect.
We wish to emphasise here that realistic anyons do not
exist in vacuum as has been assumed in the earlier sections, but actually
arise as quasiparticles in a real medium.

\endpage

\chapter{Basic Concepts in Anyon Physics}

Let us start with spin in the familiar three dimensional world. We know
that spin is an intrinsic angular momentum quantum number that labels
different particles. The three spatial components of the spin obey the
commutation relations given by
$$
[S_i,S_j] = i\epsilon_{ijk}S_k.\eqn\eoneone
$$
We shall show that these commutation relations constrain {\bf S} to be
either integer or half-integer. Let $| s,m \rangle$ be the state
with $S^2 | s,m \rangle = s(s + 1) | s,m \rangle$ and
$S_3 | s,m \rangle = m| s,m \rangle$.  By applying the
raising operator $S^{+}$, we may create the state
$$
S^{+}| s,m \rangle = [s(s + 1) - m(m + 1)]^{1/2}
  | s,m+1 \rangle = | s,m' \rangle.\eqn\eonetwo
$$
Requiring this state to have positive norm leads to
$$
s(s+1) - m(m+1) \ge 0\qquad \forall \qquad m, \eqn\eonethree
$$
which in turn leads to
$$
m \le s \qquad \forall \qquad m. \eqn\eonefour
$$
Thus, it is clear that for some value of $ m'= m +{\rm integer}$, $m'
> s$ unless $s=m'$ - $i.e.$,
$$
s-m = {\rm integer}. \eqn\eonefive
$$
Similarly, by insisting that $S^{-} = | s,m \rangle$ have positive
norm, we get
$$
s(s+1) - m(m-1) \ge 0 \qquad \forall \qquad m, \eqn\eonesix
$$
which in turn implies that
$$
m \ge -s \qquad \forall \qquad m. \eqn\eoneseven
$$
Once again, we construct the states, $S^{-}| s,m \rangle$,
$(S^{-})^2| s,m \rangle$,... and to avoid $m < -s$,
we have to set
$$
m-(-s) = {\rm integer}. \eqn\eoneeight
$$
Adding equations \eonefive\ and \eoneeight, we get
$$
2s = {\rm integer} \Longrightarrow s = {{\rm integer}\over 2}. \eqn\eonenine
$$
Thus, just from the commutation relations, we have proven that particles
in 3+1 dimensions have either integral or half-integral spin.

In two spatial dimensions, however, there exists only one axis of rotation
(the axis perpendicular to the plane). Hence, here spin only refers to $S_3$,
which has no commutation relations to satisfy. For a given magnitude of $S_3$,
it can only be either positive or negative depending upon the handedness
of the rotation in the plane. Since there are no commutation relations to
satisfy, there is no constraint on $S_3$ and hence, we can have `any' spin
in two dimensions. For completeness, note that in one spatial
dimension, there is no axis for rotation and hence, no notion of spin.

The term statistics refers to the phase picked up by a wavefunction when
two identical particles are exchanged. However, this definition is
slightly ambiguous. Does statistics refer to the phase picked up by the
wavefunction when all the quantum numbers of the two particles are
exchanged ($i.e.$, under permutation of the particles ) or the actual phase
that arises when two particles are adiabatically transported giving rise
to the exchange? In three dimension, both these definitions are
equivalent, but not so in two dimensions. We shall concentrate on the
second definition which is more crucial to physics and return to the
first definition later.

Let us first consider the statistics of two identical particles moving in
three space dimensions\LEINMYR\WU.
 The configuration space is given by the set of
pairs of position vectors $(\br_1,\br_2)$. The indistinguishability of
identical particles implies the identification $(\br_1, \br_2) \sim
(\br_2,\br_1)$
- $i.e.$, we cannot say whether the first particle is at $\br_1$ and the
second particle at $\br_2$ or the other way around. We shall also impose
the hard-core constraint, $\br_1 \not= \br_2$, to prevent intersecting
trajectories so that we can determine whether or not two particles have
been exchanged. However, as we shall see later, this constraint is
unnecessary because for all particles (except bosons), there is an
automatic angular momentum barrier, preventing two particles from
intersecting, whereas for bosons, we need not know whether the particles
have been exchanged or not, since the phase is one anyway. For convenience
in constructing the configuration space, let us define the centre of mass
(CM) and relative coordinates - $\bR = (\br_1 +\br_2)/2$ and
$\br =\br_1 - \br_2$.
In terms of these coordinates, the configuration space is  $(\bR,\br)$
with $\br \not= 0$, and  with $\br$ being identified with $-\br$. This can
be written as
$$
R_3 \otimes ({R_3 - {\rm origin} \over Z_2}). \eqn\eoneten
$$
Here $R_3$ denotes the three dimensional Euclidean space spanned by
$\bR$. The notation $(R_3 - {\rm origin})$ for $\br$ implies that the
origin $\br = 0 $ is being dropped. $Z_2$ is just the multiplicative group
of the two numbers 1 and -1. Hence, dividing $(R_3 - {\rm origin})$ by
$Z_2$ implies the identification of every position vector $\br$ in the
relative space with its negative $-\br$. To  study the phase picked up by
the wavefunction of a particle as it moves around the other particle, we
need to classify all possible closed paths in the configuration space. Notice
that the CM motion just shifts the positions of the two particles together
and is independent of any possible phase under exchange. Hence, we are only
required to classify closed paths in
$$
{(R_3 - {\rm origin})
\over Z_2} = S. \eqn\eoneeleven
$$
Instead of dealing with paths in S, for ease of visualisation, let us
construct a simpler configuration space by keeping the magnitude of $\br$
fixed , so that the tip of $\br$ defines the surface of a sphere.
Furthermore, the identification of $\br$ with $-\br$ implies that
diametrically opposite points on the sphere are identified. Thus,
configuration space is the surface of a sphere with opposite points
identified as shown in Figs.(1a,1b,1c).

%\vbox{~\vskip 2.0in
%\vbox{\hbox{\qquad\qquad\vbox{\hbox{~~~1a}}
%\qquad\qquad\qquad\qquad\qquad\qquad\vbox
%{\hbox{~~~1b}}
%\qquad\qquad\qquad\qquad\qquad\qquad\vbox
%{\hbox{~~~1c}}}}
%\vskip .1in
%\centerline{Fig.1} \vskip .25in}

\epsfbox{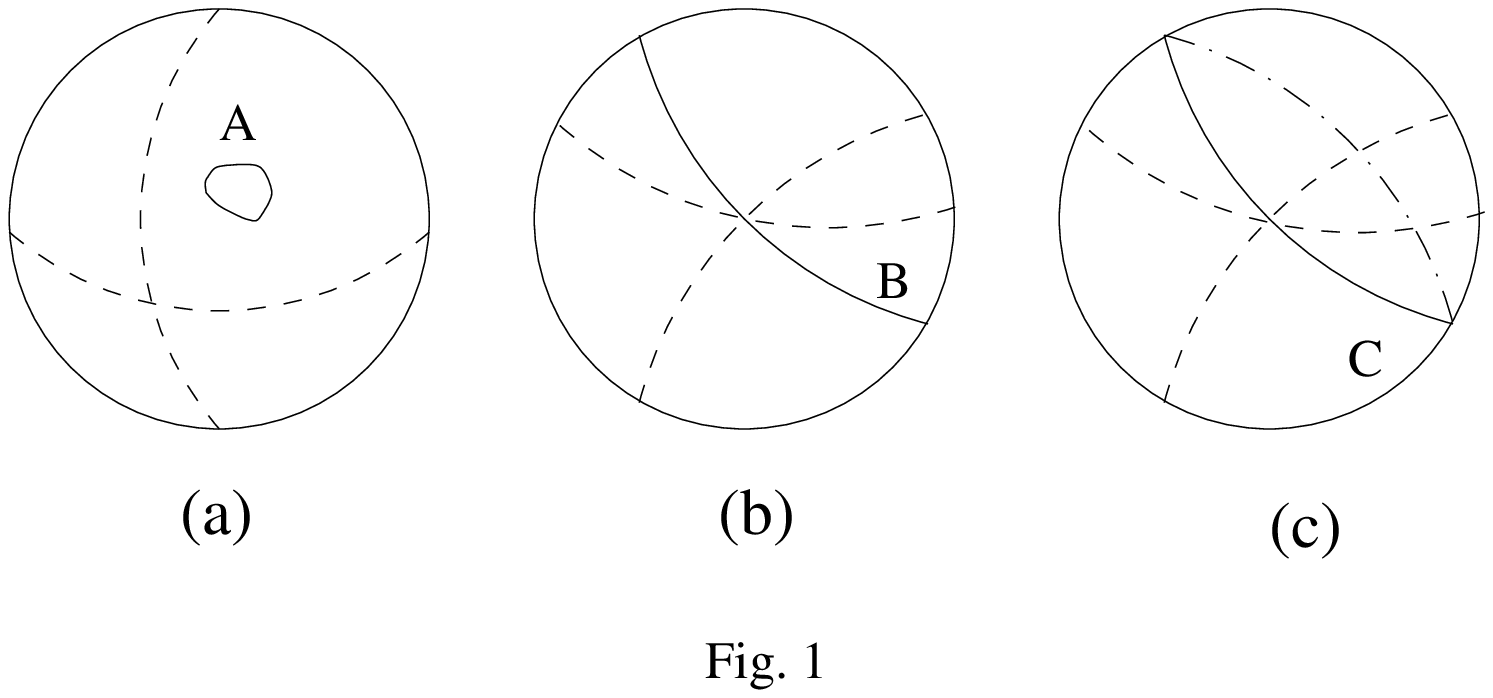}

\noindent
Now, since we have eliminated coincident points, the wavefunction is
non-singular and well-defined at all points in the configuration space. In
particular, it is non-singular on the surface of the sphere. Hence, the
phase picked up by the wavefunction under an adiabatic exchange of the two
particles is also well-defined and does not change under continuous
deformations of the path. Let us consider the possible phases of the
wavefunction when the motion of the particles is along each of
the three paths - $A$
(no exchange),
$B$  (single exchange) and $C$ (two exchanges) -
depicted in Figs.(1a, 1b, 1c). Path $A$ defines a motion of the
particles which does not involve any exchange. It is clearly
a closed path on the surface of the sphere and can be continuously
shrunk to a point. So this path cannot impart any phase to the
wavefunction. Path $B$, on the other hand, involves the exchange of two
particles and goes from a point on the sphere to its diametrically
opposite point - again a closed path.
Since the two endpoints are fixed, by no continuous
process can this path be shrunk to a point. Hence, this path can cause a
non-trivial phase in the wavefunction. However, path $C$ which involves
two exchanges, forms a closed path on the surface of the
sphere, which,
by imagining the path to be a (physical!) string looped around an orange
(surface of a sphere),  can be continuously
shrunk to a point. So once again, the wavefunction
cannot pick up any phase under
two adiabatic exchanges.
This leads us to conclude that there are only two classes of closed paths
that are possible in this configuration space - single exchange or no
exchange.
Let $\eta$ be the phase picked up by any particle under single exchange.
The fact that two exchanges are equal to no exchanges implies that
$$
\eta^2 = +1 \Longrightarrow \eta = \pm 1. \eqn\eonetwelve
$$
Hence, the only statistics that are possible in three space dimensions are
Bose statistics and Fermi statistics.

Why does this argument break down in two space dimensions? Here,
configuration space is given by
$$
R_2 \otimes {(R_2 - {\rm origin})\over Z_2}, \eqn\eonethirteen
$$
where $R_2$ is two dimensional Euclidean space.
Just as before, we ignore the CM motion and fix the magnitude of the relative
separation, so that the configuration space can be visualised as a circle
with diametrically opposite points identified (see Figs.(2a,2b,2c)).

%\vbox{~\vskip 2.0in
%\vbox{\hbox{\qquad\qquad\vbox{\hbox{~~~2a}}
%\qquad\qquad\qquad\qquad\qquad\qquad\vbox
%{\hbox{~~~2b}}
%\qquad\qquad\qquad\qquad\qquad\qquad\vbox
%{\hbox{~~~2c}}}}
%\vskip .1in
%\centerline{Fig.2} \vskip .25in}

\epsfbox{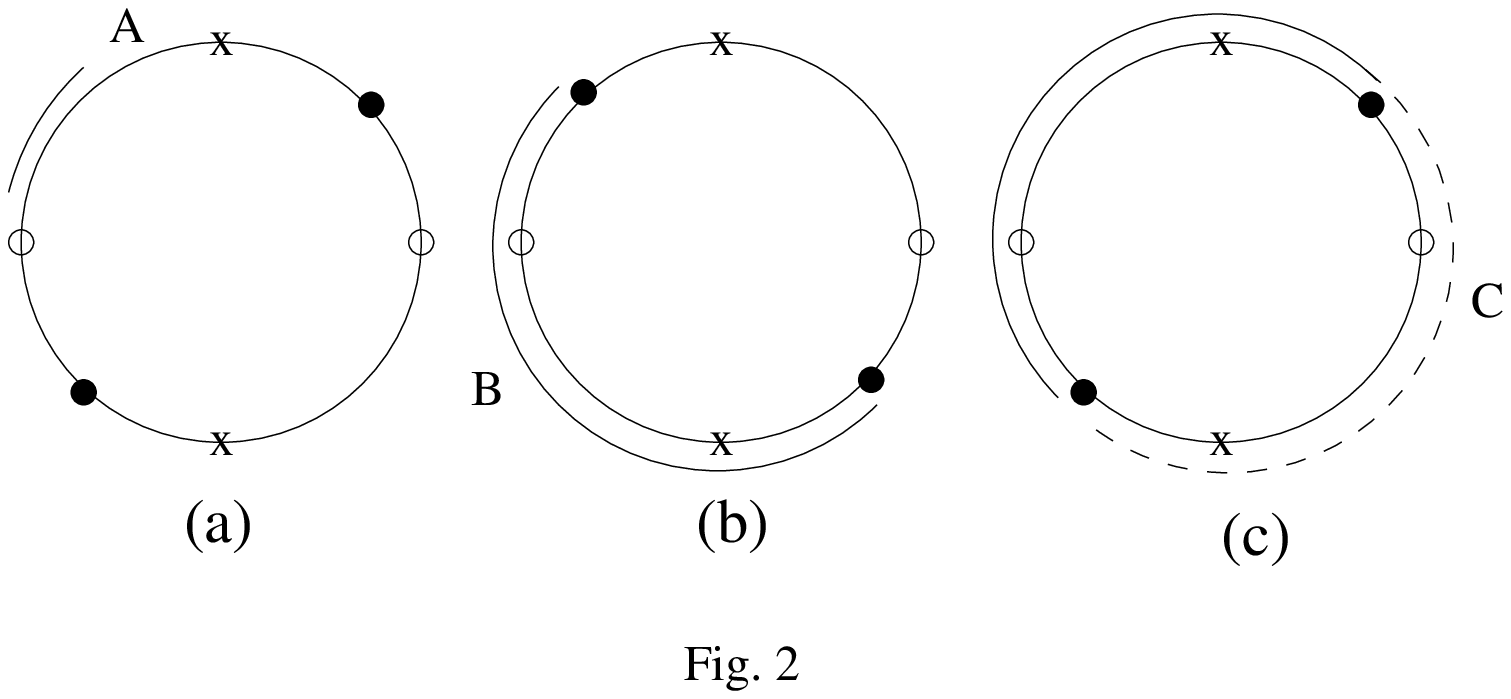}

\noindent
Here, however, several closed paths are possible.
The path $A$ that involves no
exchanges can obviously shrunk to a point since it only involves motion
along any segment of the circle and back. However, the path $B$ that
exchanges the two particles is just as obviously not contractible since
the end-points are fixed. But even the path $C$, where both the dashed and
the solid lines
are followed in the clockwise direction (or equivalently both in the
anti-clockwise direction) cannot be contracted to a point. This is easily
understood by visualising the paths as physical strings looping around a
cylinder (a circle in a plane). Thus, if $\eta$ is the phase under single
exchange, $\eta^2$ is the phase under two exchanges, $\eta^3$ is the phase
under three exchanges and so on. All we can say  is that since the modulus
of the wavefunction remains unchanged under exchange,  $\eta$ is a phase
and can be written as $e^{i\theta}$, where $\theta$ is called the
statistics parameter. This explains why we can have `any' statistics in
two spatial dimensions.

The crux of the distinction between configuration spaces in two and three
dimensions is that the removal of the origin in two dimensional space
makes the space multiply connected, whereas three dimensional space
remains singly connected. Hence, in two dimensions, it is possible to
define paths that wind around the origin. This is not possible in three
dimensions. Mathematically, this distinction is expressed in terms of the
first homotopy group $\Pi_1$, which is the group formed by inequivalent
paths (paths that are not deformable to one another),
passing through a given point in configuration space,
with group multiplication
being defined as traversing paths in succession and group inverse as
traversing a path in the opposite direction.
Thus, in two dimensions,
$$
\Pi_1(2 \,{\rm dim.\, config.\, space}) = \Pi_1{(R_2 -
{\rm origin})\over Z_2}
= \Pi_1(RP_1) = Z \eqn\eonetwelve
$$
where $Z$ is the group of integers under addition and $RP_1$ stands for real
projective one dimensional space and is just the notation for the
circumference of a circle with diametrically opposite points identified.
The equality in the above equation stands for isomorphism of the groups
so that the homotopy group of configuration space is isomorphic to the group
of integers under addition. The different paths are labelled by
integer winding numbers, so that the phases developed by the wavefunction
are of the form, $\eta^n$, with $n$ an integer , which in turn
leads to the possibility of `any' statistics in two dimensions.
In contrast, in three dimensions, we have
$$
\Pi_1(3 \, {\rm dim.\, config.\, space}) =
\Pi_1{(R_3 - {\rm origin})\over Z_2} =
\Pi_1(RP_2) = Z_2. \eqn\eonethirteen
$$
Here $RP_2$ stands for real two dimensional projective space and is the
notation for the surface of a sphere with diametrically opposite points
identified. Since, $Z_2$ has only two elements, there exist only two
classes of independent paths and thus, only two possible phases -
fermionic or bosonic - in three spatial dimensions.

The distinction between the phase of the wavefunction under exchange of
quantum numbers and the phase obtained after adiabatic transport of
particles is also now clear. Under the former definition, the phase
$\eta^2$ after two exchanges is always unity, so that $\eta = \pm 1$,
whereas the phase under the latter definition has many more possibilities
in two dimensions. Mathematically, the distinction is that the first
definition classifies particles under the permutation group $P_N$, whereas the
second definition classifies particles under the braid group $B_N$.
The permutation group ($P_N$) is the group formed by
all possible permutations of
N objects with group multiplication defined as successive permutations and
group inverse defined as undoing the permutation. It is clear that the
square of any permutation is just unity, since permuting two objects twice
brings the system back to the original configuration.
Thus, particles that transform as representations of $P_N$ can only be
fermions or bosons.
The braid group $B_N$, on the other hand, is the group of inequivalent
paths (or trajectories) that occur when adiabatically transporting $N$
objects. For example, the trajectories shown in Fig.(3)

%\vbox{~\vskip 1.5in
%\vbox{\hbox{\qquad\qquad\vbox{\hbox{~~3a}}
%\qquad\qquad\qquad\qquad\qquad\qquad\vbox
%{\hbox{~~3b}}
%\qquad\qquad\qquad\qquad\qquad\qquad\vbox
%{\hbox{3c}}}}
%\vskip .1in
%\centerline{Fig.3} \vskip .1in}

\epsfbox{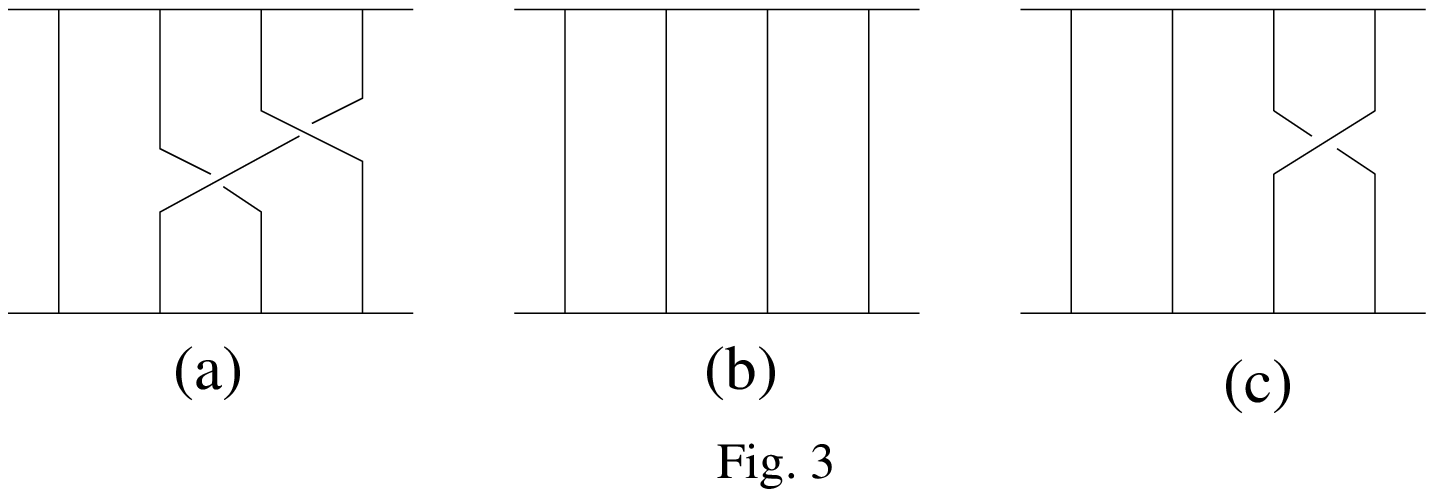}

\noindent are elements of $B_4$,
because all of them are possible paths involving four particles. Fig.(3b)
represents the identity element where none of the trajectories cross
each other. Group
multiplication is defined as following one trajectory by another as
depicted in Fig.(4)
%\vbox{~\vskip 1.5in \centerline{Fig.4} \vskip .2in}

\epsfbox{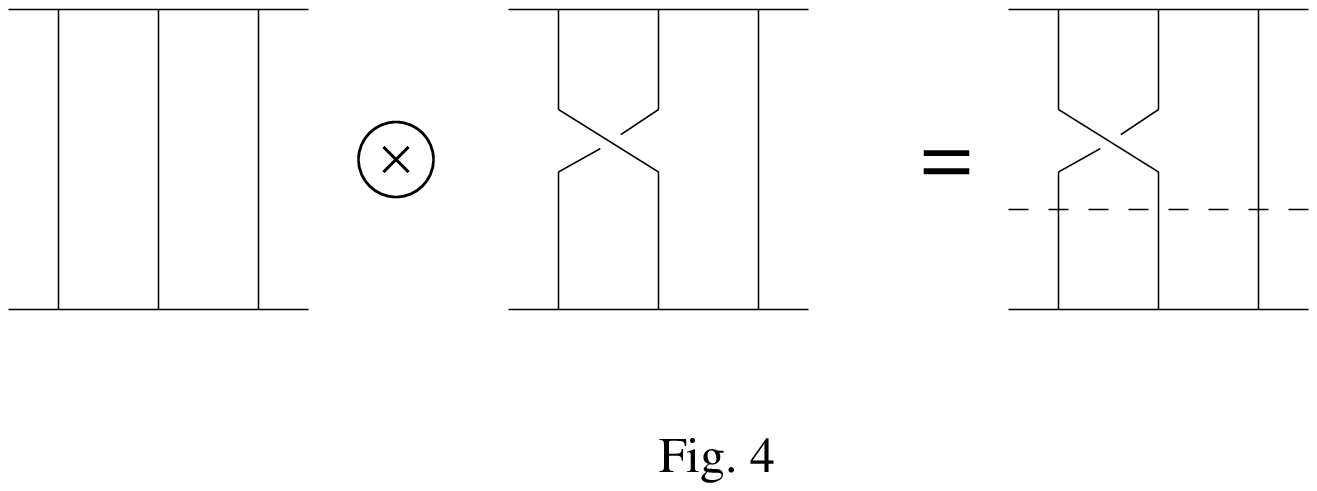}

\noindent and group inverse is defined as a reverse crossing,
(Fig.(5))
%\vbox{~\vskip 1.5in \centerline{Fig.5} \vskip .2in}
%\vskip -0.2cm

\epsfbox{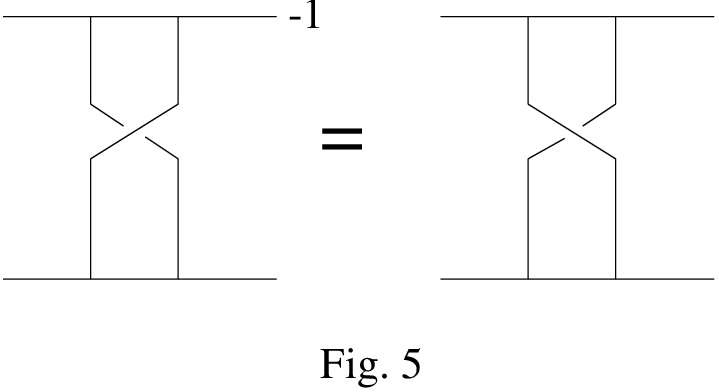}

\noindent so that the product of a trajectory and its inverse leads to the
identity as shown in Fig.(6).
%\vbox{~\vskip 1.5in \centerline{Fig.6} \vskip .2in}

\epsfbox{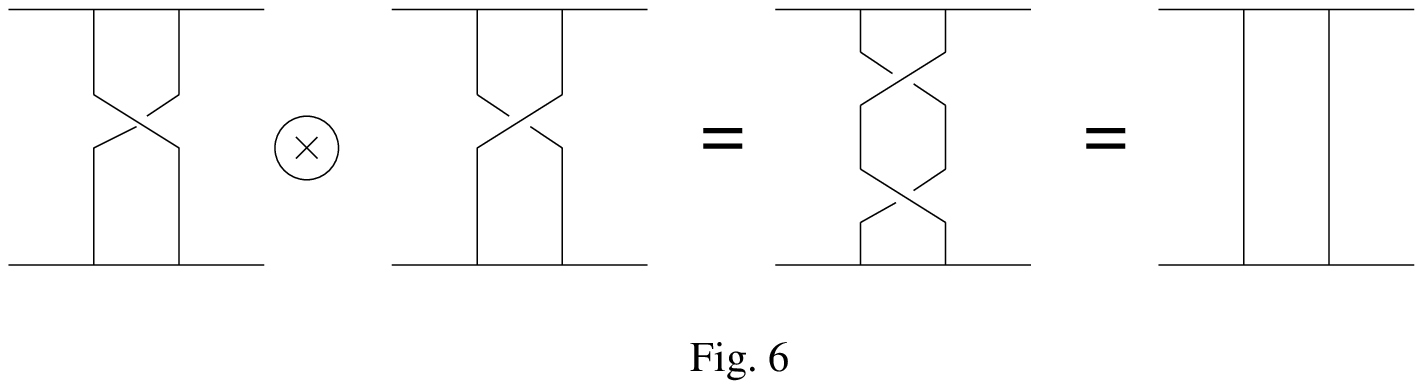}

\vfill
\eject

\noindent Here, it is pictorially clear (see Fig.(7)),
%\vbox{~\vskip 2.25in \centerline{Fig.7} \vskip .2in}

\epsfbox{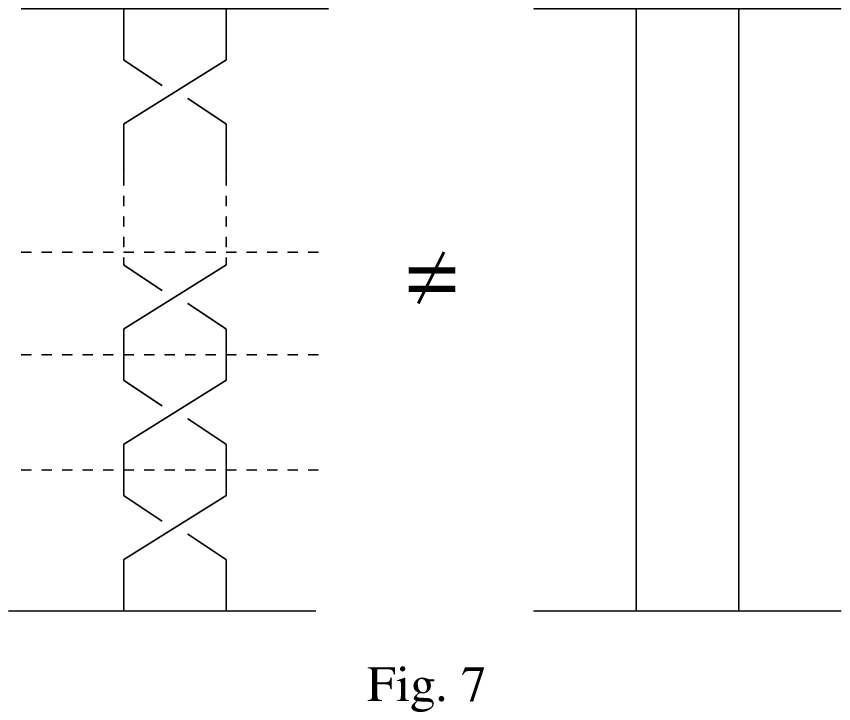}

\noindent that the square, or indeed, any power of the
trajectory representing the
adiabatic exchange of two particles is not 1. Hence, particles that
transform as representations of the braid group are allowed to pick up
`any' phase under adiabatic exchange. For completeness, we mention that
more abstractly, the braid group $B_N$ is defined as the group whose
elements (trajectories) satisfy the following two relations depicted
pictorially in Fig.(8)

\epsfbox{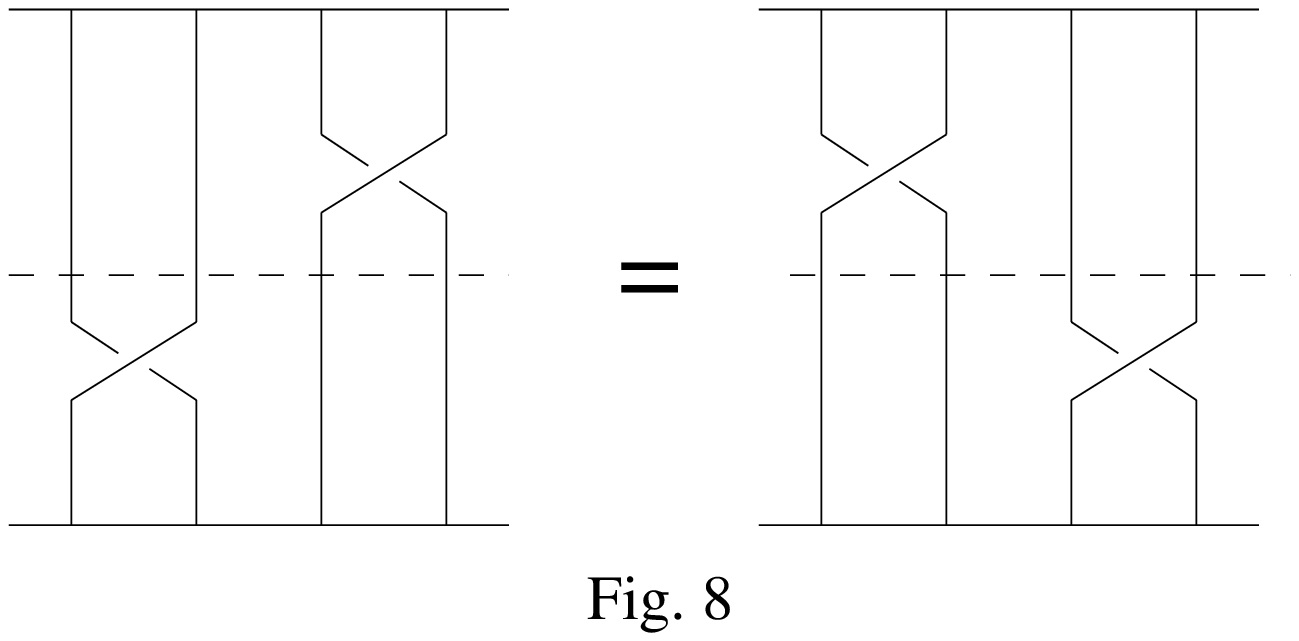}
%\vbox{~\vskip 1.75in \centerline{Fig.8} \vskip .2in}

\noindent and Fig.(9).

\epsfbox{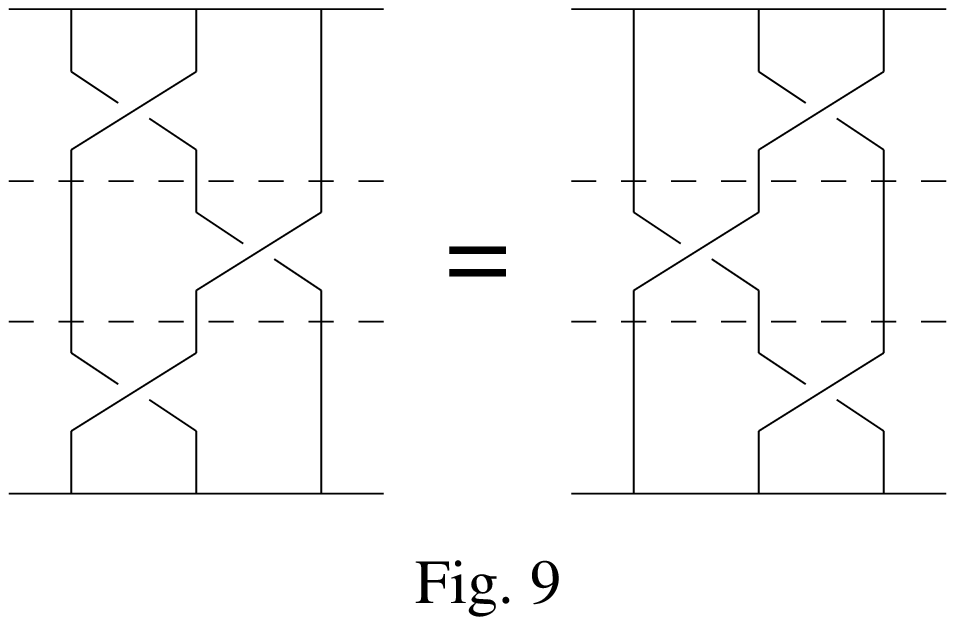}

%\vbox{~\vskip 1.75in \centerline{Fig.9} \vskip .2in}

\noindent The second relation is called the
Yang-Baxter relation.
It is clear that the braid group is a much richer group than the
permutation group and leads to a much finer classification. For example,
consider the trajectories a) and b) in Fig.(10).

%\vbox{~\vskip 2.0in
%\vbox{\hbox{\qquad\qquad\qquad\qquad\vbox{\hbox{~~~10a}}
%\qquad\qquad\qquad\qquad\qquad\qquad\qquad\qquad\vbox
%{\hbox{10b}}
%}}
%\vskip .1in
%\centerline{Fig.10} \vskip .2in}
\epsfbox{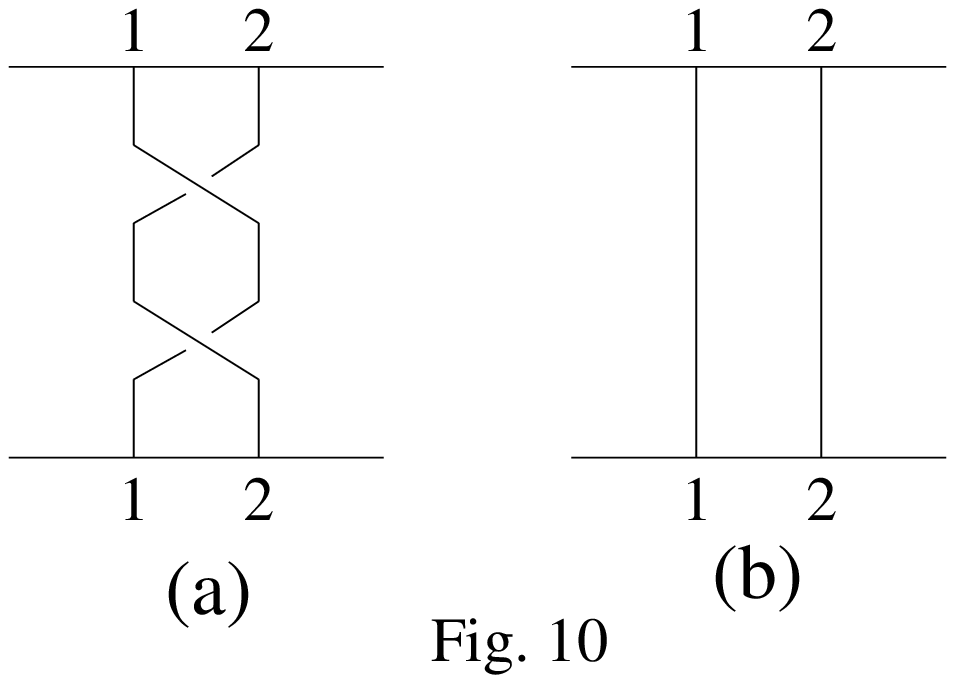}

\noindent The permutation group
cannot distinguish between these two trajectories. In both cases, there is
no permutation and hence the trajectory belongs to the identity
representation, whereas the two trajectories are distinct elements of the
braid group.

Finally notice that in one spatial dimension, two particles cannot be
adiabatically exchanged without passing through each other -$i.e.$, without
interacting. Hence any theory can be written in terms of bosons or
fermions with appropriate interactions and there is no real concept of
statistics in one dimension.

After all this abstract discussion, let us construct a simple physical
model of an anyon\WILCZEK. Imagine a spinless particle of charge $q$
orbiting around a thin solenoid along the $z$-axis, at a distance ${\bf r}$
as shown in Fig.(11).

%\vbox{~\vskip 2.0in \centerline{Fig.11} \vskip .2in}
\epsfbox{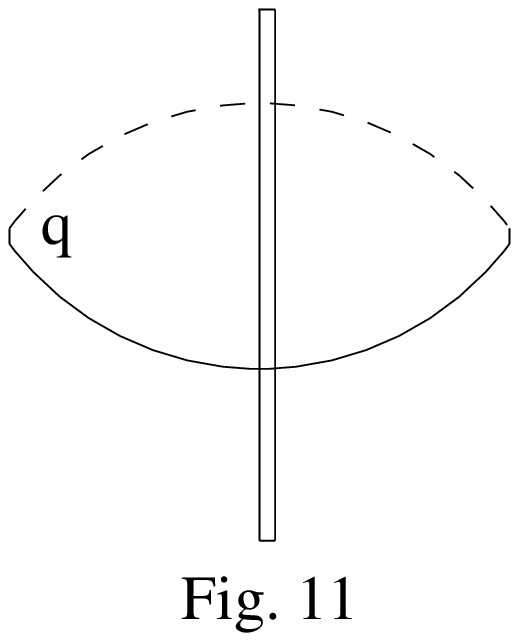}

\noindent When there is no current flowing through the solenoid,
the orbital angular
momentum of the charged particle is quantised as an integer -$i.e.,$
$$
l_z = {\rm integer}.\eqn\eonesixteen
$$
When a current is turned on, the particle feels
an electric field that can easily be computed using
$$
\int ({\bf\nabla} \times {\bf E}) d^{2}\br = -{\partial\over\partial t} \int
B d^{2}\br = - {\partial\phi\over\partial t} \eqn\eoneseventeen
$$
where $\phi$ is the total flux through the solenoid. Hence,
$$
\int {\bf E}\cdot d{\bf l} = 2\pi | {\bf r}| E_{\theta} = -{\dot \phi}
\eqn\eoneeighteen
$$
leading to
$$
{\bf E} = - {{\dot \phi}\over 2\pi | {\bf r}|} ({\hat z} \times {\hat {\bf
r}}).     \eqn\eonenineteen
$$
Thus, the angular momentum of the charged particle changes, with the rate
of change being proportional to the torque  ${\bf r} \times {\bf F}$ -$i.e.,$
$$
{\dot l_z} = {\bf r} \times {\bf F} =  \br \times q {\bf E} = - {q {\dot \phi}
\over 2\pi} \eqn\eonetwenty
$$
Hence,
$$
\Delta l_z = - {q \phi\over 2 \pi}
\eqn\eonetwentyone
$$
is the change in angular momentum due to the flux $\phi$ through the
solenoid. In the limit where the solenoid becomes extremely narrow and the
distance between the solenoid and the charged particle is shrunk to zero,
the system may be considered as a single composite object - a charged
particle-flux tube composite. Furthermore, for a planar system, there can
be no extension in the $z$-direction. Hence, imagine shrinking the
solenoid along the $z$-direction also to a point. The composite object is
now pointlike, has fractional angular momentum and perhaps can be
identified as a model for an anyon. However, this is too naive a picture.
As we shall see later, in an anyon, the charge and the flux that it
carries are related. So it is not quite right to think of the anyon as
an independent charge orbiting around an independent flux. The charge is
actually being switched on at the same time that the flux in the solenoid
is being switched on. Hence, Eq.\eonetwenty\ needs to be modified to read
$$
{\dot l_z} =  - {q(t) {\dot \phi}
\over 2\pi}. \eqn\eonetwentytwo
$$
Moreover, since $q(t) = c \phi (t)$, for some constant $c$, we get
$$
\Delta l_z = -{c\phi^2 \over 4\pi} = - {q \phi\over 4 \pi},
\eqn\eonetwentythree
$$
so that the angular momentum of a charge-flux composite, with charge
proportional to the flux, is less than what we originally computed. This
is not surprising since the original computation overestimated the charge
by keeping it fixed. Hence, our final physical model of an anyon is that
of a charge-flux composite with charge $q$ and flux $\phi$ being
proportional to each
other and with a spin given by $q\phi/4\pi$. In the next section, we shall
see that this model also exhibits fractional statistics and  complete
its identification as an anyon.

\vskip 1 cm
\centerline {Problems}
\item{1.} Fermions and bosons are one-dimensional representations of the
permutation group $P_N$ and anyons are one-dimensional representations of
the braid group $B_N$. Can one have higher dimensional representations of
$P_N$ and $B_N$? (The answer is yes.) Think about the quantum mechanics
and statistical mechanics of particles in these representations and see how
far you get.

\endpage

\chapter{Quantum Mechanics of Two Anyon Systems}

In Sec.(1), we constructed a composite object which consisted of a
spinless (bosonic) charge orbiting around a (bosonic) flux, and showed
that when the charge is proportional to the flux, this object had
fractional spin $s = q\phi/4\pi$. To determine its statistics, we need to
study the quantum mechanical system of two such objects. The Hamiltonian
for the system is given by
$$
H = {(\bpone - q\baone)^2\over 2m} + {(\bptwo - q\batwo)^2\over 2m}
\eqn\etwoone
$$
with
$$
\eqalign
{\baone &= {\phi\over 2\pi} {{\hat z} \times (\brone - \brtwo)
\over |\brone - \brtwo|^2},\cr
{\rm and} \qquad
\batwo &= {\phi\over 2\pi} {{\hat z} \times (\brtwo - \brone)
\over |\brone - \brtwo|^2},}
\eqn\etwotwo
$$
where $\baone$ and $\batwo$ are the vector potentials at the positions of the
composites 1 and 2, due to the fluxes in the composites 2 and 1
respectively. Let us work in the centre of mass (CM) and relative (rel)
coordinates - i.e., we define respectively,
$$
\bR = {\brone +\brtwo \over 2} \Rightarrow \bP = \bpone +\bptwo,
\qquad {\rm and} \qquad
\br = \brone-\brtwo \Rightarrow \bp = {\bpone -\bptwo \over 2}.
\eqn\eextra
$$
In terms of these coordinates, the Hamiltonian can be recast as
$$
H = {\bP^2\over 4m} + {(\bp - q \barel)^2\over m} \eqn\etwothree
$$
with
$$
\barel = {\phi\over 2\pi} {{\hat z} \times \br \over |\br|^2}. \eqn
\etwofour
$$
Thus, the CM motion, which translates both the particles rigidly and is
independent of statistics, is free. The relative motion, on the other
hand, which is sensitive to whether the composites are bosons, fermions or
anyons, has reduced to the system of a single charged particle of mass
$m/2$ orbiting around a flux $\phi$ at a distance $\br$. Since the
composite has been formed of a bosonic charge orbiting around a bosonic
flux, the wavefunction of the two composite system is symmetric under
exchange and the boundary condition is given by
$$
\psirel (r,\theta +\pi) = \psirel (r,\theta) \eqn\etwofive
$$
where $\psirel$ is the wavefunction of the relative piece of the
Hamiltonian in Eq.$\etwothree$ and $\br = (r,\theta)$ in
cylindrical coordinates.

Now, let us perform a (singular) gauge transformation so that
$$
\ba_{\rm rel} \longrightarrow \baprime_{\rm rel} = \ba_{\rm rel}
- \bnabla \Lambda (r,\theta)
\eqn\etwosix
$$
where $\Lambda(r,\theta) = {\phi\over 2\pi} \theta$. This gauge
transformation is singular because $\theta$ is a periodic angular
coordinate with period $\theta$ and is not single valued. In the primed
gauge,
$$
a'_{{\rm rel}\theta} = a_{{\rm rel}\theta} -
{1\over r} {\partial\over \partial\theta}
({\phi\theta\over 2\pi}) = 0 \qquad {\rm and} \qquad
a'_{{\rm rel}r} = a_{{\rm rel}r} = 0 \eqn\etwoseven
$$
-i.e., the gauge potential completely vanishes. Hence,
the Hamiltonian reduces to
$$
H = {\bP^2\over 4m} + {\bp^2\over m} \eqn\etwoeight
$$
which is just the Hamiltonian of two free particles. However, in the primed
gauge, the  wavefunction of the relative Hamiltonian has also changed. It is
now given by
$$
\psirelprime(r,\theta) = e^{-iq\Lambda} \psirel(r,\theta) =
e^{-i{q\phi\over 2\pi}\theta} \psirel(r,\theta) \eqn\etwonine
$$
which is no longer symmetric under $\br \rightarrow -\br$ since
$$
\psirelprime(r,\theta + \pi) = e^{-iq\phi/2} \psirelprime(r,\theta)
\equiv e^{-i\alpha}\psirelprime(r,\theta).
\eqn\etwoten
$$
Thus, two interacting bosonic charge-flux composites are equivalent to two
free particles whose wavefunctions develop a phase $e^{-iq\phi/2}$ under
exchange - i.e., they obey fractional statistics. This completes the
identification of charge-flux composites as anyons. Notice that the
statistics phase $\alpha = q\phi/2$ is in accordance with the spin
of the composite, which we had earlier determined to be
$q\phi/4\pi$, so that the generalised spin-statistics
theorem which relates the statistics factor $\alpha$ to the spin
$j = \alpha/2\pi$, is satisfied.

We have thus shown that anyons can either be considered as free particles
with fractional spin obeying fractional statistics, or as interacting
charge-flux composites, again with fractional spin, but obeying Bose
statistics. The free particle representation is called the anyon gauge,
and the interacting particle representation is called the boson gauge.

Let us now solve the quantum mechanical problem of two free anyons.
The Hamiltonian in Eq.$\etwoeight$ is simple, but the boundary condition in
Eq.$\etwoten$ is non-trivial. However, we
have just seen that this problem is equivalent to the problem of two
interacting charge-flux composites, described by the Hamiltonian in Eq.
$\etwothree$ with bosonic boundary conditions. The CM motion is trivial
and directly solved to yield the energy eigenvalues and eigen functions as
$$
E_{\rm CM} = {\bP^2\over 4m} \qquad {\rm and} \qquad \psi_{\rm CM} =
e^{i\bP \cdot \bR}. \eqn\etwoeleven
$$
To solve for the relative motion, we work in cylindrical coordinates,
wherein the relative part of the Hamiltonian equation takes the form
$$
[-{1\over m} ({\partial^2\over \partial r^2} + {1\over r} {\partial \over
\partial r}) + {1\over mr^2} (i{\partial \over \partial\theta} - {q\phi
\over 2\pi})^2] \psirel (r,\theta) = E_{rel} \psirel(r,\theta)
\eqn\etwotwelve
$$
with the boundary conditions in Eq.$\etwofive$.
Since the Hamiltonian is separable in $r$ and $\theta$, the wave function
factorises as $\psirel(r,\theta) = {\cal R}(r) Y_l(\theta)$.  Hence, the
angular equation reduces to
$$
(i{\partial\over\partial\theta} - {q\phi\over 2\pi})^2 Y_l(\theta) =
\lambda Y_l(\theta) \eqn\etwothirteen
$$
whose solution for $Y_l(\theta)$, consistent with the
boundary condition is given by
$$
Y_l(\theta) = e^{il\theta},\qquad l = {\rm even\, integer}, \eqn\etwofourteen
$$
which, in turn, leads to
$$
\lambda = (\lqphi)^2, \qquad l = {\rm even\, integer}. \eqn\etwofifteen
$$
When the $\lambda$ eigenvalue  is substituted in the radial equation, we get
$$
[- {1\over m}(\prsquared + \proverr) +{1\over mr^2} (\lqphi)^2 ] {\cal
R}(r) = E_{\rm rel} {\cal R}(r). \eqn\etwosixteen
$$
Notice that the net effect of the fluxtubes has been to add a
factor $q\phi/2\pi$ to the angular momentum term, thus adding to the
centrifugal barrier. This justifies our earlier statement in Sec.(1) that
all particles, except bosons, have a centrifugal barrier which prevents
intersecting trajectories.
Now, by defining $mE_{\rm rel} = k^2$, we may rewrite
the radial equation in the form
$$
[\prsquared + \proverr - {1\over r^2} (\lqphi)^2 + k^2]{\cal R}(r) = 0 ,
\eqn\etwoseventeen
$$
which is easily identified as the Bessel equation with the solution
$$
{\cal R}(r) = J_{|\lqphi|}(kr).
\eqn\etwoeighteen
$$
Hence the relative wavefunction of the composites, in the bosonic gauge,
is given by
$$
\psirel(r,\theta) = e^{il\theta} J_{|\lqphi|}(kr), \qquad l =
{\rm even\, integer}. \eqn\etwonineteen
$$
The wavefunction for two anyons can also be expressed in the anyon gauge
by performing a gauge transformation so that
$$
\psirelprime = e^{i(\lqphi)\theta} J_{|\lqphi|}(kr), \qquad
l= {\rm even\, integer}. \eqn\etwotwenty
$$
This wavefunction is obviously anyonic, since it picks up  a phase
$q\phi/2$ under $\theta \rightarrow \theta + \pi$
($i.e.$, under exchange of the
two particles).

Including the CM motion, the two anyon wavefunction can be written
as
$$
\psi(\bR,\br) = \psi_{\rm CM}(\bR) \psirel(\br) = e^{i\bP\cdot\bR}
e^{i(\lqphi)\theta} J_{|\lqphi|}(kr)
\eqn\etwotwentyone
$$
or equivalently as
$$
\psi(\bR,\br) = e^{iL\Theta} J_{L}(KR) e^{i(\lqphi)\theta} J_{|\lqphi|}(kr).
\eqn\etwotwentytwo
$$
In Eq.$\etwotwentytwo$, we have expressed the CM motion also in terms of
cylindrical quantum numbers - $L$ is the CM angular momentum and $K =
4mE_{\rm CM}$ labels the CM energy. The two particle wavefunction can be
recast in terms of the original single particle coordinates $\brone$ and
$\brtwo$ - $i.e.$, $\psi(\bR,\br) \longrightarrow \psi(\brone,\brtwo)$.
However, unless $\qphi$ is either integral or half-integral, the two
particle wavefunction cannot be factorised into a product of two suitable
one particle wavefunctions --- $\psi(\brone,\brtwo) \ne \psi_1(\brone)
\psi_2(\brtwo)$. Since handling Bessel functions is inconvenient, this
property will be demonstrated more explicitly in the next example of two
anyons in a harmonic oscillator potential. Furthermore, in principle, we
should be able to prove from this example that the energy levels of a two
anyon system  cannot be obtained as sums of one anyon energy levels. This,
again, is easier to see with discrete energy levels and will be explicitly
proven in the next example.

The second example that we shall solve explicitly is the problem of two
anyons in a harmonic oscillator potential\LEINMYR\
with the Hamiltonian given by
$$
H = {\bpone^2 \over 2m} + {\bptwo^2\over 2m} + {1\over 2} m{\omega^2}\brone^2
+ {1\over 2} m{\omega^2}\brtwo^2. \eqn\etwotwentynine
$$
As in the case of two free anyons, the problem is separable in the CM and
relative coordinates, in terms of which, the Hamiltonian can be rewritten as
$$
H = {\bP^2\over 4m} + {\bp^2\over m} + m\omega^2 \bR^2 + {1\over 4}m\omega^2
\br^2 . \eqn\etwothirty
$$
The CM motion is clearly independent of the statistics of the particles
and involves just the usual quantum mechanical problem of a single
particle of mass $2m$ in a two dimensional harmonic oscillator potential.
The energy levels and wavefunctions are obviously well-known. However, we
shall briefly recollect the familiar steps here, just to set the
field for the relative motion problem which is sensitive to the statistics
of the particles.
For the CM motion, we work in the cylindrical $(R,\Theta)$
coordinates and write
$$
H_{\rm CM} \psicm(R,\Theta) = [-{1\over 4m}(\pRsquared + \pRoverR) -
{1\over 4mR^2} \pThetasquared + m\omega^2 R^2]\psicm(R,\Theta).
\eqn\etwothirtyone
$$
The wavefunction factorises in $R$ and $\Theta$ and can be written as
$$
\psicm(R,\Theta) = e^{iL\Theta} {\cal R}_{\rm CM}(R) \eqn\etwothirtytwo
$$
with $L$ being an integer for a single valued wavefunction.
Defining $K^2 = 4mE_{\rm CM}$, the radial eigenvalue equation becomes
$$
[\pRsquared +\pRoverR -{L^2\over R^2} -4m^2\omega^2 R^2 + K^2] 
{\cal R}_{\rm CM}(R) = 0, \eqn\etwothirythree
$$
with a solution of the form
$$
{\cal R}_{\rm CM}(R) = e^{-m\omega R^2} {\sum_{n=o}^{\infty}} a_n R^{n+s}.
\eqn\etwothirtyfour
$$
By substituting this series solution in Eq.\etwothirythree, we find that
$s = |L|$ and that
$$
{a_{n+2}\over a_{n}} = {{K^2 - 4m\omega(n+|L|+1)}\over {(n+2)^2 + 2|L|(n+2)}},
\eqn\etwothirtyfive
$$
so that $n = {\rm even\, integer}$. Requiring the series to terminate
leads to
$$
E_{\rm CM} = \omega(n+|L|+1) \eqn\etwothirtysix
$$
which is the well known answer, since $n+|L| = p$ is an integer, and the
modulus factor gives the appropriate degeneracies.

The Hamiltonian for the relative motion, which does depend on the
statistics of the particles, can also be solved in the same way. In the boson
gauge, we have
$$
H_{\rm rel} \psirel =[{(\bp - q\barel)^2 \over m} + {1\over 4} m\omega^2 r^2]
\psirel = E_{\rm rel} \psirel \eqn\etwothirtyseven
$$
with $\barel = (0,\phi/2\pi r)$ and
$\psirel(r,\theta + \pi) = \psirel(r,\theta)$.
The wavefunction, once again, factorises in $r$ and $\theta$ and is given by
$$
\psirel(r,\theta) = e^{il\theta} {\cal R}_{\rm rel}(r) \eqn\etwothirtyeight
$$
However, due to the boundary condition on $\psirel$, $l$ now has to be an
even integer, analogous to the relative angular momentum quantum number for
two free anyons .
With the definition $k^2 = mE$, the radial equation is
 $$
 [\prsquared + \proverr - {1\over r^2} (l +\alpha/\pi)^2 - {1\over 4}
 m^2\omega^2 r^2 + k^2] \Rrelr = 0, \eqn\etwothirtynine
 $$
 where we have substituted $\alpha = q\phi/2$.
 As in the case of two free anyons, we see that the net effect of
statistics is to add a term to the centrifugal barrier. A series solution
of the form
$$
\Rrelr = e^{-m\omega r^2/4} \sum_{n=0}^{\infty} b_n r^{n+s} \eqn
\etwoforty
$$
can be found to Eq.\etwothirtynine, with $s = |l+\alpha/\pi|$ and
$$
{b_{n+2}\over b_{n}} = {{k^2 - m\omega(n+|l+\alpha/\pi|+1)}\over
{(n+2)^2 + 2|l+\alpha/\pi|(n+2)}},
\eqn\etwofortyone
$$
so that $n$ has to be an even integer. As before, the requirement that
the series has to terminate leads to the energy levels given by
$$
E_{\rm rel} = \omega(n+|l+\alpha/\pi|+1). \eqn\etwofortytwo
$$
Notice that $\alpha = 0$ and $\alpha = \pi$ give the usual energy levels
for bosons and fermions respectively.
The first few energy levels and their degeneracies are explicitly
given by
$$
\eqalign
{E_{0} &= (1+\alpha/\pi)\omega, \qquad {\rm deg} = 1,\cr
E_{1} &= (3-\alpha/\pi)\omega, \qquad {\rm deg} = 1, \cr
E_{2} &= (3+\alpha/\pi)\omega, \qquad {\rm deg} = 2,\cr
E_{3} &= (5-\alpha/\pi)\omega, \qquad {\rm deg} = 2,\cr
E_{4} &= (5+\alpha/\pi)\omega, \qquad {\rm deg} = 3.} \eqn\etwofortythree
$$
Thus, it is clear that the energy levels can be written as
$$
\eqalign
{&E_j = (2j+1+\alpha/\pi) \omega, \qquad {\rm deg} = j+1  \cr
{\rm and} \qquad
&E_j = (2j+1-\alpha/\pi) \omega, \qquad {\rm deg} = j,} \eqn\etwofortyfour
$$
where $j$ is an integer. The energy levels can be plotted as a function of
$\alpha$ as shown in Fig.(12).

%\vbox {~\vskip 2.5in \centerline{Fig.12} \vskip .25in}
\epsfbox{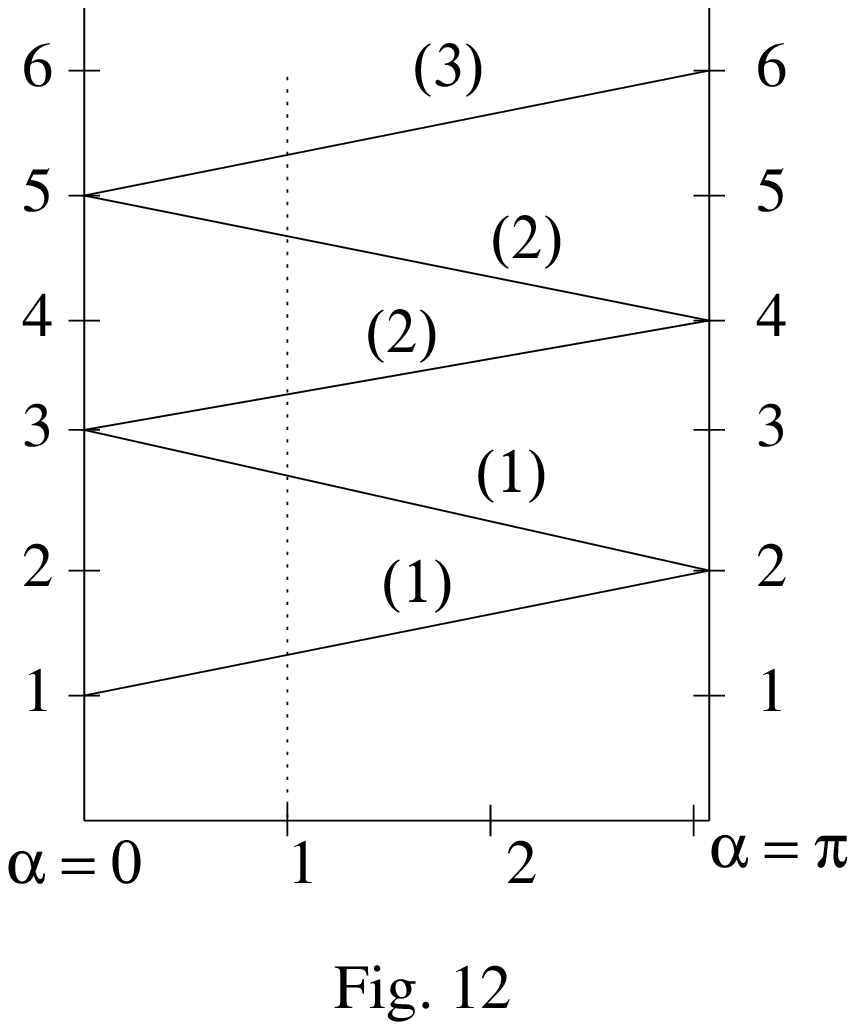}

\noindent
Here $\alpha = 0$ corresponds to bosons and $\alpha = \pi$ corresponds to
fermions. The degeneracies of the levels are mentioned within brackets.
{}From the figure, it is clear that the energies of the 2 particle system
are a monotonic function of the $\alpha$ and change continuously as
$\alpha$ changes from bosonic to fermionic value. Moreover, except at
$\alpha = 0,\pi/2,\pi$, the energy levels are not equally spaced and even
at $\alpha = \pi/2$, the spacing between energy levels is half that for
bosons and fermions.
We also note that the energy
levels do not cross each other as a function of $\alpha$ for this system.
This property, however, does not necessarily hold
even for other two anyon systems and, in general, systems with
three or more anyons
in any potential have level crossings.

Let us now combine the energy levels of the relative Hamiltonian with the
energy levels of the CM system to obtain the full two particle energy
levels -  $i.e.,$
$$
E_{2 {\rm particles}} = E_{\rm CM} + E_{\rm rel} = (2j+p+2\pm\alpha/\pi)
\omega. \eqn\etwofortyfive
$$
Now compare these energy levels with the energy levels
$$
E_n = (n+1)\omega, \qquad n = {\rm integer} \eqn\exx
$$
of a single particle in a two dimensional oscillator. For two particles,
we would naively have expected the energy levels to be of the form
$$
E_{n,m} = (n+1)\omega + (m+1)\omega \qquad n,m = {\rm integers}.
\eqn\etwofortysix
$$
This expectation is borne out only at $\alpha = 0$ and $\alpha = \pi$,
where it is clear that Eq.\etwofortyfive\ is of the form of
Eq.\etwofortysix, so that two particle energy levels can always be written
in terms of sums of single particle energy levels. However, for arbitrary
$\alpha$, the two anyon energy levels bear no simple additive or
combinatoric relation to the levels of a one anyon system. This is the
root cause of the difficulty in handling many anyon systems.

A study of the ground state wavefunction also illuminates the same point.
In the anyon gauge, the wavefunction is given by
$$
\psi(R,\Theta,r,\theta) \prop e^{-m\omega(R^2 + r^2 /4)} r^{\alpha/\pi}
e^{i\alpha\theta/\pi}
= e^{-m\omega(R^2 + r^2/4)}(\br)^{\alpha/\pi} \eqn\etwofortyfive
$$
where $\br$ in complex coordinates is $x + iy = r {\cos}\theta + ir{\sin}
\theta$. This wavefunction  can be written in the
original two particle coordinate system as
$$
\psi(\brone,\brtwo) \prop e^{-{m\omega(r_1^2 +r_2^2)/2}} (\brone - \brtwo)^
{\alpha/\pi}, \eqn\etwofortysix
$$
which, incidentally, vanishes except for bosons ($\alpha = 0$), thereby
proving that anyons obey the Pauli principle.
It is easy to see that for $\alpha = 0$ and $\alpha = \pi$, this wavefunction
can be factorised into a product of two single particle wavefunctions ---
$$
\alpha = 0 \qquad \psi(\brone,\brtwo) = e^{-m\omega r_1^2/2}
e^{-m\omega r_2^2/2} = \chi_0 (\brone) \chi_0 (\brtwo)
\eqn\etwofortyeight
$$
and
$$
\alpha = \pi \qquad
\eqalign{\psi(\brone,\brtwo) &= \brone e^{-m\omega r_1^2/2}
e^{-m\omega r_2^2/2} - \brtwo e^{-m\omega r_1^2/2}
e^{-m\omega r_2^2/2} \cr
&= \chi_1(\brone) \chi_0(\brtwo) - \chi_0(\brone)\chi_1(\brtwo)}
\eqn\etwofortyeightprime
$$
where $\chi_0$ and $\chi_1$ are the ground and first excited states of the
one particle system. For arbitrary $\alpha$, however, the two particle
wavefunction bears no simple relation to the one particle wavefunctions.
Thus, many anyon states are not products of single anyon states. This is
why even a system of free
anyons needs to be tackled as an interacting theory.

There are other simple two anyon problems that can be solved exactly, such as
the problem of two anyons in a Coulomb potential\KHARE. However, no three
anyon problem has been completely solved so far. The main hurdle in solving
systems with three or
more anyons is that the relative phases between any two anyons depends
also on the positions of all the other anyons in the system. This makes
the problem quite intractable. Only partial solutions have been obtained
so far
using various different methods. Exact solutions for a
part of the spectrum of three or more free
anyons or in a harmonic oscillator potential (or equivalently an
external magnetic field) have been found\WUWU. A semiclassical approach
to the computation of the energy levels\KIVELSON\
has also been attempted.
Perturbative approaches\MCCABE\ in the limit of small $\alpha$, 
where $\alpha$ is
the statistics parameter, have also been used to demonstrate level
crossings and the piecewise continuity of the ground state\CHITRA.
The level crossing phenomenon has been confirmed by recent
numerical computations which have been used to obtain
the first twenty odd energy levels\MURTHYETAL,
which also showed that
the analytic solutions found earlier formed a very small subset of the
total number of solutions. Finally, there exist some exact
results\DIPTIMAN\
regarding the symmetry of the spectrum as a function of the statistics
parameter $\alpha$.
However, the computation of the full
spectrum of energy levels of a three anyon system
is beyond our capacity at the present time.
Hence, we emphasize here that
by and large, the study of three and more anyon quantum 
mechanics is still an open problem.

\endpage

\centerline{Problems}

\item{1.} For a two-dimensional system of charged particles in a uniform
magnetic field $B$ (work in the gauge ${\bf A} = {B\over 2} (-y,x)$~~~),
write down the Hamiltonian in terms of the complex coordinates $z$ and
$\bar z$ where $z =(x+iy)/\ell$ and $\bar z = (x-iy)/\ell$ and $\ell$ is the
magnetic length defined to be $\ell = (\hbar c/eB)^{1/2} = {(1/eB)}^{1/2}$.

\item{a.)} Show that $[a,a^{\dagger}] = 1$ and  $[b,b^{\dagger}] = 1$
where $a = {1\over \sqrt 2} (2 {\partial\over \partial 
{\bar z}} + {z\over 2})$
and $b = {1\over \sqrt 2} (2 {\partial\over \partial z} + 
{{\bar z}\over 2})$.
\item{b.)} Express the Hamiltonian in terms of the operators $a,
a^{\dagger}, b$ and $b^{\dagger}$ and find the eigenvalues of $H$. What
are the quantum numbers required to classify the states? 
How do you see the
degeneracy of the states in terms of the quantum numbers?

\vskip 0.3in

\item{2.} Solve the problem of Mott scattering of anyons (scattering from
a $1/r$ potential). If you get stuck, look up Ref\KHARE.

\endpage

 \chapter{Many Anyon Systems:Quantum Statistical Mechanics}

 In this section, we shall study the quantum mechanical system of many
anyons using the virial expansion of the equation of state, which is valid
in the high temperature, low density limit. However, before starting on
the many anyon problem, we shall show how the
second virial coefficient of a system of free fermions or bosons is
obtained and
review in some detail, the cluster
expansion method, which leads to the expression for the  virial
coefficients in terms of the cluster integrals\HUANG.

For bosons and fermions, quantum statistical mechanics begins with the
calculation of the grand partition function. The canonical partition
function is given by
$$
Z(A,T) = \sum_{\rm states} e^{-\beta E_N} \eqn\ethreeone
$$
where $E_N$ is the energy of the $N$-particle system. In terms of
single particle energy levels $\ep$, $E_N$ can be written as
$$
E_N = \sump \ep \np \eqn\etwotwo
$$
where $\bp$ enumerates the energy levels and $\np$ is the occupation number
of each level. The occupation number $\np$ is just (0,1) for fermions and
(0,1,2,.....) for bosons, but it is constrained by the total number of
particles given by
$$
N = \sump \np. \eqn\ethreethree
$$
The canonical partition function cannot be easily evaluated because of
this constraint. However, the grand canonical partition function defined by
$$
Z_G(z,A,T) = \sum_{N=0}^{\infty} z^N Z(A,T)
= \sum_{N=0}^{\infty} z^N \sum_{\{\np\}\atop {\Sigma_{\bp} \np = N}}
e^{-\b\Sigma_p
\ep\np} \eqn\ethreefour
$$
where $\{\np\}$ represents the collection of values of $\np$ for the
different states,
invalidates the constraint, since it incorporates a sum over all $N$.
Now, from Eq.$\ethreethree$, we see  that $z^N = \Pi_p z^{\np}$
and $e^{-\b\Sigma_p \ep\np} = \Pi_p (e^{-\b\ep})^{\np}$ yielding
$$
Z_G(z,A,T) = \sumN \sum_{\{\np\}}
 \prodp (\zbp)^{\np}.
\eqn\ethreefive
$$
The whole purpose of considering the grand partition function
instead of the partition function becomes clear
when we realise that the two sums over $N$ and $\{\np\}$ reduce to
independent sums over each $\np$ -$i.e.$,
$$
Z_G(z,A,T) = \sum_{\np} \prodp (\zbp)^{\np} = \prodp \sum_n (\zbp)^n.
\eqn\ethreesix
$$
Thus, the grand partition function is given by
$$
\eqalign{
&Z_G(z,A,T) = \prodp (1+\zbp) \qquad {\rm for \quad fermions},\cr {\rm and}
\qquad
&Z_G(z,A,T) = \prodp{1\over (1 - \zbp)} \qquad {\rm for \quad bosons}, }
\eqn\ethreeseven
$$
respectively.
This method is inapplicable to anyons, because, as we have seen in Sec.(2),
the energy levels of an $N$-particle system cannot be computed in terms of
the single particle energies. Hence, the exact calculation of the
partition function for an anyon gas still remains an open problem.

However, there exist other approximation schemes under which quantum gases
are studied, one of which is called the cluster expansion (CE) method.
The classical CE involves a systematic expansion of the
interparticle potential, which is valid in the high temperature, low
density regime. The quantum CE is defined by analogy with the classical CE.
Using this expansion, corrections to the ideal gas law by systems of
interacting quantum particles can be computed.
This method is available to the
anyon gas as well, because anyons can always be thought of as interacting
bosons or fermions.

Let us briefly review the cluster expansion method before applying it to
the anyon gas. We consider a classical Boltzmann gas  with interactions.
The many particle Hamiltonian is given by
$$
H = \sum_{i}{{\bp}_i^2 \over 2m} + \sum_{i,j,...} V({\br}_i,\br_j,...)
\eqn\ethreenine
$$
and the canonical partition function for $N$ particles,
obtained by integrating over all of
classical phase space, is
$$
Z(A,T) = {1\over (2\pi)^{2N} N!} \int\dtwor \dtwop e^{-\b H}.
\eqn\ethreeten
$$
(Classically, the $N$ particles are distinguishable and taken to be
distinct. We divide by $N!$ to compensate for the overcounting.) The
momentum integrations can be performed, since the integrand can be
factorised into a product and the individual integrations are merely
Gaussian -$i.e.$,
$$
\int {d^2 {\bf p} \over (2\pi)^2} e^{-{\b\bp^2\over 2m}} = {m\over 2\pi\b}
\equiv {1\over \lambda^2} \eqn\ethreeeleven
$$
where $\lambda$ is called the thermal wavelength. Thus, the canonical partition
function, reduces to
$$
Z(A,T) = \norm \int \dtwor e^{-\b \Sigma_{i,j,..}V(\br_i,\br_j,..)}.
\eqn\ethreetwelve
$$
Now, the potential can be separated into sums of $n$-body potentials -
$i.e.$,
$$
\sum_{i,j,..} V(\br_i,\br_j,...) = \sum_{i,j} V(\br_i,\br_j) + \sum_{i,j,k}
V(\br_i,\br_j,\br_k) + ....\quad . \eqn\ethreethirteen
$$
We shall see that the lowest order correction to the ideal gas law
involves only two body interactions, but let us keep the formulation
general at this stage. The next step is to expand the integrand of
Eq.\ethreetwelve\  in powers of $e^{-\b\Sigma V} - 1$. We define
$$\eqalign{
e^{-\b V(\br_i,\br_j)} &\equiv (1 + f_{ij})\cr
e^{-\b V(\br_i,\br_j,\br_k)} &\equiv (1 + f_{ijk})}
\eqn\ethreefourteen
$$
and so on. Each of these $f_{ijk...}$ code for the
$n$-particle interactions involved in the potential.
The partition function can be written in terms of the $f_{ijk..}$ as
$$
\eqalign{
Z(A,T) &= \norm \int \dtwor \prod_{i<j} (1 + f_{ij}) \prod_{i<j<k}
(1 + f_{ijk}) \prod_{i<j<k<l} (1 + f_{ijkl}) .... \cr
&= \norm\int\dtwor[1+(f_{12} + f_{23} + ...) + (f_{12} f_{23} + f_{12} f_{14}
+ ...) + ...]\cr
&~~~~~~~~~~~~~~~ + (f_{12} f_{23} f_{13} + f_{12} f_{24} f_{14} +...) +
(f_{123} + f_{124} + ...) + ...\,.}
\eqn\ethreefifteen
$$
Now let us study, in more detail, each $l$-cluster involving interactions
between $l$ particles. A cluster integral $b_l$ is defined as
$$
b_l = {1\over l! \lambda^{2l-2} A} \int \dtwor ({\rm contribution \quad from
\quad {\it l-}clusters}) \eqn\ethreesixteen
$$
Explicitly,
$$
b_1 = {1\over A} \int d^2 {\bf r}_1 .1 = 1,
\eqn\ethreeseventeen
$$
$$\eqalign{
b_2 &= {1\over 2 \lambda^2 A} \int d^2\br_1 d^2\br_2 f_{12}\cr
    &={1\over 2 \lambda^2 A} \int d^2\br_1 d^2\br_3 f_{13}\cr
    &={1\over 2 \lambda^2 A} \int d^2\br_i d^2\br_j f_{ij},}
\eqn\ethreeeighteen
$$
 where the last equality is valid for any two arbitrary
 particles labelled $i$ and $j$, and
$$
b_3 = {1\over 6\lambda^4 A} \int d^2 \br_1 d^2 \br_2 d^2 \br_3
(f_{12} f_{13} + f_{13} f_{23} + f_{12} f_{23} + f_{12} f_{13} f_{23} +
f_{123} )
\eqn\ethreenineteen
$$
with similar contributions from any other choice of three particles.
Notice that the contributions obtained by changing the particle indices
are all equal. Also note that the computation of $b_2$ involves only
two-body interactions even if the Hamiltonian includes three or more body
interactions.

It is now clear that the partition function may be written
as a product of cluster integrals -$i.e.$,
$$
Z(A,T) \sim \norm \sum_{\{m_l\}} \prod_l (l! \lambda^{2l-2} A)^{m_l} b_l^{m_l}
\eqn\ethreetwenty
$$
with $\Sigma l m_l = N$. Here $m_l$ denotes the number of $l$-clusters and
$\{m_l\} = (m_1,m_2,...)$ denotes the collection of $m_l$.
However, we still need the combinatoric factor, which specifies the number
of times a given cluster integral appears in the product. Let us compute
this combinatoric factor.
Firstly, since there are $N$ particles in the system, naively we expect
$N!$ clusters. Hence, the
$R.H.S$ of Eq.$\ethreetwenty$ is multiplied by $N!$.
However, particles in the same cluster are indistinguishable; so we have
to divide by $(l!)^{m_l}$. Moreover, any two $l$-clusters
are indistinguishable.
Hence, for $m_l\quad l$-clusters, we have to divide the $R.H.S$ by $m_l!$.
Putting all this together, the partition function is given by
$$\eqalign{
Z(A,T) &= \norm \sum_{\{m_l\}} \prod_l (l! \lambda^{2l-2} A)^{m_l}
{N!\over (l!)^{m_l} m_l!} b_l^{m_l} \cr
&= \sum_{\{m_l\}} \prod_l {1\over m_l!} ({A b_l \over \lambda^2})^{m_l}}
\eqn\ethreetwentyone
$$
with $\Sigma l m_l = N$. (The simplest way, to convince yourself that this
expression for the partition function in terms of the cluster integrals
and all the combinatoric factors is correct, is to work out the partition
function for a few particle systems explicitly.)
As before, the constraint on the number of particles makes this partition
function difficult to evaluate. Hence, just as was done for fermions and
bosons, we move on to the grand partition function where the total number
of particles $N$ is also summed over.  The expression for the grand
partition function for this system is given by
$$\eqalign{
Z_G(z,A,T) &= \sumN z^N Z(A,T) \cr
 &=\prod_l \sum_{m_l =0}^{\infty} {1\over m_l!}
 ({A b_l z^l\over \lambda^2})^{m_l} \cr
 &= e^{\Sigma_l  ({A b_l z^l\over \lambda^2})}.}
 \eqn\ethreetwentytwo
$$
 Notice that here again, the unrestricted sum over $N$
 and the sum over $\{m_l\}$
 has been translated to an
unrestricted sum over each of the $m_l$. The equation of state
obtained from this partition function is given by
$$
{PA\over KT} = {\rm ln} Z_G = {A\over \lambda^2} \sum_l b_l z^l.
\eqn\ethreetwentythree
$$
Furthermore, the average number of particles $<N> = N$ is given by
$$
N = z{\partial\over \partial  z} {\rm ln} Z_G =
{A\over \lambda^2} \sum_l l b_l z^l.
\eqn\ethreetwentyfour
$$
Combining Eq.$\ethreetwentythree$ and Eq.$\ethreetwentyfour$, we get
$$
{PA\over NKT} = {\Sigma_l b_l z^l \over \Sigma_l l b_l z^l} =
{b_1 z + b_2 z^2 + ...\over b_1 z + 2 b_2 z^2 + ...}.
\eqn\ethreetwentyfive
$$
Since $b_1 = 1$, to first order in $z$, we find that
$$
{PA\over NKT} = 1 - b_2 z + O(z^2).
\eqn\ethreetwentysix
$$
Now, the virial coefficients of the system are defined by
$$
{PA\over NKT} = {P\over \rho KT} = \sum_{l=1}^{\infty} a_l
(\rho\lambda^2)^{l-1}
\eqn\ethreetwentyseven
$$
where $\rho$ is the density of particles. Thus, we have a
power series expansion of
the equation of state in terms of the inverse temperature $\lambda$ and
the density $\rho$. This is clearly a useful concept at high temperatures
and low densities, where only the first few terms in the series are likely
to be relevant.
Substituting for $(\rho\lambda^2)$ from Eq.$\ethreetwentyfour$, we see that
$$\eqalign{
{P\over \rho KT} &= \sum_{l=1}^{\infty} a_l (\Sigma_{l'=1}^{\infty} b_{l'}
l' z^{l'})^{l-1} \cr
&=a_1 + a_2 (z + 2 b_2 z^2 + .....)+...}
\eqn\ethreetwentyeight
$$
to lowest order in $z$. Comparing this equation with
Eq.$\ethreetwentysix$, we see that
$a_1 = 1$ and $a_2 = -b_2$. So to find the second virial coefficient,
we only need to compute the two-cluster integral, which, in turn, depends
only on two-body interactions.

The quantum cluster expansion is defined by analogy with the classical
cluster expansion. The quantum partition function is given by
$$\eqalign{
Z &= \sum_{\alpha} \int \dtwor \psi_{\alpha}^{*}(\br_1,...\br_N)
e^{-\b H} \psi_{\alpha}(\br_1,...\br_N) \cr
&= \norm \int \dtwor W_N (\br_1,....\br_N),}
\eqn\ethreetwentynine
$$
where $\{\psi_{\alpha}\}$
is a complete set of orthonormal wavefunctions for the system labelled by
the quantum number $\alpha$ and
the last equality defines $W_N(\br_1,...\br_N)$.
By comparing this equation with the
definition of the classical partition function in Eq.$\ethreetwelve$, we
see that
$$
W_N^{\rm cl}(\br_1,...\br_N) = e^{-\b\Sigma_{i,j} V(\br_i,\br_j)}
=\prod_{i<j}(1+f_{ij})
\eqn\ethreethirty
$$
(Since we shall only compute the second virial coefficient and we have
seen that the second virial coefficient depends only on two body
interactions, we have specialised to the
case of two-body interactions alone.)
Now, let us define new quantities $U_l(\br_1,...\br_l)$ through
$$\eqalign{
W_1(\br_1) &= U_1(\br_1), \cr
W_2(\br_1,\br_2) &= U_1(\br_1) U_1(\br_2) + U_2 (\br_1,\br_2), \cr
W_3(\br_1,\br_2,\br_3) &= U_1(\br_1) U_1(\br_2) U_1(\br_3) + U_1(\br_1)
U_2(\br_2,\br_3) + U_1(\br_2) U_2(\br_1,\br_3) \cr
&~~~~~+ U_1(\br_3) U_2(\br_1,\br_2)
+ U_3(\br_1,\br_2,\br_3),}
\eqn\ethreethirtyone
$$
and so on. From Eq.$\ethreethirty$, we see that the classical limits of the
$U_l$ functions can be identified as follows ---
$$\eqalign{
W_1^{\rm cl}(\br_1) = 1 &\Rightarrow U_1^{\rm cl}(\br_1) = 1, \cr
W_2^{\rm cl}(\br_1,\br_2) = 1.1 + f_{12} &\Rightarrow
U_2(\br_1,\br_2) = f_{12}, \cr
W_3^{\rm cl}(\br_1,\br_2,\br_3) = 1.1.1 +1.(f_{12} + f_{13} + f_{23})
 &+(f_{12}f_{13} +f_{12}f_{23} +f_{13}f_{23}) + f_{12}f_{23}f_{13}\cr
 \Rightarrow U_3^{\rm cl}(\br_1,\br_2,\br_3) = (f_{12}f_{13} &+ f_{12}
 f_{23} + f_{13}f_{23}) + f_{12}f_{13}f_{23},}
 \eqn\ethreethirtytwo
 $$
 and so on - $i.e.$, the $U_l(\br_1,...\br_l)$ are the quantum analogs of
the $l$-clusters in the classical case. Hence, for quantum statistical
mechanics, the classical cluster integrals can be replaced by the quantum
cluster integrals given by
$$
b_l = {1\over l!\lambda^{2l-2}A} \int \dtwor U_l(\br_1,...\br_l).
\eqn\ethreethirtythree
$$

To calculate the second virial coefficient for any system, we need $b_2$,
since we have already found that $a_2 = -b_2$ , and to find $b_2$, we need
to compute $W_2(\br_1,\br_2)$ which is a property of the two-body system.
Let the Hamiltonian for the two-body system be
$$
H = -{1\over 2m}({\bf \nabla}_1^2 + {\bf \nabla}_2^2)
+ v(|\br_1 - \br_2|)
\eqn\ethreethirtyfour
$$
with eigenvalues $E_{\alpha}$ and
eigenfunctions $\psi_{\alpha} (\br_1,\br_2)$.
We transform to the CM ($\bR$) and relative ($\br$) coordinates to solve
the problem. In terms of these coordinates,
$$
\psi_{\alpha}(\br_1,\br_2) \longrightarrow \psi_{\alpha}(\bR,\br) =
{e^{i\bP\cdot\bR}\over {\sqrt A}} \psi_{n}(\br) \eqn\ethreethirtyfive
$$
and $E_{\alpha} = \bP^2/4m +\epsilon_n$. Here, the quantum number $\alpha$
has been split into $(\bP,n)$ where $\bP$ refers to the continuum quantum
numbers of the CM system and $n$ is the quantum number labelling the
energy levels of the relative Hamiltonian. $\epsilon_n$ is found by
solving the eigenvalue equation of the relative Hamiltonian given by
$$
[-{1\over m} {\bf \nabla}^2 + v(\br)]\psi_n(\br) =
\epsilon_n (\br)\psi_n(\br).
\eqn\ethreethirtysix
$$
The definition of $W_2(\br_1,\br_2)$ in Eq.$\ethreetwentynine$ leads to
its identification in this system as
 $$\eqalign{
 W_2(\br_1,\br_2) &= 2\lambda^4 \sum_{\alpha} \psi_{\alpha}^*(\br_1,\br_2)
 e^{-\b H} \psi_{\alpha} (\br_1,\br_2) \cr
 &= {2 A \lambda^4 \over (2\pi)^2} \int d^2\bP \sum_n {e^{-\b \bP^2/4m-\b
 \epsilon_n}  \over A} |\psi_n(\br)|^2 \cr
 &= 4\lambda^2 \sum_n |\psi_n(\br)|^2 e^{-\b \epsilon_n} }
 \eqn\ethreethirtyseven
 $$
 which, in turn, identifies
 $$
 U_2(\br_1,\br_2) = 4\lambda^2 \sum_n |\psi_n(\br)|^2 e^{-\b\epsilon_n} -1.
 \eqn\ethreethirtyextra
 $$
 Hence, the definition of the quantum cluster integral in
 Eq.$\ethreethirtythree$ leads to
 $$
 b_2 = -a_2 = {1\over 2 A \lambda^2} \int d^2 \bR d^2 \br
 [4\lambda^2 \sum_n |\psi_n(\br)|^2 e^{-\b\epsilon_n} -1].
 \eqn\ethreethirtyeight
 $$
 The two individual terms in the integrand above give rise to area
divergences when integrated over $\br$ and $\bR$. It is only their
difference that is finite in the thermodynamic limit. (Even the difference
is not always finite. The usual condition for finiteness of the difference
is that the interaction between particles should be short-ranged.)
Often, it is more convenient to compute the second cluster integral as a
difference between $b_2$ for the system under study and $b_2^0$ for some
reference system, where $b_2^0$ is known exactly - $e.g.$, $b_2^0$ could
be computed when $v(\br) = 0$. Then , for normalised relative
wavefunctions ,-$i.e.$, when $\int d^2\br |\psi_n(\br)|^2 = 1$,
we get
$$
b_2 -b_2^0 = 2\sum_n (e^{-\b\epsilon_n} - e^{-\b\epsilon_n^0}).
\eqn\ethreethirtynine
$$

For the ideal Fermi and Bose gases, $b_2^0$ can be directly identified
from the equation of state. For the Fermi gas, from Eq.$\ethreeseven$, we
see that the equation of state is given by
$$
{PA\over KT} = {\rm ln} Z_G = {A\over (2\pi)^2} \int d^2\bp \, {\rm log}
(1 + \zbpm),
\eqn\ethreefortyone
$$
since for free fermions $\ep = \bp^2/2m$. The $R.H.S$ can now be expanded
in a power series in $z$ from which the $b_l$ may be identified - $i.e.$,
$$\eqalign{
{PA\over KT} &= {A\over
\lambda^2} (z - {z^2\over 2^2} + {z^3\over 3^2}  - ...) \cr
   &= {A\over \lambda^2} \sum _l b_l^0 z^l .}\eqn\ethreefortytwo
$$
Hence, $b_2^0$(fermions) is identified as $b_{2f}^0 = -1/4$.
Similarly, for the Bose gas,
$$\eqalign{
{PA\over KT} &= - {A\over (2\pi)^2} \int d^2\bp \, {\rm log} (1
- \zbpm)        \cr
&= {A\over \lambda^2} (z + {z^2\over 2^2} + {z^3\over 3^2} + ....) }
\eqn\ethreefortythree
$$
so that $b_2^0$(bosons) is identified as $b_{2b}^0 = 1/4$.

We had earlier noted that the free anyon gas is already an interacting
system, so a direct evaluation of the partition function is not possible.
However, the second virial coefficient can be computed as long as we know the
energy levels of the two anyon system. In Sec.(2), we explicitly
computed the energy
levels of two anyons in a harmonic oscillator potential. Let us use those
results to find the second virial coefficient of an anyon gas in
the same harmonic oscillator potential\SMHO. Ultimately, we shall take the
limit where the oscillator potential vanishes to obtain the virial
coefficient of the free anyon gas. We shall choose our reference system to
be that of free bosons ($\a = 0$).
{}From Eq.(2.39), we find that
$$
\eqalign{
b_2 - b_{2b}^0 &= 2 \sum_{j=0}^{\infty} [(j+1)e^{-\b(2j+1+\a/\pi)\omega}
+ j e^{-\b(2j+1-\a/\pi)\omega}
- (j+1)e^{-\b(2j+1)\omega} - j e^{-\b(2j+1)\omega}] \cr
&= 4 [\cosh (\a/\pi -1)\b\omega - \cosh \b\omega ]
\sum_{j=0}^{\infty} j e^{-2j\b\omega} \cr
&= {\cosh (\a/\pi -1)\b\omega - \cosh \b\omega \over \sinh^2 \b\omega}.}
\eqn\efortyfour
$$
To find the virial coefficient of a system of free anyons, we take the limit
$\omega\rightarrow 0$, which yields
$$
b_2 - b_{2b}^0
= {1\over 2} [({\a \over \pi})^2 - {2\a\over \pi}] ,
\eqn\ethreefortyseven
$$
so that substituting  $b_{2b}^0 = 1/4$, we get
$$
b_2 = {1\over 4} [2({\a \over \pi})^2 - {4\a\over \pi} +1 ] .
\eqn\ethreefortyeight
$$
Notice that this equation is only valid for $0\le \a < 2\pi$, since $\a$ being
a periodic variable specifying the statistics is periodic in $2\pi$.
Hence, the virial coefficient is non-analytic in $\a$ and has a cusp
whenever $\a = 2\pi j$ for $j$ an integer. By defining $\d$ such that $\a
= 2\pi j +\d$, we see that
$$
b_2 = {1\over 4}  [2({\d \over \pi})^2 - {4\d\over \pi} +1 ]
\eqn\ethreefortynine
$$
is valid for any $\a$. The virial coefficient can be plotted\AROVAS\
as a function
of $\a$ as shown in Fig.(13).

%\vbox {~\vskip 2.5in \centerline{Fig.13} \vskip .25in}
\epsfbox{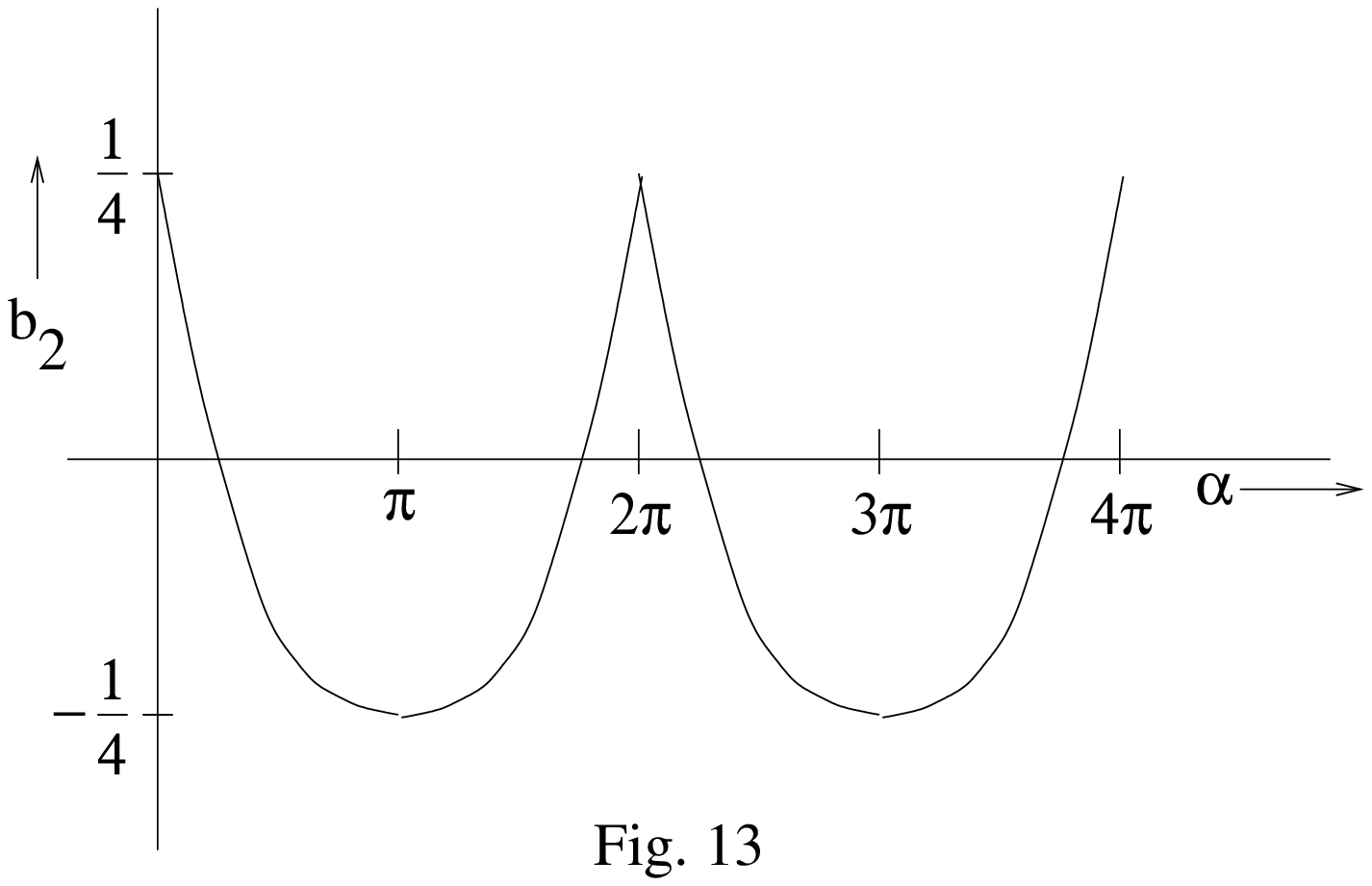}

\noindent
{}From the figure, it is clear that the values of the
virial coefficient of an anyon
gas interpolates between the values of that for fermions and bosons.

Our result for the virial coefficient of an anyon gas
is not obvious, though it is satisfying - just as spin and
statistics of anyons are intermediate between spin and statistics of
fermions and bosons, so are the virial coefficients.  In fact, it is not
even clear why the cluster expansion method works, because, when anyons
are considered as interacting fermions or bosons, even the two-body
interaction between particles is actually long range. However, the result
that we have obtained here for the anyon gas is consistent, because the
same answer has also been obtained using other regularisation schemes.
Here, we used the harmonic oscillator potential as a regulator to
discretise the energy levels and then took the limit where the oscillator
potential vanishes to obtain the virial coefficient, whereas the original
calculation of the virial coefficient involved a box normalisation\AROVAS.

The next logical step in this programme would be to compute the third
virial coefficient. But, this would require the knowledge of
three-body interactions. Since, as was mentioned in Sec.(2), the three
anyon problem has not been completely solved in any potential so far,
the computation of the third virial coefficient too is an unsolved
problem. However, as in the case of three anyon quantum mechanics, partial
results have been obtained\THIRDVIRIAL.

\vskip 1cm
\centerline{Problems}
\item{1.}In class, we studied the virial coefficient of a gas of anyons
using a harmonic oscillator potential as a regulator. Show that the same
answer is obtained using box regularisation. (This was how it was
originally done in Ref.\AROVAS.)
\endpage

\chapter{Many Anyon Systems : Mean Field Approach}

 The basic idea of the mean field approach is to replace the effect of
many particles by an `average' or `mean' field and to accomodate
deviations from the mean field as residual short range interactions. In
the context of anyons, the mean field approach involves replacing the
flux-tubes carried by the charges by a uniform magnetic field with the
same flux density\FHL\ANYSUP.
It is clear that this approximation is valid when the
density of flux-tubes (equivalently particles) is high and fluctuations
are small, -$i.e.$, it is a high density, low temperature expansion.

The many anyon Hamiltonian can be obtained by generalising the two anyon
Hamiltonian in Eq.\etwoone\ to $N$ particles as
$$
H= \s1N \pqa \eqn\efourone
$$
with
$$
\ba_i = \ptp \sij {{\hat z}\times \rij \over \mrij}.  \eqn\efourthree
$$
Thus the charge in each anyon sees the vector potential due to the
flux-tubes in all the other anyons. We can also compute the magnetic field
at the position of the $i^{th}$ charge. However, a naive computation leads
to
$$
b_i = {\bf \nabla} \times \ba_i = 0, \eqn\efourfour
$$
which is not surprising since $\ba_i$ can also be written as a gradient -
$i.e.$,
$$
\ba_i = \ptp \sij {\bf \nabla}_i \theta_{ij} \eqn\efourfive
$$
where $\theta_{ij}$ is the angle made by the vector $\rij$ with an
arbitrary axis.
But using a regularisation scheme \EXACT\ with
$$
\ba_i = \lim_{\epsilon \rightarrow 0} \ptp \sij {{\hat z}
\times \rij \over \mrij +\epsilon^2},          \eqn\efoursix
$$
we can show that
$$
b_i = {\bf \nabla} \times \ba_i = \phi \sij \delta \rij.  \eqn\efourseven
$$
Not surprisingly, there is no magnetic field at the position of the $i^{th}$
charge, due to any other particle, unless the two particles coincide.
However, in the mean field approach, the flux-tubes are replaced by a
constant magnetic field $b$ with the same flux density. Let us assume
that the density of anyons per unit area is given by $\rhobar$. Then the
flux density of the anyons is given by
$$
\int b_i d^2 \br = \phi (\rhobar -1) \simeq \phi \rhobar
\eqn\efoureight
$$
when the integral is over a unit area and when $\rhobar$ is sufficiently
large. Hence, the appropriate uniform magnetic field to be used in the mean
field approach is just
$$
b = \phi \rhobar \eqn\efournine
$$
so that we have a system of charges moving in a constant magnetic field
as illustrated in Fig.(14).

%\vbox{~\vskip 2.5in
%\vbox{\hbox{\vbox{
%\hbox{\qquad\qquad$b_i$ at charges}\hbox{\qquad\qquad
%${{\rm flux}\over {\rm area}}
%= \phi \rhobar$}}
%\qquad\qquad\qquad\qquad\qquad\qquad\qquad\qquad\vbox
%{\hbox{ uniform $\bbar$}\hbox{$ {{\rm flux}\over{\rm area}} = \phi \rhobar$}}
%}}
%\vskip .1in
%\centerline{Fig.14} \vskip .25in}
\epsfbox{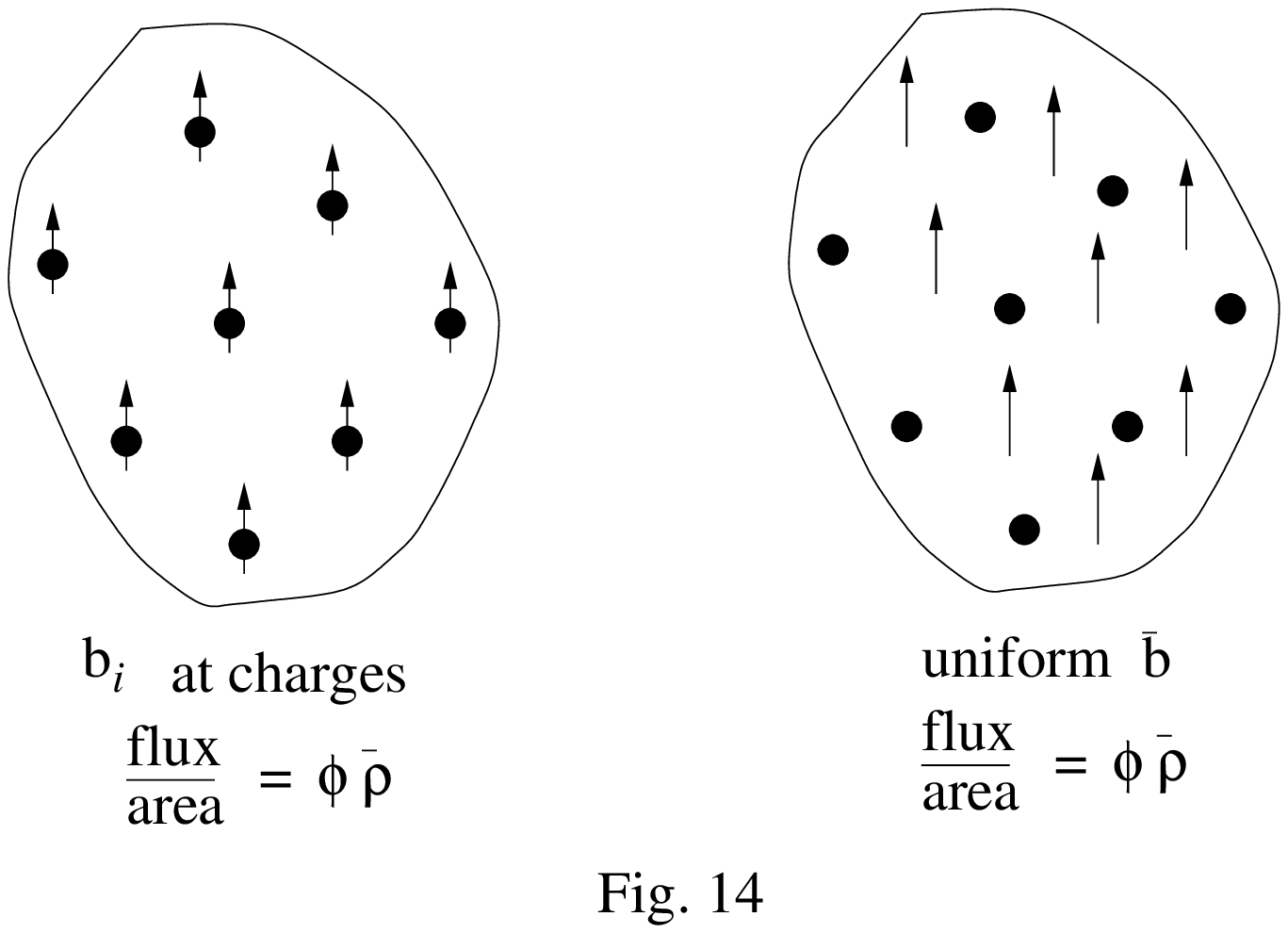}

Let us now study the quantum mechanical motion of a single particle in a
constant magnetic field. The Hamiltonian is given by
$$
H = {(\bp - q\ba)^2\over 2m}\eqn\efoureighta
$$
with the magnetic field $B = {\bf \nabla}\times {\bf A}$ = constant.
${\bf A}$ can be
chosen in many different ways (corresponding to different gauges), all of
which lead to the same $B$. We shall work in the Landau
gauge given by
$$
A_x = -By, ~~~A_y = 0. \eqn\efourninea
$$
(Another illustrative gauge (given as a problem in Chapter 2) 
is the symmetric gauge $A_x = -By/2,A_y = Bx/2$,
which is useful in the study of the Fractional Quantum Hall phenomenon.)
Therefore, the appropriate Schrodinger equation that governs the motion of
the particle is given by
$$
[{(p_x + qBy)^2 \over 2m} + {p_y^2 \over 2m}] \psixy = \Exy \psixy.
\eqn\efourten
$$
Here the $p_x$ and $p_y$ labels on the wavefunction and energy eigenvalues
are the momentum quantum numbers. Since there is no explicit
$x$-dependence in the Hamiltonian, the motion in the $x$-direction is free
and the wavefunction can be chosen to be of the form
$$
\psixy = e^{i p_x x} \chi(y) \eqn\efoureleven
$$
where $\chi(y)$ satisfies Eq.\efourten\ where, however, $p_x$ is now
interpreted as the eigenvalue of the $x$-momentum operator. By defining
$qB/m = \omega$ ($\omega$ is the cyclotron frequency of a charged
particle moving in the magnetic field $B$), and the magnetic length
$l = \sqrt{1/qB} = \sqrt {1/m\omega}$, Eq.\efourten\ may be rewritten as
$$
{\omega\over 2}[p_y^2 l^2 +  ({y \over l} + p_x l)^2] \chi(y) =
\Exy \chi(y). \eqn\efourthirteen
$$
This is just the the Schrodinger equation of a shifted harmonic oscillator
in the $(y,p_y)$ co-ordinates. Hence, the energy eigenvalues are discrete
and given by
$$
E_{p_x,n} = (n + 1/2) \omega, \eqn\efourfourteen
$$
where we have replaced the continuum label $p_y$ in the subscript of $E$
by the discrete
numbers $n$. These discrete energy eigenvalues labelled by the integers $n$
are called Landau levels. The corresponding eigenfunctions  are given by
$$
\chi(y) = {1\over \pi^{1/4}} {1\over \sqrt {2^n n! l}}
e^{-{(y/l+p_x l)^2 \over 2}} H_n (y/l + p_x l)
\eqn\efourfifteen
$$
where the $H_n$ are Hermite polynomials. Notice that the energy
eigenvalues are independent of $p_x$, which only effects a shift in the
origin of the oscillator. Since for motion in a plane, $p_x$ is
unrestricted, the Landau levels are infinitely degenerate. However, the
degree of degeneracy becomes finite when the motion in the plane is
restricted to a finite box with area $A = L_x L_y$. This degeneracy is
easily computed. For motion in a one dimensional box of length $L_x$, $p_x$
is  quantised as
$$
p_x = 2\pi n_x /L_x. \eqn\efoursixteen
$$
Furthermore, the allowed values of $p_x$ are restricted by the condition
that the centre of the oscillator has to lie between $0$ and $L_y$. Hence,
$$
{2\pi n_x\over L_x} l^2 <L_y \eqn\efourseventeen
$$
which, in turn, implies that the number of allowed values of $n_x$, or
equivalently, the degeneracy of the Landau levels per unit area is given by
$$
{n_x\over L_x L_y} = {1\over 2\pi l^2} = {qB\over 2\pi}. \eqn\efoureighteen
$$

Let us now return to the many anyon problem which had earlier been reduced,
in a mean field approach,
to the problem of fermions or bosons moving in a uniform magnetic field.
We shall choose our anyons to be fermions\ANYSUP, rather than bosons,
with attached flux-tubes, mainly because the problem of bosons in a
magnetic field is itself unsolved. Hence, the analysis of fermions in a
magnetic field (leading to anyon superconductivity) is easier. The other
reason is that many properties of anyons appear to show a cusp in their
behaviour in the bosonic limit. This was seen in Sec.(2), in the problem of
two anyons in a harmonic oscillator potential, as well as in Sec.(3), where
the second virial coefficient of the anyon gas was computed.
Hence, the
anyon gas may not have a smooth limit as the statistics parameter goes
to zero if we start by perturbing from the bosonic end.
However, the results that we shall obtain here by considering anyons to be
fermion with attached flux-tubes have also been argued by starting with
anyons as bosons\WENZEE\ albeit with some approximations and hand-waving. 
Ultimately, whether we start with anyons as
bosons or fermions is a matter of choice, and for each specific property
of the anyon gas, one or the other approach may be easier.

In the mean field approach,
we need to solve the
problem of fermions moving in a uniform magnetic field
$b = \phi \rhobar = 2\alpha\rhobar/q$. (Remember that in Sec.(2),
we had shown that the statistics factor $\alpha$ is given by $q\phi/2$.)
The degeneracy of the Landau levels in this field is given by
$$
{\rm deg} = {qb\over 2\pi} = {\alpha \rhobar\over\pi}.
\eqn\efournineteen
$$
Now, let us choose the statistics parameter $\alpha$ to be of the form
$\alpha = \pi/n$, where $n$ is any integer, so that the degeneracy is simply
$$
{\rm deg} = {\rhobar\over n}. \eqn\efourtwenty
$$
Since $\rhobar$ is the density of particles and each level can contain
$\rhobar/n$ particles,  clearly $n$ Landau levels are completely filled.
The next available single particle state is in the next Landau level which
is an energetic distance $\omega = qb/m$ away. In the many-body or
condensed matter parlance, this is called having a gap to single particle
excitations. But if $n$ is not an integer, then the last Landau level will
only be partially filled and there will be no gap to single particle
excitations. Hence, the parameter fractions $\alpha = \pi/n$ appear to be
special ($e.g.$, like the magic numbers in the shell models of atomic and
nuclear physics) and hence the states formed at these fractions should be
particularly stable.

To prove that the states formed at these special fractions are
superconducting, we have to study the effect of adding a real magnetic
field $B$ to the fictitious magnetic field $b$ and check whether a Meissner
effect exists. The argument differs slightly depending on the relative
signs of $B$ and $b$, and we shall consider both the cases separately.

When the real magnetic field is aligned parallel to the fictitious
magnetic field, they add and increase the degeneracy of the Landau level -
$i.e.$,
$$
{\rm deg} = {q(b+B)\over 2\pi}. \eqn\efourtwentyone
$$
But the number of particles per unit area $\rhobar = nqb/2\pi$ remains
unchanged. Hence, now the highest Landau level is only partially filled.
Let us denote its filling fraction by $(1-x)$. From the conservation of
density of particles, we have
$$
(n-1){q(b+B)\over 2\pi} + {q(b+B)\over 2\pi}(1-x) = \rhobar = {nqb\over 2\pi}
\eqn\efourtwentytwo
$$
from which we see that
$$
(b+B)x = Bn. \eqn\efourtwentythree
$$
Also from the energy eigenvalues in Eq.\efourfourteen,
we see that the total energy of $\rhobar$ particles is given by
$$
E = \qbptp \sum_{j=0}^{n-2}(j+{1\over 2}) \omega + \qbptp (n-{1\over 2})
(1 - x) \omega,
\eqn\efourtwentyfour
$$
which can be simplified to give
$$
E = {q(b+B)\over 2\pi} {q(b+B)\over m}[{n^2\over 2} - (n - {1\over 2})x].
\eqn\efourtwentyfive
$$
Substituting for $x$ from Eq.\efourtwentythree, we get
$$
E = {q^2 n^2\over 4\pi m}[b^2 +{bB\over n} - B^2 (1 - {1\over n})]
\eqn\efourtwentysix
$$
For small external magnetic fields $B$, the energy relative to the
ground state with no magnetic field is clearly positive and grows linearly
with $B$.
Thus, the anyon gas is a perfect diamagnet and tends to expel any
external flux.

When the external magnetic field $B$ is in the opposite direction to
statistical magnetic field $b$, the degeneracy of Landau levels decreases
- $i.e.$,
$$
{\rm deg} = {q(b-B)\over 2\pi}. \eqn\efourtwentyseven
$$
So some of the $\rhobar$ particles have to occupy the $(n+1)^{th}$
Landau level.
Let us denote the filling fraction of the highest level by $x$. Then from
conservation of particles, we have
$$
{n q (b-B)\over 2\pi} +{q(b-B)x\over 2\pi} = {nqb\over 2\pi}
\eqn\efourtwentynine
$$
leading to
$$
(b-B)x = Bn. \eqn\efourtwentynine
$$
The total energy of the system is given by
$$
E = {q^2n^2\over 4\pi m} [b^2 + {bB\over n} - B^2(1+{1\over n})]
\eqn\efourthirty
$$
Notice that the linear term remains the same for $B$ parallel and
anti-parallel to $b$. Hence, once again for small $B$, the energy
of this state relative to the state with $B=0$ is positive and grows
linearly with $B$, emphasizing the need to repel any external magnetic field.

We have just demonstrated the Meissner effect by showing that the anyon
gas finds it energetically favourable to exclude any external magnetic field.
Thus, the anyon gas is a superconductor.
The effect of adding an external magnetic field to the anyon gas can be
depicted schematically as shown in Fig.(15). (In the figure, crosses
denote particles and circles denote holes.)

%\vbox{~\vskip 2.5in
%\vbox{\hbox{\vbox{\hbox{${\bar \rho}$=5,~{\rm Ext. mag. field} = 0}
%\hbox{ ~~~deg. of L.L = 5}
%\hbox{~~~~~~~~~~~~15a}}
%\qquad\vbox
%{\hbox{${\bar \rho}$=5,~{\rm Ext. mag. field} = +B}
%\hbox{ ~~~deg. of L.L = 6}
%\hbox{~~~~~~~~~~~~15b}}
%\qquad\vbox
%{\hbox{${\bar \rho}$=5, ~{\rm Ext. mag. field} = $-$B}
%\hbox{ ~~~deg. of L.L = 4}
%\hbox{~~~~~~~~~~~~15c}}}}
%\vskip .1in
%\centerline{Fig.15} \vskip .25in}
\epsfbox{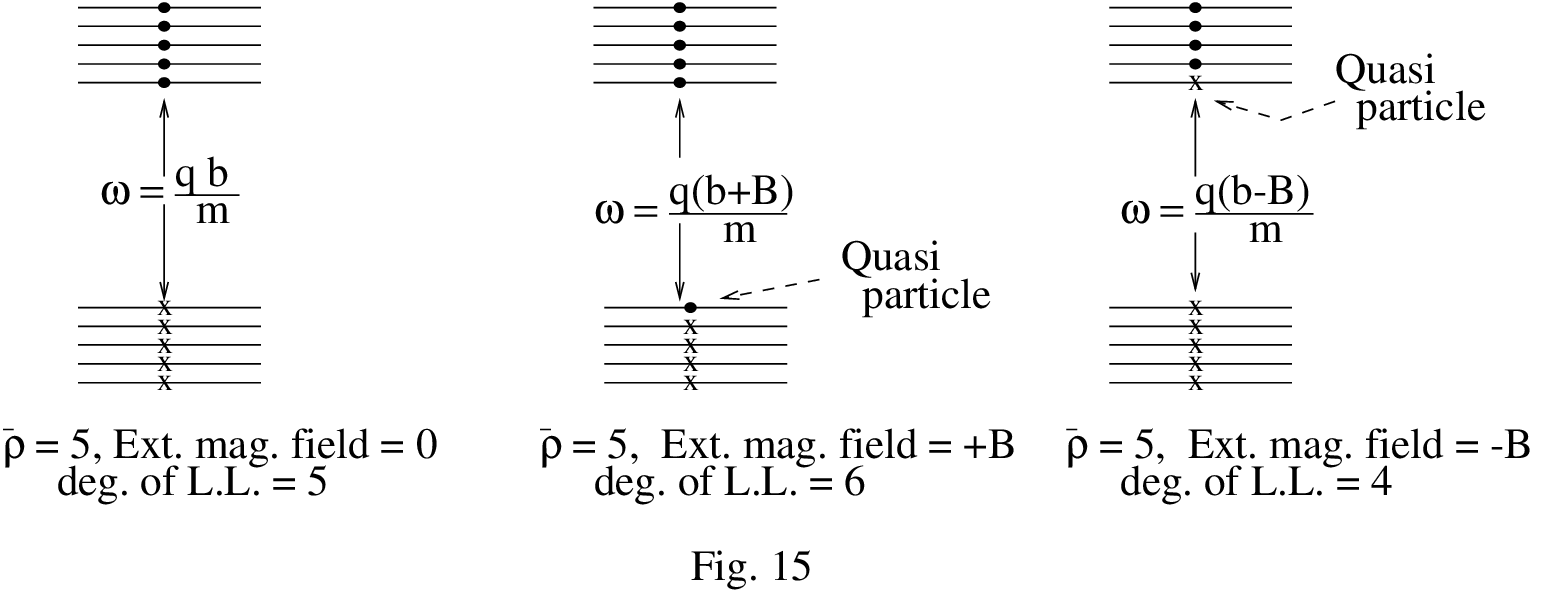}

Compare this with the schematic depiction (see Fig.(16)) of the creation of a
quasiparticle-quasihole pair in the same system which costs an energy $qb/m$.

%\vbox{~\vskip 2.5in
%\vbox{\hbox{\qquad\qquad
%\vbox
%{\hbox{$~~~~~~~~~~{\bar \rho}$=5}\hbox{ deg. of L.L = 5}}
%\qquad\qquad\qquad\qquad\qquad\qquad\qquad\qquad\vbox
%{\hbox{$~~~~~~~~~~{\bar \rho}$=5}\hbox{ deg. of L.L = 5}}}}
%\vskip .1in
%\centerline{Fig.16} \vskip .25in}
\epsfbox{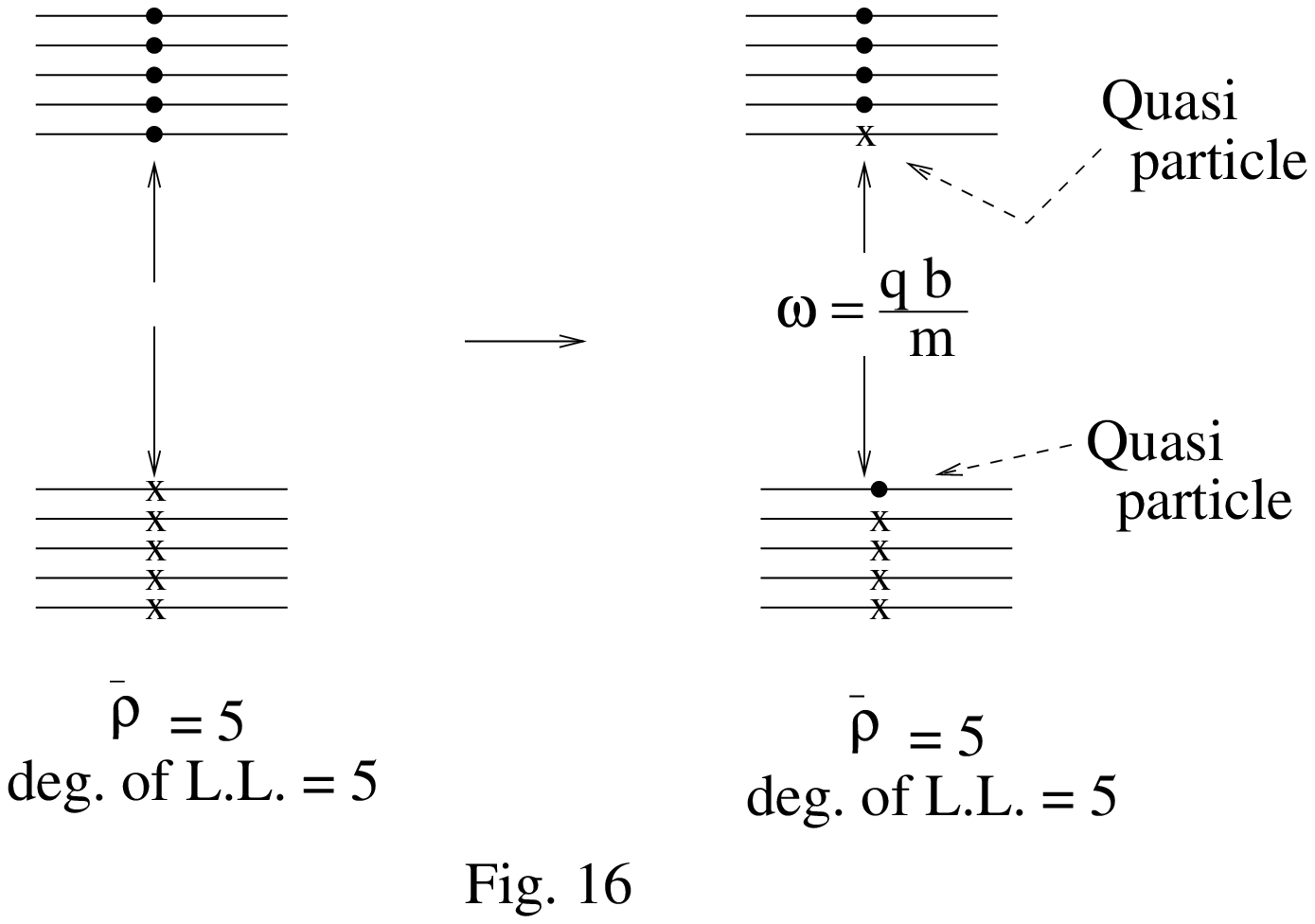}

It is clear that the production of quasiparticle and quasiholes are
closely related to the presence of a real magnetic field, because if the
quasiparticle and quasihole were spatially separated, then the quasiparticle
excitation is analogous to the situation depicted in Fig.(15c) with a real
magnetic field $-B$ and the quasihole excitation is analogous to the
situation in Fig.(15b)
with the field $+B$. (For $\rhobar$ sufficiently large, the
difference between $\rhobar$ and $\rhobar\pm 1$ is negligible.)
With this connection, the Meissner effect can also be argued in the
following way. Since for anyons, the magnetic field
and hence degeneracy of Landau levels is tied to the
density, to accomodate any external magnetic field which changes the
degeneracy, particles (or holes) have to be excited across the gap.
This costs energy and hence, penetration by magnetic fields is
unfavourable. Conversely, we also see that if particles do not fill Landau
levels, there must be a real magnetic field to account for the mismatch
between density of particles and degeneracy of Landau levels. Hence,
this argument also implies that every
quasiparticle excitation in the system
is accompanied by a real magnetic field, so that
in anyon superconductors, charged quasiparticle excitations and vortex
excitations are indistinguishable, in contrast to usual superconductors where
there are two types of excitations.

To actually prove superconductivity in the anyon gas, we need to show
that the collective  excitation in the system is massless. This can be
done by including fluctuations (residual interactions) about the mean
field state\FHL\ANYSUP\ in a random phase  approximation.
Relativistic field theory models have also been used to obtain the
massless mode\BANKSLYKKEN.
However, even
without going into the details of the calculation, there exists a simple
heuristic argument to indicate the presence of the massless collective
mode. Consider a very long wavelength density fluctuation (a collective
excitation). Then $\rhobar$, though varying, is approximately constant
over macroscopic lengths. Within each such macroscopic area, $b =
\alpha\rhobar/\pi = \rhobar/n$   is constant. Hence, locally the system
always has $n$ filled Landau levels. Therefore, such an excitation neither
requires any particle to be excited into a higher Landau level nor does
it require any energy. Hence, such a wave is massless - $i.e.$, we have
proved the existence of a massless collective mode.

Much further work has been accomplished in this field. Important questions
like the persistence of superconducting currents at finite temperatures
have been addressed\FTAS, though the results are not yet conclusive.
The connection between superconducting anyon states and FQHE states have
been explored\ASFQHE\ and for interacting anyons,
new superconducting states have been found\RATSTAT\ at statistics parameter
fractions $\alpha = \pi/\nu$, where $\nu$ is a fraction at which FQHE occurs.
Also, the all-important question of experimental predictions and tests of
anyon superconductors have been studied\ASTESTS.
On the experimental front, the one robust experimental prediction of all
anyon models of superconductivity has been violation of the discrete
symmetries parity $P$ and time reversal $T$. But recent experimental
results\PTEXPT\ (though somewhat controversial) appear to disfavour bulk
$P$ and $T$ violation. Hence, the latest theoretical works\LAYER\CSL\ have
concentrated on studying the effects of layering on anyon
superconductivity, in order to understand whether $P$  and $T$ violation
survive layering. However, even if present day high $T_c$ superconductors
are unlikely candidates of anyon superconductors, efforts to understand
many anyon systems and, in particular, their superconducting behaviour
remains undiminished.

\endpage

\centerline{Problems}
\item{1.} The standard mean field approach to anyon superconductivity
starts from fermions in a magnetic field. But if we start with
anyons as bosons plus attached flux-tubes, then the mean field problem to
be solved is that of bosons (albeit hard-core because of the hard-core
nature of anyons) in a magnetic field.  Can you solve this problem?
(This is as yet unsolved, but for starters, look up Ref.\JAINRAO. )

\endpage

\chapter{Anyons in Field Theory}

Bosons and fermions are introduced in a second quantised field theory
formalism through field operators that obey local commutation
or anticommutation rules.
However, anyon field operators , being representations of the braid group
instead of the permutation group, cannot obey local commutation rules.
This makes it hard to construct anyon field theories in the canonical way.
Path integral quantisation is no easier, since the phase picked up by any
anyon moving along a particular path depends on the position of all the
other anyons in the system. Hence, unlike the case with bosons or
fermions, paths cannot be weighted by a unique weight factor.
But the study of anyons in a field theoretic formulation was rendered
possible by the introduction\AROVAS\
of an elegant idea called the
Chern-Simons construction. Here, anyons were introduced as interacting
fermions or bosons.  It is this formulation (as opposed to more abstract
formulations\FROHLICH\ ) that we shall study in this section.

Let us consider any field theory in $2+1$ dimensions described by a
Lagrangian $L$ and having a conserved current $j_{\mu}$ - $i.e.$,
$\partial^{\mu} j_{\mu} = 0$. We can manufacture a gauge field $a_{\mu}$
and add to the Lagrangian
$$
\Delta L = j_{\mu} a^{\mu} - {\mu\over 2} \epsilon_{\mu\nu\alpha} a^{\mu}
\partial^{\nu} a^{\alpha}. \eqn\efiveone
$$
Is this an allowed extension of the Lagrangian?
Firstly, if $a_{\mu}$ is to be a gauge field, $\Delta L$ has to be gauge
invariant. Under $a_{\mu} \rightarrow a_{\mu} - \p_{\mu} \Lambda$,
$$
\int \Delta L \,d^3 x \rightarrow  \int \Delta L \,d^3 x
 - \int j_{\mu} (\p^{\mu} \Lambda) \, d^3 x + {\mu\over 2} \int
 \emna (\p^{\mu} \Lambda)  (\p^{\nu} a^{\alpha}) \,d^3 x.
 \eqn\efivetwo
$$
The third term obviously vanishes (upto surface terms) on integrating by
parts due to the anti-symmetry of $\emna$. The second term also
certainly vanishes on
integrating by parts at the equation of motion level, since $\p_{\mu}
j^{\mu} = 0$. However, for any explicit current, we can, in fact,
construct a gauge invariant coupling of the current with a local gauge
field. For example, when $j_{\mu}$ is a fermionic current,
the gauge invariant coupling is just $\jmam = {\bar \psi} \gamma_{\mu}
\psi a^{\mu}$. But when $j_{\mu}$ is a scalar current, $\jmam$ is replaced
by $D_{\mu} \phi D^{\mu} \phi $. Hence, the $\Delta L$ in Eq.\efiveone\ is
certainly an allowed gauge invariant extension of any Lagrangian.

The term
$(\Delta L)_{CS} = \CS $  is called the Chern-Simons ($CS$) term.
The $CS$ term and its non-abelian generalisations
are interesting field theories in their own right and have been studied
for years\CSREFS\ in other contexts.
More recently, they have shot into prominence as
prototypes of `topological' or `metric independent' field theories\WITTEN.
However, for our purposes here, we shall only recollect the following
salient features of Chern-Simons theories. An abelian $CS$ field theory is
described by
$$
L = -{1\over 4} F_{\mu\nu} F^{\mu\nu} + {\mu \over 2} \emna A^{\mu}
\p^{\nu} A^{\alpha}, \eqn\efivefour
$$
where $F^{0i} \equiv \p_0 A_i - \p_i A_0 = E_i$ and $ (1/2) \e0ij
F^{ij} = B$. It is clear that the electric field $E_i$ is a two component
vector and the magnetic field $B$ is a pseudoscalar, and that these are
the only non-zero components of $F^{\mu\nu}$,
since we are in $2+1$ dimensions. The parameter $\mu$ in Eq.\efivefour\
has the dimensions of a mass and is the gauge invariant mass term for the
gauge field. This can be seen by explicitly computing the propagator.
Furthermore, the $CS$ term is odd under the discrete symmetries parity $P$
($x \rightarrow -x, y \rightarrow y$) and time reversal $T$ ($t
\rightarrow -t$). Thus, the Lagrangian in Eq.\efivefour\ describes a
$U(1)$ gauge theory with a massive photon. A gauge invariant mass for the
photon has been introduced at the expense of the violation of $P$ and $T$.

With this introduction to $CS$ or topological field theories, let us get
back to the study of Eq.\efiveone. Notice that the Lagrangian
in Eq.\efiveone\
does not include the usual kinetic piece, the  $F_{\mu\mu}F^{\mu\nu}$
term for the gauge field. Also, from the equation of motion, we get
$$
j_{\mu} = \mu \emna \p^{\nu} a^{\alpha}, \eqn\efivefive
$$
which, when coupled with a gauge condition allows for a solution for the
gauge field in terms of the current $j_{\mu}$. ( We shall see this
explicitly later in this section.) Hence, the motion of the gauge
field $a_{\mu}$  is completely determined by the current $j_{\mu}$ and
has no independent dynamics. In this respect, $CS$ theories of
relevance to anyons (or metric-independent field theories) differ from
most of the $CS$-models studied earlier.
However, the gauge field does affect the statistics of the current carrying
particles. Integrating the zeroth component of Eq.\efivefive\ over all
space, we get
$$
\eqalign{
\int j^0 \, d^2\br &= \mu \int \e0ij (\p^i a^j)\, d^2\br \cr
\Rightarrow \quad q &= \mu \phi.} \eqn\efivesix
$$
Thus, every charge $q$ of the current $j_{\mu}$ is accompanied by a flux
$\phi$ and is an anyon. This mechanism of attaching fluxes to charges is
called the $CS$ construction. The current $j_{\mu}$ can be a fermion
number current ($j_{\mu} = {\bar \psi} \gamma_{\mu} \psi$) in which case
fermions turn into anyons, or a bosonic current ($j_{\mu} = \phi ^{\dagger}
\p_{\mu}\phi - (\p_{\mu}\phi)^{\dagger}\phi$)
so that bosons turn into anyons, or even a topological
current so that topological objects like solitons and vortices turn into
anyons.

Let us compute the statistics of these particles in the field theory
context. In the path integral formalism, when two such objects (flux
carrying charges) are exchanged, the phase is given by $e^{iS_{\rm ex}}$ where
$\Sex$ is the action involved in exchanging them.

%\vbox{~\vskip 1.0in \centerline{Fig.17} \vskip .2in}
\epsfbox{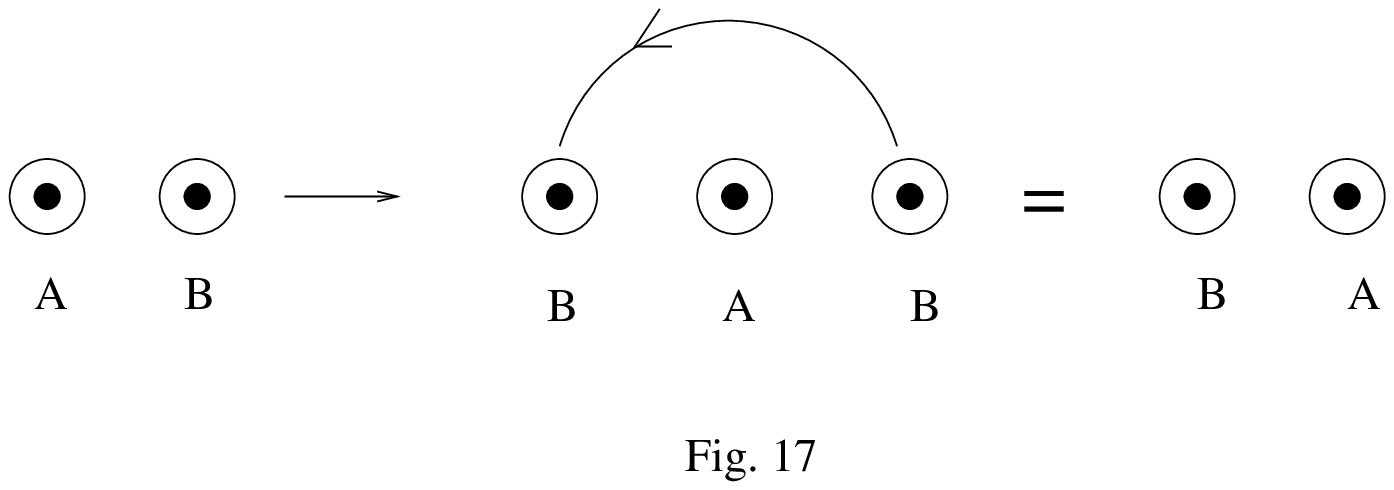}

\noindent
Now, it is clear that $S_{\rm ex} = (1/2) S_{\rm rot}$, where $\Srot$ is
the action for taking particle $B$ all around particle $A$.

%\vbox{~\vskip 1.0in \centerline{Fig.18} \vskip .2in}
\epsfbox{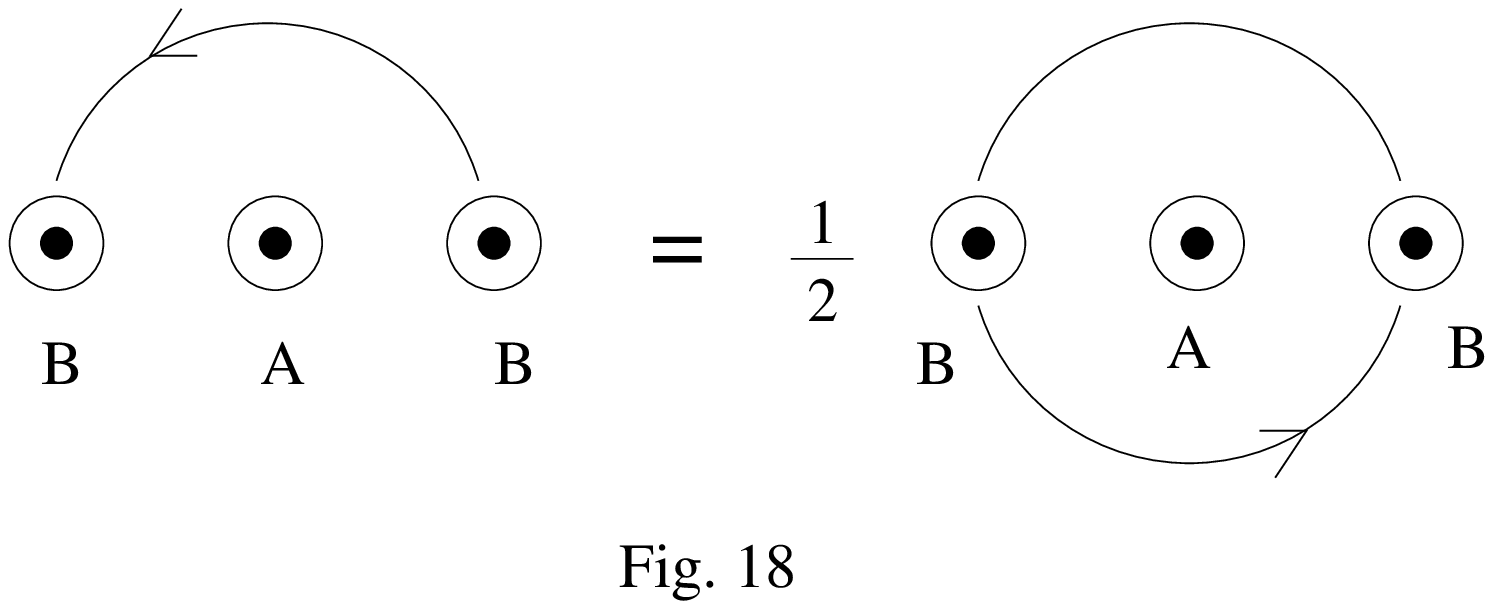}

\noindent
$\Srot$, in turn, is twice the Aharanov-Bohm action $\SAB$ required to
take a single charge around a single flux, because $\Srot$  involves
taking both a charge and a flux around a charge and a flux. Therefore,
$$
\Sex = \Srot/2 = \SAB. \eqn\efiveseven
$$
The Aharanov-Bohm phase is easily computed since we know the
`electromagnetic' Lagrangian (Eq.\efiveone\ ) that governs the motion of a
charged particle in the field of a flux-tube and is given by
$$
\eqalign{
\Sex = \SAB &= \int \Delta L \, dt \cr
&= \int j_i a^i \, dt + {\mu\over 2} \int \e0ij a_i \p_0 a_j \,dt}
\eqn\efiveeight
$$
(in the $a_0 = 0$ gauge). Since $j_i = q v_i$, the first term in
Eq.\efiveeight\ gives
$$
\int j_i a^i \, dt = q \int {\bf a} \cdot d{\bf l} = q\phi.
\eqn\efivenine
$$
Also, from the equation of motion in Eq.\efivefive, we see that the motion
of the gauge field is related to the current as
$$
\mu \e0ij \p^0 a^j = - j_i. \eqn\efiveten
$$
Substituting Eq.\efiveten\ in Eq.\efiveeight, we see that
$$
\Sex = q\phi/2 \equiv \alpha.   \eqn\efiveeleven
$$
Thus, the statistics of a charge with a flux induced on it by a $CS$
construction is $\alpha = q\phi/2$, consistent with  what we had earlier
derived for the charge-flux composite in the Hamiltonian formulation.

The relation between the Lagrangian $CS$ field theory formulation and the
Hamiltonian formulation for many anyons introduced in Sec.(4), can be
explicitly demonstrated\ANYSUP\
by carrying out the canonical transformation
between the Lagrangian and Hamiltonian formulations for a gas of anyons.
We start with the Lagrangian formulation of a gas of anyons (represented
by point particle bosonic charges with attached flux-tubes) given by
$$
L = \sum_{\alpha} [{m\over 2} {\dot {\bf r}}_{\alpha}^2 +
q a_0 ({\bf r}_{\alpha})
+ q {\dot{\bf r}}_{\alpha}\cdot{\bf a}({\bf r}_{\alpha})] -
{\mu\over 2} \int \emna a^{\mu} \p^{\nu} a^{\alpha} \, d^2\br
\eqn\efivetwelve
$$
where $\alpha$ is the particle index and the coupling between the particle
with charge $q$ and the gauge field $a_{\mu}$ is as in
standard electromagnetism.
However, the usual kinetic term of electromagnetism
$F_{\mu\nu} F^{\mu\nu}$ is now replaced by
the $CS$ term, in accordance with Eq.\efiveone.
This Lagrangian may be rewritten as
$$
L = \sum_{\alpha} {m\over 2} \xss + \int a_0 ( j_0 - \mu \e0ij \p_i
a_j)\,d^2\br + q\sum_{\alpha} \xs\cdot {\bf a} + {\mu\over 2} \int \e0ij a_i
{\dot a}_j \, d^2\br \eqn\efivethirteen
$$
where
$$
j_0 = q \sum_{\alpha}\delta^2 ({\bf r} - {\bf r}_{\alpha})
\eqn\efivethirteene
$$
represents the point particle charge density of the anyons. Notice that
apart from the first term, all the terms are linear in $a_0$ or time
derivatives. Also, since there is no kinetic term for $a_0$, the equation
of motion with respect to $a_0$ yields the constraint
$$
{\p L \over \p a_0} = 0 \Rightarrow j_0 = \mu \e0ij \p_i a_j = \mu b,
\eqn\efiveeighteen
$$
so that the field strength $f_{ij}$ is completely determined by $j_0$. But
for an abelian gauge field theory, the entire gauge invariant content of
the gauge field $a_i$ is contained in the field strength $f_{ij}$. Hence,
if we eliminate the extra degrees of freedom in $a_i$ by a gauge
condition, we can actually solve for $a_i$ in terms of $j_0$, so that the
motion of the gauge field is entirely determined by the fields forming
$j_0$ and has no independent dynamics. The $CS$ gauge field has been
introduced merely to attach flux-tubes to charges and turn them into
anyons.

To go to the Hamiltonian formulation, we require the canonical momenta
given by
$$
\pa = {\p L\over \p \xs}  = m\xs + q\ba
\eqn\efivefourteen
$$
and
$$
p_i = {\p L\over \p {\dot a}_i} = - {\mu\over 2} \int \e0ij a^j \, d^2\br.
\eqn\efivefifteen
$$
The Hamiltonian can now be written as
$$
\eqalign{
H &= \sum_{\alpha} \pa {\dot\br}_{\alpha} + p_i {\dot a}_i - L \cr
  &= \suma (m \xs +q \ba) \cdot \xs - {\mu\over 2} \int \e0ij
{\dot \ba}_i \ba_j \, d^2\br - \suma {m\over 2} \xss - q\suma \xs\cdot\ba
+{\mu\over 2} \int \e0ij {\dot \ba}_i \ba_j \, d^2\br \cr
 &= \suma {m\over 2} \xss =
 \suma{(\pa - q\ba)^2 \over 2m}, }
 \eqn\efiveseventeen
$$
which, in terms of the velocity is just the Hamiltonian of free particles.
So the classical equations of motion are identical to the equations of
motion for free particles. What has been altered, however, is the relation
between canonical velocity and momenta (see Eq.\efivefifteen\ ). Hence,
the quantum commutation relations are no longer the same. This is the
significance of introducing the $CS$ term without the usual kinetic piece
and is consistent with our original introduction of $a_{\mu}$ in Sec.(4),
as just a way of enforcing anyon boundary conditions in a different gauge.

Let us now solve for $a_i$ in terms of $j_0$ from Eq.\efiveeighteen\ and the
additional gauge condition
$$
\p_i a^i = 0. \eqn\eeighteene
$$
The solution to these two equations, for the $j_0$ given in
Eq.\efivethirteene,
is given by
$$\eqalign{
a_i(\br) &= {1\over 2\pi\mu} \int d^2 \brprime \e0ij
{(\br - \brprime)_j\over |\br - \brprime|^2}
j_0(\brprime) \cr
&= {q\over 2\pi\mu} \suma \e0ij {(\br - \brprime_{\alpha})\over |\br -
\brprime_{\alpha}|^2}}
\eqn\efivenineteen
$$
which can be easily checked. The gauge condition is obviously satisfied
due to the antisymmetry of $\e0ij$. Also, by regularising the denominator of
$a_i$  as explained in Sec.(4), we can explicitly check that
$$
b = {\bf \nabla} \times \ba = {1\over 2\pi\mu} \int j_0(\brprime) 2\pi \delta
(\br - \brprime)\, d^2\brprime = {j_0(\br)\over \mu}.
\eqn\efivetwenty
$$
Notice that the many body Hamiltonian in Eq.\efiveseventeen, along with
the solution for $a_i$ in Eq.\efivetwenty, is precisely the many body
Hamiltonian (Eq.\efourone) and gauge field $a_i$ (Eq.\efourthree) that
was used in Sec.(4), where it was obtained by generalising the two anyon
Hamiltonian. Thus, we have established the equivalence of the Lagrangian
$CS$ formulation and the many body Hamiltonian formulation  used in the
earlier sections\ANYSUP.

Let us now study a study a specific example of a field theory
- the abelian Higgs model with  a $CS$ term\EXAMPLE\ - whose
solitons (classical solutions of the equations of motion) are `anyonic'.
We shall first construct the topologically non-trivial vortex solutions of
the abelian Higgs model and then show that they are charged and have
fractional spin in the presence of a $CS$ term.  The Lagrangian for this
model is given by
$$
L = - {1\over 4} F_{\mu\nu}F^{\mu\nu} + {1\over 2} (\p_{\mu} -iqA_{\mu})
\phi^{*} (\p^{\mu} +iqA^{\mu}) \phi - c_4(\phi^*\phi - {c_2\over 2c_4})^2
+ {\mu\over 4} \emna F^{\mu\nu} A^{\alpha} .
\eqn\efivetwentyfive
$$
This model has a $U(1)$ gauge symmetry. However, when $c_2>0$, the
potential energy $V(\phi)$ is minimised when
$$
V(\phi) = c_4 (\phi^* \phi - {c_2\over 2c_4})^2 = 0
\Rightarrow |\phi | = \sqrt{{c_2\over 2c_4}} = v,
\eqn\efivetwentysix
$$
where $v$ is the vacuum expectation value. Hence, the $U(1)$ symmetry is
spontaneously broken by the vacuum.

The usual vacuum has
$$
\phi(r,\theta) = v, \bA(r,\theta) = 0 \quad {\rm and} \quad A_0(r,\theta) = 0,
\eqn\efivetwentyseven
$$
which is a solution of the equations of motion and perturbation theory is
built up by expanding the fields in modes around this vacuum. But besides
the vacuum solution, this theory also possesses topologically non-trivial
finite energy solutions\NIELSEN. To have finite energy solutions, all we
need to ascertain is that
$$
|\phi(r \rightarrow \infty,\theta)| = v,
\bA(r \rightarrow \infty, \theta) = 0 \quad {\rm and} \quad
A_0 \rtheta = 0, \eqn\efivetwentyeight
$$
so that the energy integral does not diverge. However, the condition for
the scalar field is satisfied even if the scalar field has a non-trivial
phase at infinity. So a solution of the equations of motion which
satisfies
$$\eqalign{
\phi \rtheta &= v e^{in\theta} \cr
\Rightarrow \p_{\theta} \phi (\rtheta) = &{1\over r} {\p\over \p \theta} v
e^{in\theta} =i {n\over r}}
\eqn\efivetwentynine
$$
has finite potential energy. Finiteness of the kinetic energy of the
scalar field also implies that
$$
\int (\p_i -iqA_i)\phi^* (\p^i -iqA^i)\phi \ d^2{\bf r} =
{\rm finite}, \eqn\efivethirty
$$
which, in turn, gives the conditions on the asymptotic behaviour of $\bA$
and $A_0$ as
$$
A_{\theta} \rtheta = {n\over qr}, A_r = 0 \quad {\rm and} \quad A_0 =0.
\eqn\efivethirtyone
$$
Solutions that satisfy Eq.\efivetwentynine\ and Eq.\efivethirty\
are topologically non-trivial vortices. They are called
vortex solutions because they carry flux -
$i.e.$, using these solutions, we see that the flux is given by
$$
\int B\,d^2\bx = \int \bA\cdot d{\bf l} = \int ({n\over qr}) rd\theta =
{2\pi n\over q}, \eqn\efivethirtytwo
$$
where $n$ is called the vorticity.

These solutions are topologically stable because solutions with different
values of $n$ are not deformable into each other without changing the
scalar field configuration throughout the infinite boundary of space.
This argument is easily understood pictorially. In Fig.(19), we have the
vacuum solution with $\phi = v$ everywhere, including at spatial infinity,

%\vbox{~\vskip 2.0in \centerline{Fig.19} \vskip .2in}
\epsfbox{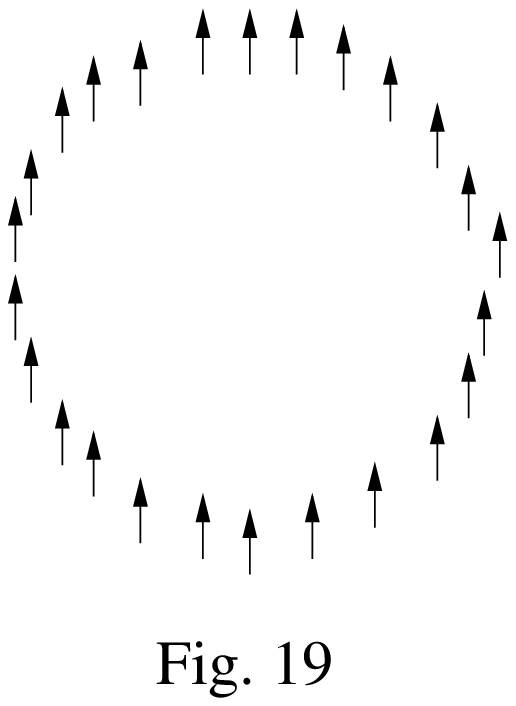}

\noindent
whereas in Fig.(20), we have the one-vortex solution with $\phi(\infty,
\theta) = ve^{i\theta}$.

%\vbox{~\vskip 2.0in \centerline{Fig.20} \vskip .2in}
\epsfbox{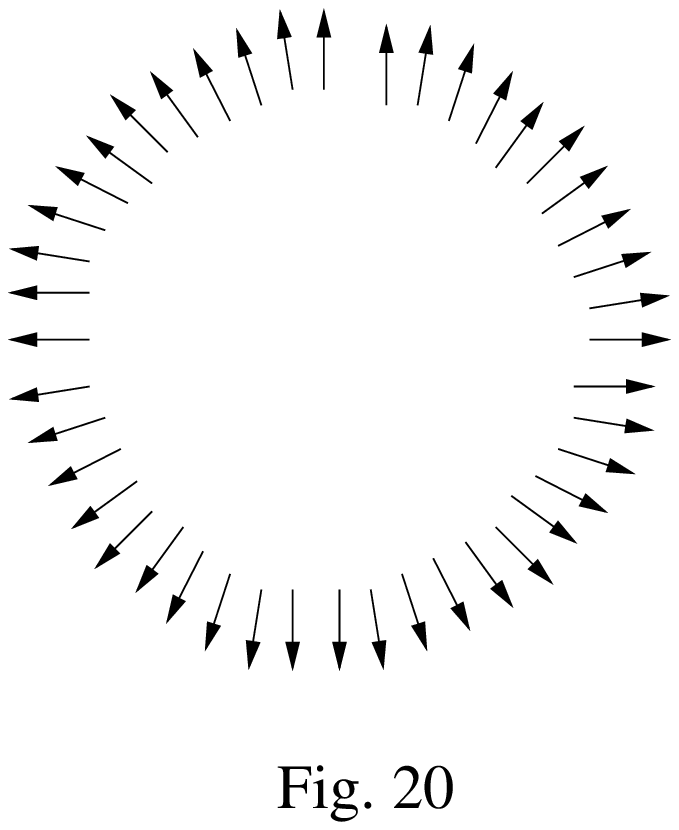}

\noindent
 Both the vacuum and the one-vortex configurations
satisfy $V(\phi) = 0$. But to deform one solution to another, we need to
go through configurations where $\phi(\infty,\theta)$ is neither $v$ nor
$ve^{i\theta}$  and which have $V(\phi)\ne 0$. Since this has to occur
throughout the infinite boundary of two dimensional space, this will cost
an infinite energy.
In general, from Eq.\efivetwentyfive, we see that the vacuum expectation
value $v$ takes values on a circle (since $\phi$ is complex and $\phi^2 =
v^2$). Hence, the configuration space of the vacuum is denoted by $S^1_c$.
The boundary of real (2-dimensional) space is also a circle and is denoted
by $S^1_r$. Hence, the boundary condition expressed in Eq.\efivetwentynine\
denotes a mapping from $S^1_r$ to $S^1_c$ and the vorticity $n$ represents
the number of times $\phi$ encircles  $S^1_c$ when $\br$
encircles $S^1_r$ once.  Each of these solutions falls in a different
topological class and is
stable, since it requires an infinite energy to change the
configuration of $\phi$  throughout the infinite boundary.

Now, let us introduce a $CS$ term for the gauge field. We had
earlier seen that in the presence of the
$CS$ term, the flux and charge get
related. Hence, now, these vortex solutions also possess a charge given by
$$
Q = \mu \phi = \mu({2\pi n\over q}).
\eqn\efivethirtythree
$$
Since these charged vortices carry both charge and flux, by our
earlier arguments in Sec.(1), they
are anyons with spin $j= Q\phi/4\pi = \pi n^2
\mu/q^2$ and statistics phase $\alpha = Q\phi/2 = 2\pi^2 n^2/q^2$.
Notice that here an explicit kinetic term for the gauge field has been
included, so that the gauge field is really a dynamical degree of freedom.
Hence, this model is not really relevant to anyons. However, even when the
usual kinetic piece is switched off, the model continues to exhibit
anyonic solutions\DILEEP.

This was just one example of a field theory whose solitonic excitations
are anyons. Another example is the $O(3) \quad \sigma$-model with the Hopf
term, whose skyrmionic excitations are anyons\ASFQHE. In all such
models, the basic ingredient is the $CS$ term (or equivalently the Hopf term)
which relates the charge and the flux.

The $CS$ term can also be extended to non-abelian theories with the
appropriate Lagrangian being given by\CSREFS\
$$
L = {1\over 2} {\rm tr} F_{\mu\nu} F^{\mu\nu} - {\mu\over 2} \emna {\rm
tr} (F^{\mu\nu} A^{\alpha} -{2\over 3} g A^{\mu} A^{\nu}A^{\alpha}) .
\eqn\efivetwentythree
$$
This Lagrangian can be shown to be gauge invariant provided $4\pi\mu/
g^2$ = integer. As in the abelian theory, $\mu$ is the gauge invariant
mass for the gauge boson and the $CS$ term is odd under parity and time
reversal. The pure non-abelian $CS$ Lagrangian
without the usual kinetic piece has
been recently related to problems in topology and knot theory, integrable
models in statistical mechanics, conformal field theories in 1+1
dimensions and 2+1 dimensional  quantum gravity. (For references to the
original papers, see Ref.\ASFQHE.) The remarkable feature of pure
$CS$ theories that makes it amenable to exact solutions is its general
covariance, --- the Lagrangian is not only Lorentz invariant, but it is also
generally covariant without any metric insertions.
Hence, correlation functions of the pure $CS$ theory depend only on the
topology of the manifold and not on details such as the metric on the
manifold. Besides theoretical interest in non-abelian theories, there have
also been speculations\MOOREREAD\ that non-abelions,  --- generalisations of
anyons that form non-abelian representations of the braid group, --- may
play a role in the even denominator FQHE.
However, since non-abelian $CS$ theories are not directly related to
anyons, we shall not pursue this topic any further here.

\endpage

\centerline{Problems}

\item{1.} Similar to vortices and charged vortices in $2+1$
dimensions, in $3+1$ dimensions we have monopoles and dyons. The
Georgi-Glashow model in $3+1$ dimensions, which consists of an
$SO(3)$ gauge field $A_{\mu}^a$ interacting with an isovector Higgs field
$\vec \phi$  is one of the simplest examples. The Lagrangian for this
model is given by
$$
L = - {1\over 4} {\rm tr} F_{\mu\nu} F^{\mu\nu} + {1\over 2} {\rm tr} ~
D_{\mu} \phi D^{\mu} \phi - {\lambda\over 4} ({\vec\phi}^2 - a^2)^2,
$$
where $F_{\mu\nu}^a = \partial_{\mu} A_{\nu}^a - \partial_{\nu} A_{\mu}^a
- g \epsilon_{abc} A_{\mu}^b A_{\nu}^c$ and $(D_{\mu}\phi)^a = \partial_{\mu}
\phi^a - g \epsilon_{abc} A_{\mu}^b \phi^c$. Can
you construct monopole solutions for this field theory 
by analogy with the vortex solution 
in Eqs. (5.25) and (5.26)? 
\item{b.} Why are monopoles topologically stable?
\item{c.} Can you guess the generalisation of the Chern-Simons term that
induces charge on the monopoles (converting them to dyons)? \hfill\break
Some useful references are Refs.\GODDARD\ and ~~\DYWITTEN.

\endpage

\chapter{Anyons in the Fractional Quantum Hall Effect}

Any study of anyons would be incomplete without an account of its most
outstanding success --- its application to the FQHE. This is the
only physical system where there exists incontrovertible evidence for the
existence of anyons, because quasi-particle excitations over the FQHE
ground state have been explicitly shown to obey fractional statistics.
However, since familiarity with the Quantum Hall system is not in the
repertoire of the average graduate students, we
shall introduce the background material in some detail.

Let us first remind ourselves of the Hall Effect. Here, electrons in a
plane show transverse conductivity when a magnetic field is applied
perpendicular to the plane. The Hall geometry is depicted in Fig.(21).

%\vbox{~\vskip 1.5in \centerline{Fig.21} \vskip .2in}
\epsfbox{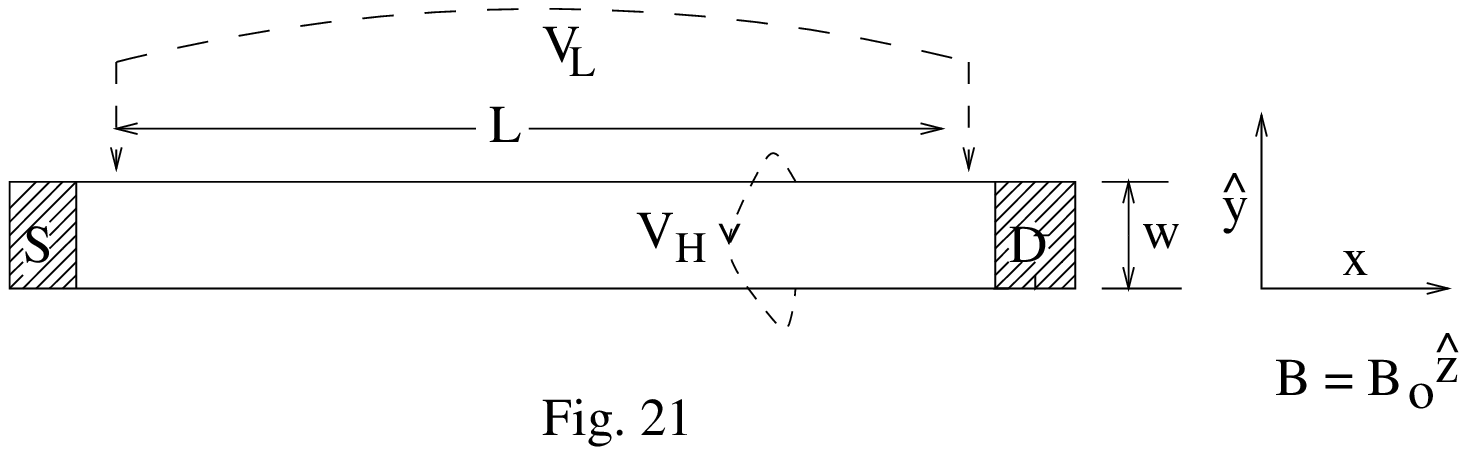}

\noindent
Electrons are allowed to move from a source $S$ to a drain $D$ causing a
current $I$, which is measured, as are the longitudinal and transverse
voltage drops $V_L$ and $V_H$. The existence of a non-zero $V_H$ can be
explained just by classical electrodynamics using the equation
$$
{\bf F} = e{\bf E} + e{\bf v} \times {\bf B} \eqn\esixone
$$
for the electrons with charge $q = e$.
For the geometry in Fig.(21), we have the
equations
$$
{\dot v}_x = {eE_x\over m} + {ev_yB_0\over m},\quad
{\dot v}_y = -{ev_xB_0\over m}
\eqn\esixtwo
$$
whose solution is
$$
v_x = v_0 e^{i\omega t},\quad
v_y = -{E_x\over B_0} + iv_0 e^{i\omega t}
\eqn\esixthree
$$
where $v_0$ is the initial velocity and $\omega = eB_0/m$ is the frequency
of the cyclotron motion. The constant term in $v_y$ represents the drift
velocity ${\bf v}_d$ and the Hall current is given by
$$
{\bf j}_H = \rho e {\bf v}_d \Rightarrow
j_{Hx} = 0,
j_{Hy} = -{\rho e E_x\over B_0},
\eqn\esixfour
$$
where $\rho$ is the density of charge carriers.
By changing the Hall geometry, we can also find
$$
j_{Hx} =  {\rho e E_y \over B_0}, \quad j_{Hy} = 0,
\eqn\esixfive
$$
when the electric field is along the $y$-direction.
Hence, if we define a conductivity matrix $\sigma_{ij}$ by
$$
\sigma_{ij} = \pmatrix{\sigma_{xx}&\sigma_{xy}\cr
\sigma_{yx}&\sigma_{yy}\cr}
\eqn\esixsix
$$
then classically, $\sigma_{xy} = -\sigma_{yx} = -\rho e/B_0$ and
$\sigma_{xx} = \sigma_{yy} = 0$. The resistivity matrix $\rho_{ij} =
\sigma_{ij}^{-1}$ and Eq.\esixsix\ leads to
$$
\rho_{xy} = -\rho_{yx} = {B_0\over e\rho } \quad
{\rm and} \quad \rho_{xx} = \rho_{yy} = 0.
\eqn\esixseven
$$
We have not taken collisions between electrons into account in deriving
the above classical result. However, a more realistic picture requires
collisions and by incorporating them, we get the semiclassical equations
of motion given by
$$\eqalign{
<{\dot v}_x> &= {eE_x\over m} + {q<v_y> B_0 \over m} - {<v_x>\over \tau} \cr
{\rm and}\qquad
<{\dot v}_y> &= -{e <v_x> b_0 \over m} - {<v_y> \over \tau},}
\eqn\esixsevenprime
$$
where $<{\bf v}>$ is the average velocity of the electrons and $\tau$ is the
average time between collisions. In equilibrium, $<{\dot {\bf v}}> = 0$
and we get
$$
v_x = {\tau e /m \over (\omega^2 \tau^2 + 1)} E_x, \quad
v_y = - {\omega \tau^2 e /m \over \denom} E_x \eqn\esixeight
$$
where, as before, $\omega = e B_0 /m$. Note that unlike Eq.\esixfive\
which does not give the correct $B_0 \rightarrow 0$ limit, (Hall current
goes to infinity instead of vanishing), here $\omega \rightarrow 0$ as
$B_0 \rightarrow 0$ and  the Hall current vanishes.
Thus, in the semiclassical limit, the elements of the conductivity matrix
are
$$
\sxx = \syy = {\rho e^2 \tau /m\over \denom} \quad {\rm and}
\quad \syx = -\sxy = -{\rho e^2 \tau^2 \omega /m \over \denom}
\eqn\esixnine
$$
and the elements of the resistivity matrix are
$$
\rxx = \ryy = {m\over \rho e^2 \tau} \quad {\rm and} \quad
\ryx = -\rxy = -{m\omega\over \rho e^2} = {B_0\over e\rho}.
\eqn\esixten
$$
Notice that the off-diagonal elements of the resisitivity are unchanged from
their classical values.

Now, when the Hall resistance was experimentally measured\VONKLITZING\ in
1980, at low temperatures (0 - $2\deg K$) and in strong magnetic fields (1
- 20 {\it Tesla }), a surprising result was found. $\sigma_H$ showed 
plateaux (as shown schematically in Fig.22)
instead of varying linearly with $1/B_0$ as expected classically or
semiclassically.

%\vbox{~\vskip 3.0in \centerline{Fig.22} \vskip .2in}
\epsfbox{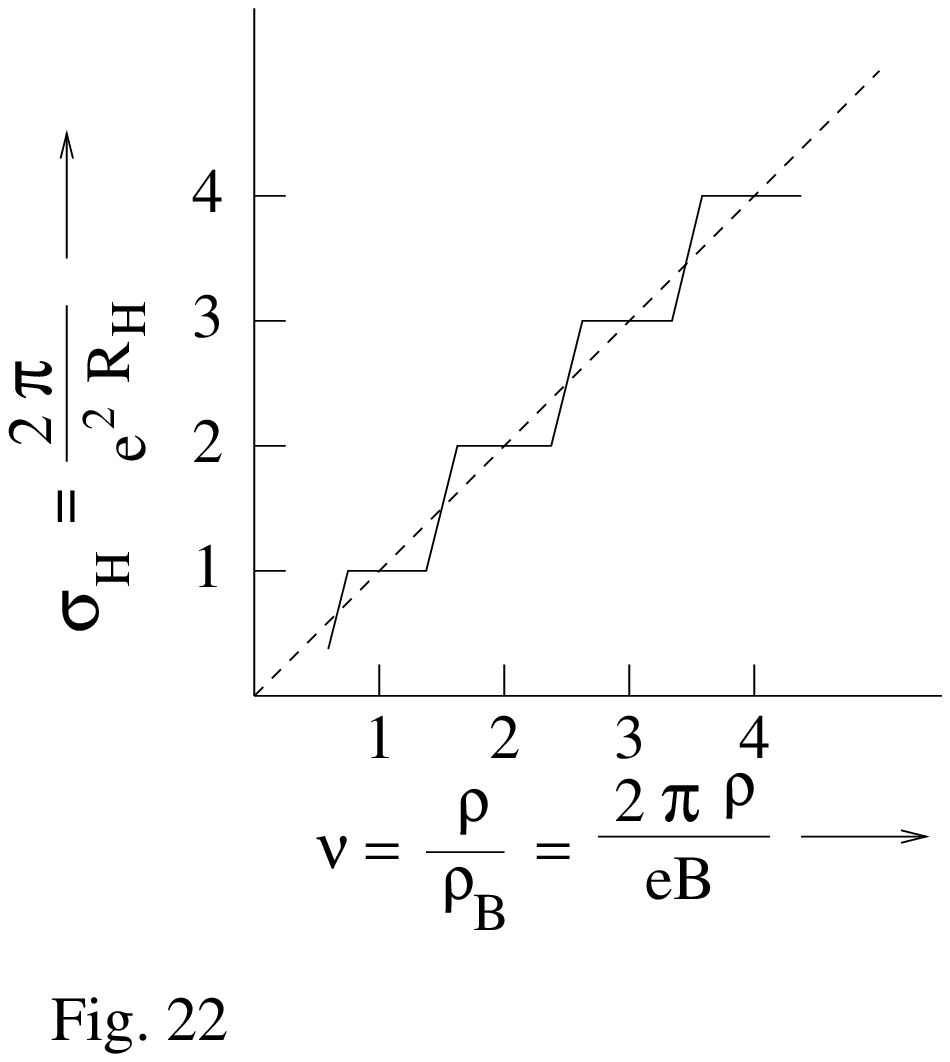}

\noindent Furthermore, the longitudinal resistance, instead of
being constant as expected by Eq.\esixnine, vanished at the plateaux and
peaked in between as shown (schematically) in Fig.23.

%\vbox{~\vskip 2.5in \centerline{Fig.23} \vskip .2in}
\epsfbox{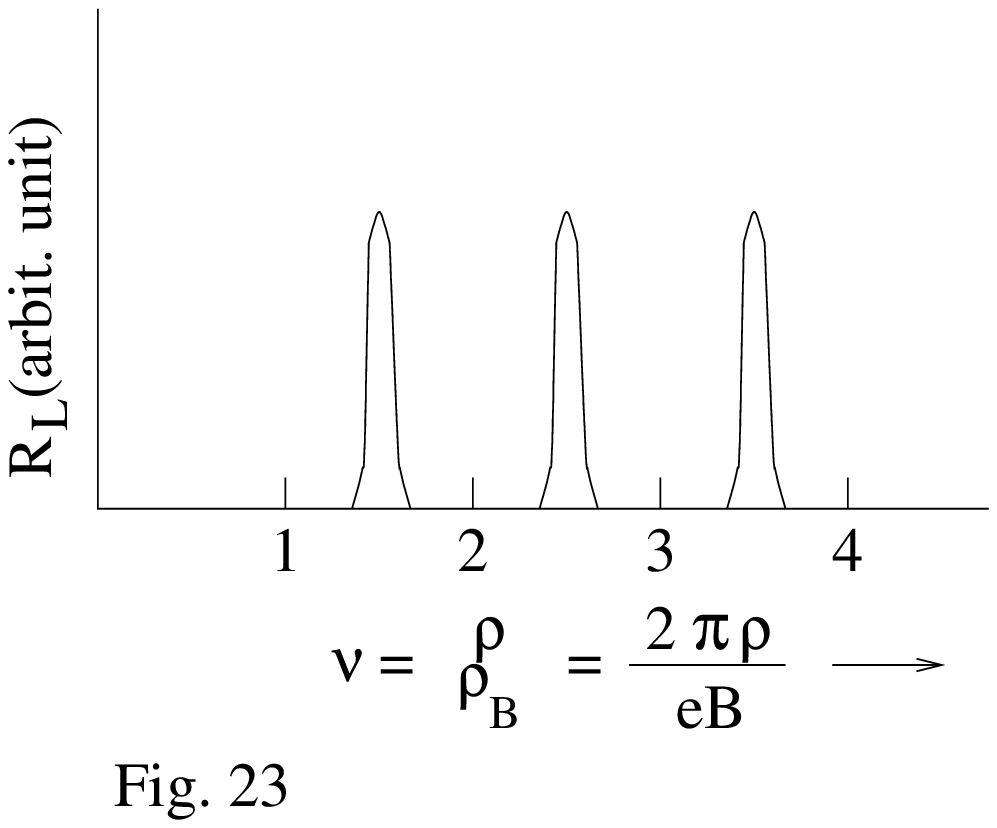}

\noindent
Even more surprisingly, though the behaviour of the Hall resistance at the
edges of the  steps was non-universal, the $R_H$ at the plateaux was
unexpectedly constant and reproducible to the accuracy of one part in
$10^5$. This effect was called the Integer Quantum Hall Effect (IQHE),
because the midpoints of the plateaux occured at integer values of
a ratio $\nu =
\rho/\rho_B$ (where $\rho_B = eB_0/2\pi$).
Since $\rho_B$ is the degeneracy of each Landau level, this
ratio which expressed the fraction of filled Landau levels was called the
filling factor.
The filling factor could either be thought of
as a measure of $B_0$ at fixed $\rho$ or as a measure of $\rho$ at fixed
$B_0$.

The remarkable accuracy of the
experimentally measured  resistances is explained by the following
observation. Usually the resistances $R_H = V_H/I$
and $R_L = V_L/I$ are related to the
resistivities by geometric factors of length or width.
However, for the Hall geometry, when $L>>W$ and the voltage drops are
measured sufficiently far from  the actual edges so that the applied
current density  is uniform, we see that
$$
R_H = {V_H \over I} =  {E_y W\over j_x W} = \rxy.
\eqn\esixtenprime
$$
Hence, geometric factors which can never be measured to accuracies of one
in $10^5$
have cancelled out leaving the transverse resistance
equal to the transverse resistivity. In fact, the quantisation
of the experimentally measured Hall resistance is so
accurate, that the Quantum Hall system is now used for the most precise
determination of the fine structure constant.

The midpoint values of the plateaux can be easily understood by studying
the quantum mechanical problem of non-interacting electrons in a
transverse magnetic field. In Sec.(4), we studied the problem of an
electron in a transverse magnetic field and found that the energy levels
are given by
$$
E_n = (n+1/2) {e  B_0\over m}
$$
with the degeneracy
$$
{\rm Deg} = \rho_B = {e B_0\over 2\pi},
$$
where we have substituted $\omega = e B_0/m$ in Eqs.\efourfourteen\ and
\efoureighteen. Hence, when the filling factor $\nu=\rho/\rho_B$ is an
integer, it is clear that precisely an integer number of Landau levels are
filled.  As mentioned in Sec.(4), this implies a gap to single particle
excitations, given by $\omega = e B_0/m$. Also, unlike the case in
Sec.(4), here $B_0$ is independent of $\rho$ and there is no argument for
a massless collective excitation. On the contrary, explicit calculations
show that the collective excitation is massive. So the system is
particularly stable when the density (or equivalently the magnetic field)
is such that $\nu$ is an integer.  From Eq.\esixten, we see that for these
values of the density,
$$
R_H = {B_0\over e\rho} = {B_0\over e\rho_B}
{1\over {\rm integer}} = {2\pi\over e^2} {1\over {\rm integer}},
\eqn\esixeleven
$$
in accordance with the experimental values of $R_H$ at the midpoint of
the plateaux. Thus, the densities of the electrons at the midpoint of the
plateaux have been identified with fully filled Landau levels and the
correct Hall conductivity is predicted for these densities.

However, to understand why the conductivity remains fixed even when $B_0$
(or $\rho$) is changed from the midpoint value is harder.
At a hand-waving level, we can argue that because of the stability of the
system when $\nu$ is an integer, even when the field $B_0$ is changed
slightly, the system prefers to keep the average density fixed such that
$\nu$ is an integer and accomodate the deviation in $B_0$ as a local
fluctuation. These local fluctuations do not contribute to the
conductivity because they get `pinned' or `localised' by impurities in the
sample.  Hence, the conductivity stays fixed at the value that it had for
$\nu$ = integer. At a slightly more rigorous level, the idea is that weak
disorder in the system (due to impurities and imperfections in the sample)
leads to the formation of some localised states ,
whereas other states are extended. Current can only be carried by extended
states. Hence, if the density of states as a function of the energy had
the pattern shown in Fig.(24),

%\vbox{~\vskip 2.0in \centerline{Fig.24} \vskip .2in}
\epsfbox{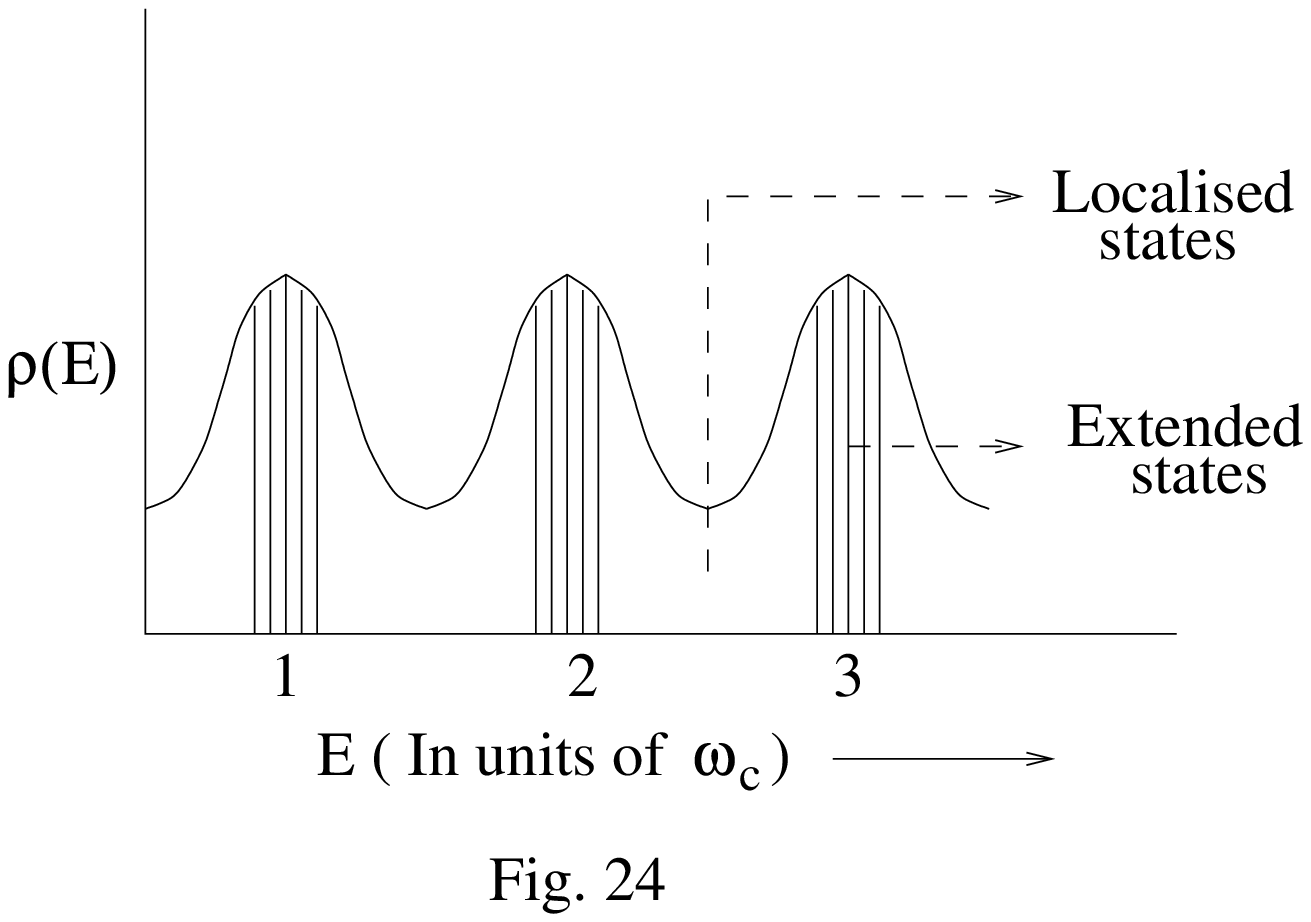}

\noindent
then as the Fermi level spans each localised region, the current will
remain constant (plateau in the Hall conductivity), while when it spans
the extended states, the current (and hence, conductivity) will increase.
To study in more detail the kinds of disorder potentials that can lead to
the density of states diagram in Fig.(24) would lead us too far afield.
For further details, we refer the reader to Ref.\FQHEBOOK.

In 1984, in stronger magnetic fields and cleaner samples, the FQHE
was seen\TSUI, where the midpoints of the plateaux were found to occur at
fractional values of the filling factor such as 1/3, 1/5, etc.   This was
unexpected because the single particle quantum mechanics analysis that was
used for the IQHE predicted that at $\nu$ = fraction, the system would be
highly degenerate and not at all stable. So, the very existence of the
effect showed that inter-electron interactions which were ignored in the
earlier study must be important.
Hence,  the appropriate
Hamiltonian for the FQHE is expected to be
$$
H = \sum_i {(p_i - e A_i)^2\over 2m} +  \sum_i V({\bf r}_i) +
{1\over 2}\sum_{i\ne j} V({\bf r}_i - {\bf r}_j)
\eqn\esixeightp
$$
where $V({\bf r}_i - {\bf r}_j)$ denotes the inter-electron Coulomb repulsion
and $V(\br_i)$ is a neutralising background potential.
Unlike the IQHE problem, this Hamiltonian cannot be solved exactly even in
the absence of disorder.

There have been several approaches to the theoretical understanding of the
FQHE, mainly because no theory has provided a complete picture yet. Here,
we shall concentrate on two approaches, both of which appear to use anyon
ideas in a fundamental way --- the trial wavefunction approach and the
Landau-Ginzburg-Chern-Simons field theory approach.

The trial wavefunction approach was initiated by Laughlin\LAUGHLIN\ who
proposed that the ground state of the FQHE system is described by the
wavefunction given by
$$
\psi(z_1,z_2,....z_N) = \prod_{i<j}^N (z_i - z_j)^m \exp \eqn\esixninep
$$
for the fraction $\nu = 1/m$. Here $z_i = x_i +i y_i$ is the position of
the $i^{th}$ particle and $m$ has to be an odd integer to satisfy the
criterion that the particles are electrons. He arrived at this
wavefunction from certain general principles such as $a.$) the wavefunction
should be antisymmetric under exchange, $b.$)  the wavefunction should
comprise of single particle states in the lowest Landau level, $c.$) the
wavefunction should be an eigenstate of total angular momentum, and
d.) inspired guesswork. 
This ansatz for the ground state wavefunction has had enormous success,
mainly because no other ansatz has been found with lower energy.

To understand the significance of this wavefunction, let us write
$$
|\psi|^2 = e^{-\beta \phi} \eqn\esixninepp
$$
so that
$$
\phi = -{2m\over\beta} \sum_{i<j}^N {\rm ln}{|z_i - z_j|\over l} + {1\over
2\beta} \sum_i^N {|z_i|^2\over l^2}.  \eqn\esixnineppp
$$
By setting the fictitious temperature $\beta = 1/m$, we see that $\phi$
can be interpreted as the potential energy of a two dimensional gas of
classical particles of charge $`m'$ repelling each other through a
logarithmic interaction and being attracted to the origin by a uniform
(opposite) charge density $\rho_U = 1/2\pi l^2$. This potential energy is
minimised ($i.e.$, $|\psi|^2$ is maximised)
by a uniform distribution of the charge `m' particles and is
`electrically' neutral everywhere, when the average density of the charge
`m' particles is precisely equal to $\rho_U$, which in turn implies that
the average density of the electrons $\rho$ is equal to $1/2\pi m l^2$.
Hence, at these densities, the many electron wavefunction peaks when the
coordinates $z_i$ are uniformly distributed
and is expected to be energetically
favourable,  since the Coulomb repulsion in Eq.\esixeightp\ is minimised.
For other values of $\rho$, the classical gas has excess charge `m'
particles either near the origin or near the sample boundary and hence
fails to be uniformly distributed. So, the appropriate $\psi$ peaks at a
configuration of the $z_i$'s that does not
describe a uniform distribution of electrons and suffers from a high
repulsive Coulomb energy. Hence, the Laughlin picture is that as the
density is reduced from $\rho_B = e B/2\pi$ where one Landau level is
filled, whenever $\rho = \rho_m = \rho_B/m \Rightarrow \nu = 1/m$, the
energy is minimised and the state is stable.

The Laughlin state at $\nu =1/m$ can be shown to have massive
fractionally charged
quasiparticle and quasihole single particle excitations, which obey
anyon statistics.   To see this,
let us insert adiabatically one unit of flux through an imaginary thin
solenoid piercing the many body state. The intermediate Hamiltonian as the
flux is being evolved through the solenoid is given by
$$
H = \sum_i {({\bf p}_i - e {\bf A}_i -e {\bf A}'_i)^2\over 2m}
+  \sum_i V({\bf r}_i) +
{1\over 2}\sum_{i\ne j} V({\bf r}_i - {\bf r}_j) \eqn\esixtwelvep
$$
where ${\bf A}'_i$ is the gauge potential of the flux through the
solenoid.  As the flux changes, the wavefunction also changes so as to be
an eigenstate of the Hamiltonian. At the end of the process, however, the
Hamiltonian has returned to the original Hamiltonian, since one unit of
flux through an infinitely thin solenoid can be gauged away. (Remember
that the term $e {\bf A}'_i$ in the Hamiltonian can be traded  for a phase
$e^{ie\phi}$ in the wavefunction in the `anyon' gauge. But when the flux
$\phi = \phi_0 =2\pi/e$, the phase is just 1 and hence irrelevant.) But
the state has not returned to the original state.   As flux is
adiabatically added, every single particle state
$$
z^m e^{-|z|^2/4}
\rightarrow z^{m+1} e^{-|z|^2/4} \eqn\esixthirteenp
$$
-$i.e.$, its angular momentum increases by one. The state with the
highest angular momentum moves over to the next Landau level and a new
state appears at $m=0$.
If a single unit of flux is removed adiabatically, then
$$
z^m
e^{-|z|^2/4}  \rightarrow z^{m-1} e^{-|z|^2/4}. \eqn\esixfourteenp
$$
Hence, a state from the next Landau level moves down and the $m=0$ state
disappears. Thus, the effect of adding
(or removing) one
quantum of flux and then gauge transforming is to increase (or decrease) the
angular momentum of the single particle states by one unit.  But if we
assume that the original state described by the Laughlin wavefunction is
non-degenerate, (since no other states with the same energy have been
found), then the new state, after evolution of the flux, has to describe an
excited state of the original Hamiltonian, with a higher energy
eigenvalue. This automatically proves that the quasiparticle or
quasihole excitations have a gap.

The electric charge  of these
excitations can also be easily computed. Let us assume that the flux has been
evolved through a very thin solenoid.  Faraway from the solenoid, we
expect that this state will be indistinguishable from the ground state,
except that every level of the single particle states has moved over to
the next level. Hence, if the flux point is surrounded by a large circle,
then the charge that has entered or left the circle is just the average
charge per state, provided the total charge and the total number of states
are uniformly distributed. In position space, single particle states are
labelled by the position vector $\br$ and are uniformly distributed with a
degeneracy of $eB/2\pi$. By the plasma analogy, charge is also uniformly
distributed in real space, with the density $e\rho_m = e^2B/2\pi
m$ for $\nu = 1/m$. Thus, the charge per state is just $\pm e/m$ which is
identified as the charge of the excitation.

The statistics of these quasiparticles and quasiholes can  be
explicitly found\ASW\ by a Berry phase calculation. However, a much
simpler way is to notice that these excitations have a flux $2\pi/e$ and a
charge $e/m$ and hence are anyons with fractional spin and statistics
given by
$$
j = {q\phi\over 4\pi} = {e/m \times 2\pi/e \over 4\pi} = {1\over 2m}
\quad {\rm and} \quad
\alpha = {q\phi\over 2} = {e/m\times 2\pi/e\over 2} =
{\pi \over m}.\eqn\esixfifteenp
$$
Eq.\esixthirteenp\ implies that an explicit
ansatz for the wavefunction of a quasihole excitation can be written as
$$
\psi(z_0,z_1,...z_N) = \prod_i (z_i - z_0)
\prod_{i<j} (z_i - z_j)^m \exp  \eqn\esixsixteenp
$$
where $z_0$ is the position of the infinitely thin solenoid.  This ansatz
is expected to be good except very near the solenoid.  The wavefunction
for the quasielectron involves derivative operators and is not as
straighforward to understand, at least within the Laughlin scheme.

Laughlin's wavefunctions only explained the plateaux at the fractions $\nu
= 1/m$ where $m$ was an odd integer. But, experimentally, plateaux were
seen at many other rational fractions $\nu = p/q$ where $q$ was an odd
integer. To explain the other fractions, the hierarchy scheme was
evolved\HALDHALP. The idea was that quasiparticle or quasihole excitations
over the Laughlin state themselves behaved like particles in a magnetic
field and could form new correlated many body states which could represent
the FQHE state at other fractions.

The hierarchy scheme is easily understood in the anyon language.
Firstly, notice that the wavefunction for a
quasihole excitation remains analytic (see Eq.$\esixsixteenp$).  Also,
quasiholes obey anyonic statistics. Hence, the simplest possible
wavefunction for a collection of many quasiholes of generic charge $qe$
and statistics $\alpha'$
(using Laughlin's
arguments to arrive at his wavefunction, but now remembering that
the particles are anyons) is given by
$$
\psi(z_{01},z_{02},....z_{0M}) = \prod_{i<j}^M (z_{0i} - z_{0j})^{2k+\alpha'}
 \expone. \eqn\esixseventeenp
$$
This wavefunction looks exactly like Laughlin's wavefunction except that $m$ is
replaced by $2k+\alpha'$ to account for the changed statistics and in the
exponent $l^2$ is replaced by $e l^2/|q|$ to account for the charge of the
quasiholes which is $qe$. The same plasma analogy now suffices to find the
density of quasiholes $\rho_{qh}$ for which this wavefunction describes a
uniformly distributed electrically neutral plasma and is hence likely to
be an energetically favoured wavefunction. The plasma has `charge'
($2k+\alpha'$)   particles repelling each other and being attracted to the
origin by a uniform `charge' density $\rho = |q|/2\pi el^2$. So by our
earlier argument, we shall have a uniform density of quasiholes when
$$
\rho_{qh} = {1\over (2k+\alpha')} \times
{|q|\over 2\pi el^2 } . \eqn\esixeighteenp
$$
Now, we know that the charge of the quasiholes is $qe = -e/m$ and their
statistics parameter $\alpha = \pi/m \Rightarrow \alpha' =
1/m$. (Remember that we are using complex notation in Eq.\esixseventeenp.
Hence, under exchange of particles, $(z_{\rm rel})^{\alpha'} =
(re^{i\theta})^{\alpha'} \rightarrow (r e^{i(\theta+\pi)})^{\alpha'} =
(z_{\rm rel})^{\alpha'} e^{i\pi\alpha'}$ ).
Using this, the filling fraction of the
quasiholes is given by
$$
\nu_{qh} = {\rho_{qh}\over \rho_B} = {1\over \dis m(2k+{1\over \dis m})}.
\eqn\esixnineteenp
$$
To find the equivalent density of electrons, notice that at a given
density of electrons, the total charge remains fixed, whether it is
counted as quasihole charges or electron charges.
Hence, the total charge carried by the quasiholes is given by
$$
q\rho_{qh} = -{e\over m} \rho_{qh} = - {e\over \dis
m^2(2k+{1\over \dis m})}
\eqn\esixtwentyp
$$
which, in turn, is equal to $e\rho_{eqh}$,
where $\rho_{eqh}$ is the equivalent density of electrons.
So the total density of electrons and hence, the filling fraction
is given by
$$
\nu = {1\over m} - {1\over \dis m^2(2k+{1\over
\dis m})} = {1\over
\dis m+{1\over \dis 2k}}. \eqn\esixtwentyonep
$$
For $m=3$ and $k=1$, $\nu=2/7$ and for $m=3$ and $k=2$, $\nu =4/13$. Thus,
we have obtained other fractions with odd denominators.
This process can now be iterated. We can consider excitations over
the $\nu =2/7$ or the $\nu = 4/13$ state and form new correlated states
with those excitations. A little algebra shows that by repeating this
procedure, we get filling fractions which can be written as
$$
\nu ={1\over \dis m+{1\over \dis 2k_1 +{1\over \dis 2k_2 + \cdots +{1\over
\dis 2k_S}}}}.\eqn\esixtwentythreep
$$
Quasielectron excitations have a statistics factor $2k-\alpha$ instead of
$2k+\alpha$ and have the same sign of charge as the electrons.
Repeating the same analysis as above for quasielectron excitations gives
$\nu$ as in Eq.\esixtwentythreep\ except that all the plus signs are
replaced by minus signs. In general, when we allow for both quasiparticle and
quasihole excitations over each state, the possible filling fractions are
given by
$$
\nu ={1\over \dis m+{\alpha_1\over \dis 2k_1 +{\alpha_2\over \dis 2k_2
+ ...+ {\alpha_S \over \dis 2k_S}}}}.
\eqn\esixtwentyfourp
$$
where $\alpha_i$
are either +1 or -1 depending on whether quasiparticle or quasihole
excitations are involved. All rational fractions with odd denominators are
obtained once in this way. Also for the FQHE system, the hierarchy, as the
above scheme to generate the fractions is called, works in the sense that
for any fraction that has been observed, all the other fractions that lie
before it in the hierarchy have also been observed.

The problem with the hierarchy scheme is that some fractions ($e.g.$, $\nu
= 5/13$, at the third level of hierarchy) have not been observed. Also,
for other fractions like $\nu = 6/13$ which is seen at the fifth level of
hierarchy starting from the $\nu = 1/3$ state, the number of
quasiparticles of various types is so much more than the number of
original electrons, that it is hard to understand how the explanation of
the original electrons forming a uniformly distributed $\nu = 1/3$ state
survives. A way out of this predicament was suggested by Jain\JAIN, who
could directly obtain the wavefunctions for all the odd denominator
fractions that have been seen. The starting point of his approach was to
note that the FQHE was phenomenologically very similar to the IQHE, and
hence, the theories of both the phenomena should also be related.

Let us start with an IQHE state, say, the $p$-filled Landau
level state represented by
$$
\nu = {\rho\over \rho_B} = p \, \Rightarrow \,B_0 = {2\pi\rho\over p e},
\eqn\esixtwentyfivep
$$
and then attach fluxtubes of strength $\phi = 2k\phi_0$ to each electron.
We know that this will convert fermions to anyons, but provided $k$ is an
integer, the statistical parameter $\alpha = e\phi/2 = 2k\pi$ is
irrelevant (we have chosen the statistical charge to be the same as the
real electric charge, which is always possible) and the electrons remain
fermions. These fermions with attached fluxtubes are called `composite
fermions'. Now, within the mean field approach (valid for a high density
of fermions), the flux of the fluxtubes can be spread out and we have
ordinary fermions moving in an effective magnetic field
$$
B_{\rm eff} =
\pm B_0 + 2k\phi_0 = \pm {2\pi\rho\over p e} + {4\pi k\over e},
\eqn\esixtwentysixp
$$
so that the filling factor $\nu$ for ordinary electrons is given by
$$
{1\over \nu} = {\rho_B\over\rho} = {e B_{\rm eff}/2\pi\over \rho}
= \pm{1\over p} + 2k = {2kp\pm 1\over  p}.
\eqn\esixtwentysevenp
$$
Thus, the basic idea is that the FQHE for ordinary electrons occurs
because of the IQHE for composite electrons --- $i.e.$, the same kind of
correlations between electrons are responsible both for the IQHE and
the FQHE. When $p = 1$, this procedure yields the Laughlin fractions $\nu
= 1/(2k\pm 1)$.

This idea can also be used to write down trial wavefunctions
for arbitrary odd denominator fractions.  For instance, to obtain the
Laughlin wavefunctions, we start with the wavefunction for one filled
Landau level given by
$$
\psi_1 (z_1,...z_N) = \prod_{i<j} (z_i - z_j) e^{-\Sigma_i |z_i|^2 /4l^2}
\eqn\esixtwentyeightp
$$
and then adiabatically introduce fluxtubes with $\phi = 2 k \phi_0$ at the
site of each electron. We know that adiabatic evolution of a flux unit at
any point $z_0$ involves the factor $\Pi_i (z_i - z_0)$ from
Eq.\esixsixteenp. Hence, insertion of $2k$ fluxtubes at the positions of all
the electrons leads to the wavefunction given by
$$\eqalign{
\psi_{2k+1} (z_1,....z_N) &= \prod_{i<j} (z_i - z_j)^{2k} \psi_1
(z_1,...z_N) \cr
&=\prod_{i<j} (z_i - z_j)^{2k+1} e^{-\Sigma_i |z_i|^2/4l^2}}
\eqn\esixtwentyninep
$$
which is precisely the Laughlin wavefunction for the filling fractions
$\nu = 1/(2k+1)$. The same procedure can also be used to write down
wavefunctions for the other fractions $\nu = p/(2kp+1)$ as
$$
\psi_{\nu} = \prod_{i<j} (z_i - z_j)^{2k} \psi_p (z_1,...z_N,{\bar
z}_1,...{\bar z}_N)
\eqn\esixthirtyp
$$
where $\psi_p (z_1,...z_N,{\bar z}_1,...{\bar z}_N)$ is the wavefunction
for $p$ filled Landau levels. But the form of the wavefunctions at $\nu =
p/(2kp-1)$ is not as obvious.
Explicit quasiparticle and quasihole wavefunctions
can also be written down very easily
in this formalism. Just as the FQHE wavefunctions are obtained from the
IQHE wavefunctions $\psi_p$, the quasiparticle or quasihole wavefunctions
for the FQHE are obtained from the quasihole wavefunctions $\psi_p^+$ or
quasielectron wavefunctions $\psi_p^-$ of the IQHE, by multiplying them
by the factor $\Pi_{i<j} (z_i -z_j)^{2k}$. The quasihole wavefunction for
the $\nu = 1/(2k+1)$  state
in this formalism, coincides
with the Laughlin quasihole for the same state, but the quasielectron
wavefunction involves higher Landau levels and differs
from Laughlin's ansatz and other trial wavefunctions in the literature.
However, it appears  much more natural and less arbitrary
and works quite well numerically. The
quasiparticle and quasihole charges and statistics can also be computed as
was done for Laughlin's wavefunctions and the hierarchy wavefunctions and
they coincide with the earlier results for any fraction $\nu = p/q$.

The wavefunction approach is a microscopic approach and has been fairly
successful in explaining the phenomenon of the FQHE. However, it is not
completely satisfactory, because it fails to illuminate all the symmetries
and does not provide a complete understanding of the problem.
To give an analogy, it is as if soon after superconductivity was
discovered, the $N$-body projected $BCS$ wavefunction in the coordinate
representation was directly written down. Although it was
correct, a complete understanding of the phenomenon of superconductivity
would not have been possible without
discovering the Landau-Ginzburg theory and the phenomenon of Cooper
pairing which actually led to the $BCS$ theory. Hence, for the FQHE system
too, there have been several attempts
to find  analogues of the Cooper pair that condenses and a
Landau-Ginzburg theory.
It is only in the last couple of years
that these attempts have begun to bear fruit and an effective field theory
called the Landau-Ginzburg-Chern-Simons (LGCS) theory has
been evolved\READ. This approach uses anyon ideas and the Chern-Simons
construction introduced in Sec.(5) to write down an effective action for
the FQHE problem, whose saddle point solutions correspond to FQHE states.
The key to this approach is the realisation that within the mean field
approximation, real magnetic field is indistinguishable from the
fictitious or statistical magnetic field introduced by a Chern-Simons
term. Hence, real magnetic flux can be moved on to fermions and identified
as statistical flux, so that the fermions get converted to anyons.  The
interesting result that was discovered in this approach was that precisely
at those values of the magnetic field where the FQHE occured, all of the
real magnetic flux could be transferred onto the fermions so that they
could be transmuted to bosons.  Pictorially, this procedure can be
represented as shown in Fig.(25).

%\vbox{~\vskip 2.0in \centerline{Fig.25} \vskip .2in}
\epsfbox{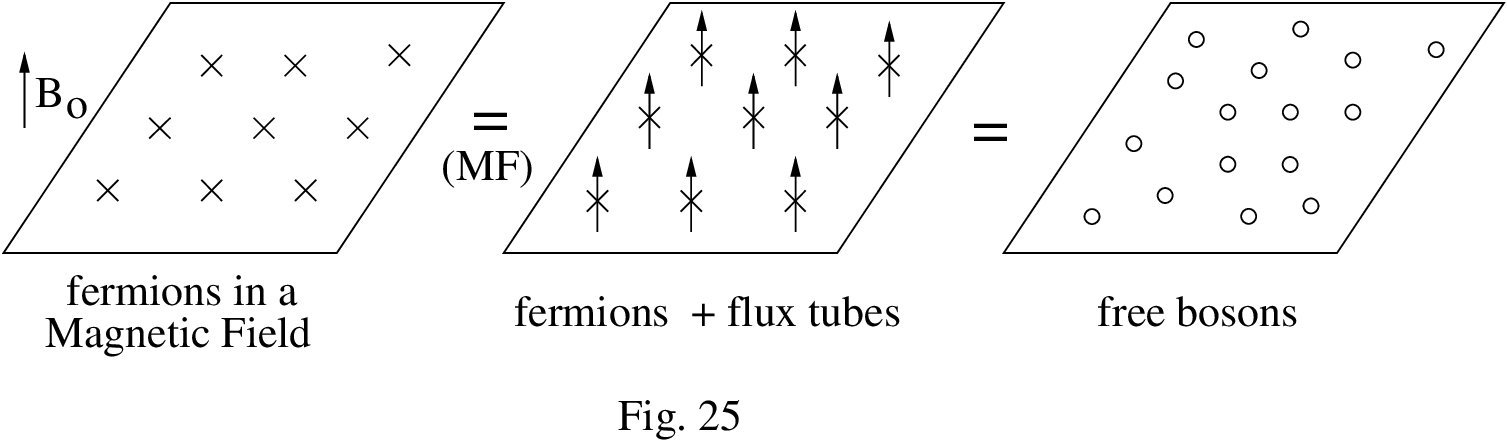}

\noindent
An equivalent picture (which is the one that we shall implement formally)
is to consider the fermions as bosons with attached flux tubes. Then the
FQHE occurs precisely when the external magnetic field cancels (in a mean
field sense) the effect of the flux tubes.  Once again, this can be
depicted as shown in Fig.(26).

%\vbox{~\vskip 2.0in \centerline{Fig.26} \vskip .2in}
\epsfbox{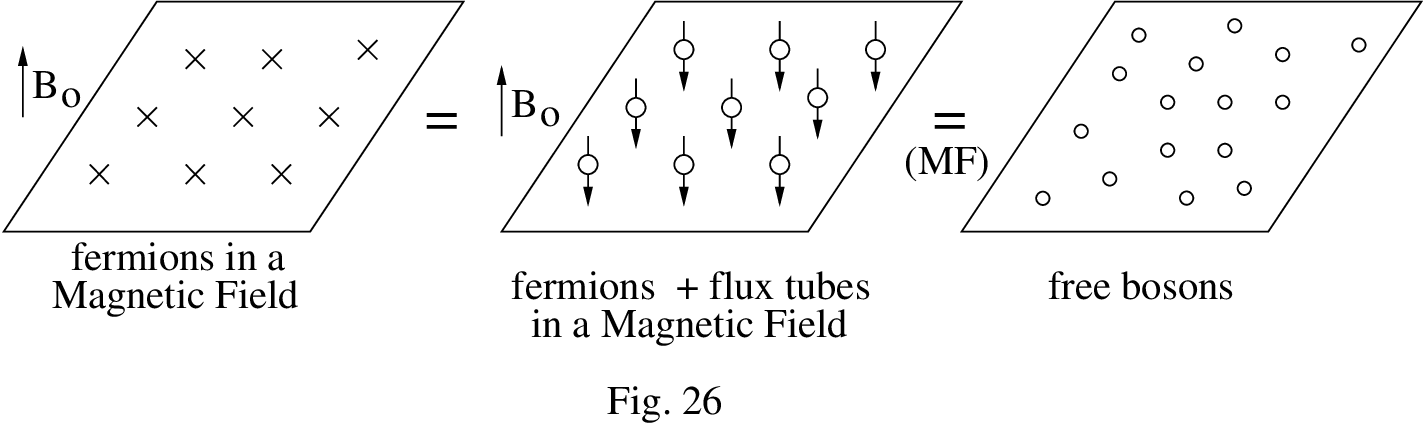}

\noindent
In either picture, FQHE occurs when the fermions turn into free bosons on
the average. Thus, the FQHE (or formation of an incompressible fluid) is
equivalent to the formation of a Bose condensate. The stability
of the FQHE states is `explained' by the well-known fact that
bosons can lower their energy by Bose condensing.

Let us now derive the above explanation more formally, starting from the
microscopic Hamiltonian  for electrons in an external electromagnetic
potential given by
$$
H = {1\over 2m} \sum_{i=1}^N (\bpi - e \bA(\bri))^2 + \sumi e A_0 (\bri) +
\sumij V\brij,  \eqn\esixeleven
$$
where $V\brij$ is the repulsive interelectron potential.
(In writing this Hamiltonian, we have dropped the background neutralising
potential, which, of course, is always present.) The solution
$\psi (\brone,\brtwo,...\brN)$ to the Schrodinger equation
$$
H\psiN = E \psiN \eqn\esixelevena
$$
has to be antisymmetric under the exchange of any two coordinates, since
the particles are fermions. These fermions can also be written as bosons
interacting with an appropriate Chern-Simons gauge field. Hence, an
equivalent formulation of the problem is given by
$$
H' = {1\over 2m} \sumi (\bpi - e A(\bri) -e \bai)^2 + \sumi e A_0 + \sumij
V\brij, \eqn\esixtwelve
$$
with
$$
H'\phiN = E \phiN. \eqn\esixtwelvep
$$
$\phi$ is now symmetric under the exchange of coordinates, and $\bai$ in
Eq.$\esixtwelve$ is given by
$$
\bai = {1\over e} {\alpha\over\pi} \sum_{j\ne i} {{\hat z} \times \brij
\over \brmodij^2}
= {1\over e} {\alpha\over \pi} \sum_{j\ne i} {\bf \nabla}_i \theta_{ij},
\eqn\esixthirteen
$$
where $\theta_{ij}$ is the angle made by $\brij$ with an arbitrary axis.

To prove that the two systems are really equivalent, all we need to show
is that the factor $\alpha/e \pi$ in Eq.\esixthirteen\
is precisely the factor needed to
convert fermions to bosons.  Consider a unitary transformation on the
wavefunction and the Hamiltonian in Eq.$\esixtwelvep$ ---$i.e.$,
$$
\phiN \longrightarrow \phitN = U \phiN = e^{-{i\alpha\over \pi}
\Sigma_{i<j}\theta_{ij}}\phiN \eqn\esixsixteen
$$
and
$$
H'\longrightarrow UH'U^{-1} = H.
\eqn\esixseventeen
$$
The transformed Schrodinger equation is given by
$$\eqalign{
UH'U^{-1} U\phiN &= E U\phiN \cr
\Rightarrow  H\phitN &= E\phitN.}
\eqn\esixeighteen
$$
Furthermore, since $\theta_{ij}$ is the angle made by $\brij$ with an
arbitrary axis,
$$
\theta_{ij} = \theta_{ji} + \pi .
\eqn\esixnineteen
$$
Hence, if $\phi$ is symmetric under exchange, ${\tilde \phi}$ picks up the
phase $e^{i\alpha}$ and is fermionic whenever $\alpha = (2k+1)\pi$. For
these values of $\alpha$, the systems described by Eqs.$\esixeleven$ and
$\esixelevena$ on the one hand and Eqs.$\esixtwelve$ and $\esixtwelvep$ on
the other, are equivalent ---$i.e.$, $\phitN = \psiN$. This is not really
surprising, since the unitary transformation in Eq.$\esixsixteen$ is just
the generalisation  to many particles
of the gauge transformation used in Sec.(2), to go from
the fermion gauge to the anyon gauge.
$\alpha = (2k+1)\pi$ is the value of the statistics parameter for fermions
to become bosons.

We can also explicitly check that the Hamiltonian in Eq.$\esixtwelve$ can
be obtained from a Chern-Simons Lagrangian. From Eq.$\esixthirteen$, we
can show that
$$
b(\bri) = {\bf \nabla} \times {\bf a} ={2\alpha\over e} \sum_{j\ne i}
\delta \brij = {2\alpha\rho\over e} \eqn\esixtwentyone
$$
where $\rho$ is the density of particles. Comparing with Eq.$\efivesix$, we
see that this is precisely the form of the equations of motion derived
from a Chern-Simons Lagrangian. Hence, the appropriate action that
describes the effective field theory of the FQHE problem is given by
$$\eqalign{
S &= S_a + S_{\phi} \cr
\quad {\rm where }\quad
S_a &= \int d^3 x \{ {e^2\over 4\alpha} \epsilon^{\mu\nu\rho} a_{\mu}
\partial_{\nu} a_{\rho}\} \cr
\quad {\rm and}\quad
S_{\phi} &= \int d^3 x \{\phi^{\dagger}
[i\partial_t -e(A_0 +a_0)]\phi +\mu\phi^{\dagger}
\phi +{1\over 2m}[\phi^{\dagger}(-i{\bf \nabla} - e \bA - e\ba)^2 \phi]\}\cr
&- {1\over 2} \int d^2 \brone d^2 \brtwo
\{\phid(\brone)\phid(\brtwo) V(\brone - \brtwo)
\phi(\brone)\phi(\brtwo)\}.}
\eqn\esixtwentytwo
$$
This field theory formalism now enables us to look for minima of the
action  (which correspond to the usual mean field theories) and
incorporate quantum corrections by expanding around the minima.

Let us first consider the case where there exists a non-zero magnetic
field, but no electric field so that
$$
A_0 = 0, \quad {\rm and} \quad \epsilon^{ij}\partial_i A_j =
-B_0 = {\rm constant}.
\eqn\esixtwentythree
$$
Here, the action in Eq.$\esixtwentytwo$ is minimised by choosing
$$
\phi = \sqrt{\rho}, \, \bA = -<\ba>, \quad {\rm and} \quad a_0 = 0,
\eqn\esixtwentyfour
$$
where the density of particles $\rho$ is a constant. ($\phid \phi = \rho$
is enforced by choosing the chemical potential $\mu$ suitably).  Since
the statistical magnetic field $b$ is related to $\rho$
(Eq.$\esixtwentyone$) and Eq.$\esixtwentyfour$ relates $b$ to the external
magnetic field $B_0$, this minimisation is only possible when
$$
B_0 = b = {2\alpha\rho\over e} \eqn\esixtwentyfive
$$
or when
$$
\nu = {\rho\over \rho_B} = {\rho\over e B_0/2\pi} ={\pi\over\alpha} =
{1\over (2k+1)} \eqn\esixtwentyfivep
$$
where $\rho_B = eB/2\pi$ = degeneracy of each Landau level and we have
substituted $\alpha = (2k+1)\pi$. Thus, the action is minimised at the
densities for which the filling fraction $\nu = 1/$odd integer -$i.e.$,
the Laughlin fractions. To prove that the vacua at these fractions are
incompressible, we also need to show that all excitations over these vacua
are massive. For the quasiparticle excitations, we have already seen in
Sec.(5) that the CSLG theory or the abelian Higgs model with a CS term has
charged vortex solutions that have finite (non-zero) energies and
fractional spin.  Collective excitations, which are fluctuations of
$(\bA+\ba)$ (which could be massless, in principle, as for anyon
superconductivity) are also massive because of the spontaneous breakdown
of the U(1) symmetry caused by the vacuum expectation value of $\phi$.
Hence, it appears reasonable to identify the classical minima of the
effective field theory with the Laughlin states.

Let us now calculate the Hall conductance by applying an external electric
field $E_i = -\partial_i A_0$ along with the external magnetic field $B_0 =
-\epsilon^{ij}\partial_i A_j$. The observed current can be obtained
from the action by using
$$
j_i = {\partial S\over \partial A_i} = {\partial S_{\phi}\over\partial
A_i} = {\partial S_{\phi}\over \partial a_i}
\eqn\esixtwentysix
$$
since $A_i$ and $a_i$ enter $S_{\phi}$ symmetrically. The equations of
motion with respect to $a_i$ are given by
$$
{\p S\over \p a_i} = 0 \Rightarrow {\p S_{a}\over \p a_i} + {\p
S_{\phi} \over \p a_i} = 0.
\eqn\esixtwentyseven
$$
Substituting for $\p S_{\phi}/\p a_i$ in Eq.$\esixtwentysix$, we see that
$$
j_i = -{\p S_a\over \p a_i} = {e^2\over 2\alpha} \epsilon^{0ij} (\p_0 a_j
- \p_j a_0). \eqn\esixtwentyeight
$$
Now, from the minimum energy ansatz in Eq.$\esixtwentyfour$
for a static magnetic field
and the ansatz $E_i = \p_i A_0$ for the external electric field,
we have
$$
\p_0 a_j = -\p_0 A_j =0 \quad {\rm and} \quad \p_j a_0 = -\p_j A_0 = E_j.
\eqn\esixthirty
$$
Thus, the current can be written as
$$
j_i = {e^2\over 2\pi(2k+1)} \epsilon^{ij} E_j,
\eqn\esixthirtyone
$$
so that we obtain the Hall conductance as
$$
\sigma_{xy} = {e^2\over 2\pi(2k+1)} \eqn\esixthirtytwo
$$
which agrees with the semiclassical answer for the Hall conductivity
$\sigma_{xy} =e\rho/B$  given in Eq.$\esixfive$, since $\rho/B =
e\nu/2\pi$ and $\nu = 1/(2k+1)$.  Hence, the LGCS theory `explains' the
stability of the FQHE states at the Laughlin fractions  and gives the
right Hall conductivity at these densities.

The charm of the LGCS theory lies in the fact that besides reproducing the
phenomenology of the FQHE, it also provides a formalism for addressing
questions like the existence of an order parameter and off-diagonal long
range order in the system.  The full implications of the LGCS theory are
yet to be understood, although the theory has been taken considerably
further\LGCSTWO.  This is an active field of research and many more questions
remain, both to be formulated and answered.

\vskip 1cm
\centerline{Problems}
\item{1.} One of the points which was glossed over in the lecture was the
question of how states get `pinned' or `localised' by impurities. Just to
get a feel for this point, solve the quantum mechanical problem of
a single fermion in a magnetic field and in the presence of an impurity 
scattering potential $V_I = \lambda \delta(x-x_0) \delta(y-y_0)$. (If you
get stuck, look up Ref.\PRANGE.)

\item{2.} A subtlety. \hfill\break
Fermions with attached flux tubes of strength $2k\phi_0$ remain fermions
because the statistical parameter $\alpha = ({e\over 2}) \phi = ({e\over 2})
2k ({2\pi\over e}) = 2k\pi$ is defined only modulo $2\pi$. So why are
`composite fermions' different from ordinary fermions and lead to
different physics (FQHE vs. IQHE)?

\endpage
\chapter{Conclusion}

We conclude this review by pointing out that these lectures are merely
meant to serve as an eye-opener to the ever-expanding field of anyon
physics. Besides, the several `known' unanswered questions in the field,
there probably remain many more unexplored and unexpected connections
between CS theories and other topological and non-topological field
theories. Applications of anyon ideas to other phenomena in condensed
matter physics also remain a distinct possibility. Our hope is to inspire
many more readers to join the `anyon bandwagon'.

\ack
I am grateful to D. M. Gaitonde, D. Sen and A. Sen for several useful
conversations. I also wish to thank D. M. Gaitonde, D. P. Jatkar, A.
Khare, D. Sen and A. Sen for reading the manuscript carefully.

\endpage

\refout

\end